\documentclass[prd,nofootinbib,showpacs,preprint]{revtex4}
\usepackage{amsmath}
\usepackage{graphicx}
\graphicspath{{Figs/}}
\usepackage{float}
\usepackage{dcolumn}
\usepackage{bm}
\usepackage{amssymb}
\usepackage[usenames,dvipsnames]{color}
\usepackage{slashed}
\usepackage[dvipdfm,colorlinks,citecolor=blue]{hyperref}
\begin{document}
\title{Discriminating Majorana Neutrino Textures in the light of Baryon Asymmetry}
\author{Manikanta Borah}
\email{mani@tezu.ernet.in}
\author{Debasish Borah}
\email{dborah@tezu.ernet.in}
\author{Mrinal Kumar Das}
\email{mkdas@tezu.ernet.in}
\affiliation{Department of Physics, Tezpur University, Tezpur-784028, India}

\begin{abstract}
We study all possible texture zeros in the Majorana neutrino mass matrix which are allowed from neutrino oscillation as well as cosmology data when the charged lepton mass matrix is assumed to take the diagonal form. In case of one-zero texture, we write down the Majorana phases which are assumed to be equal and the lightest neutrino mass as a function of the Dirac CP phase. In case of two-zero texture, we numerically evaluate all the three CP phases and lightest neutrino mass by solving four real constraint equations. We then constrain texture zero mass matrices from the requirement of producing correct baryon asymmetry through the mechanism of leptogenesis. Adopting a type I seesaw framework, we consider the CP violating out of equilibrium decay of the lightest right handed neutrino as the source of lepton asymmetry. Apart from discriminating between the texture zero mass matrices and light neutrino mass hierarchy, we also constrain the Dirac and Majorana CP phases so that the observed baryon asymmetry can be produced. In two-zero texture, we further constrain the diagonal form of Dirac neutrino mass matrix from the requirement of producing correct baryon asymmetry.
\end{abstract}

\maketitle

\section{INTRODUCTION}
\label{sec:intro}
Origin of sub-eV scale neutrino masses and large leptonic mixing is one of the biggest unresolved mysteries in particle physics. Due to the absence of right handed neutrinos in the standard model (SM), neutrinos remain massless at renormalizable level. Several beyond standard model (BSM) frameworks have been proposed to explain tiny neutrino masses observed by neutrino oscillation experiments more than a decade ago \cite{Fukuda:2001nk,Ahmad:2002jz,Ahmad:2002ka,Bahcall:2004mz}. More recently, the experiments like T2K \cite{Abe:2011sj}, Double Chooz \cite{Abe:2011fz}, Daya-Bay \cite{An:2012eh} and RENO \cite{Ahn:2012nd} have not only confirmed the earlier measurements but also discovered a small but non-zero reactor mixing angle. Two different sets of latest global fit values for $3\sigma$ range of neutrino oscillation parameters given in \cite{Gonzalez-Garcia:2014bfa} and \cite{Forero:2014bxa} are shown in table \ref{tab:data1} and \ref{tab:data2} respectively.
\begin{center}
\begin{table}[htb]
\begin{tabular}{|c|c|c|}
\hline
Parameters & Normal Hierarchy (NH) & Inverted Hierarchy (IH) \\
\hline
$ \frac{\Delta m_{21}^2}{10^{-5} \text{eV}^2}$ & $7.02-8.09$ & $7.02-8.09 $ \\
$ \frac{|\Delta m_{31}^2|}{10^{-3} \text{eV}^2}$ & $2.317-2.607$ & $2.307-2.590 $ \\
$ \sin^2\theta_{12} $ &  $0.270-0.344 $ & $0.270-0.344 $ \\
$ \sin^2\theta_{23} $ & $0.382-0.643$ &  $0.389-0.644 $ \\
$\sin^2\theta_{13} $ & $0.0186-0.0250$ & $0.0188-0.0251 $ \\
$ \delta_{CP} $ & $0-2\pi$ & $0-2\pi$ \\
\hline
\end{tabular}
\caption{Global fit $3\sigma$ values of neutrino oscillation parameters \cite{Gonzalez-Garcia:2014bfa}}
\label{tab:data1}
\end{table}
\end{center}
\begin{center}
\begin{table}[htb]
\begin{tabular}{|c|c|c|}
\hline
Parameters & Normal Hierarchy (NH) & Inverted Hierarchy (IH) \\
\hline
$ \frac{\Delta m_{21}^2}{10^{-5} \text{eV}^2}$ & $7.11-8.18$ & $7.11-8.18 $ \\
$ \frac{|\Delta m_{31}^2|}{10^{-3} \text{eV}^2}$ & $2.30-2.65$ & $2.20-2.54 $ \\
$ \sin^2\theta_{12} $ &  $0.278-0.375 $ & $0.278-0.375 $ \\
$ \sin^2\theta_{23} $ & $0.393-0.643$ &  $0.403-0.640 $ \\
$\sin^2\theta_{13} $ & $0.0190-0.0262$ & $0.0193-0.0265 $ \\
$ \delta_{CP} $ & $0-2\pi$ & $0-2\pi$ \\
\hline
\end{tabular}
\caption{Global fit $3\sigma$ values of neutrino oscillation parameters \cite{Forero:2014bxa}}
\label{tab:data2}
\end{table}
\end{center}
Although the $3\sigma$ range for the leptonic Dirac CP phase $\delta_{CP}$ is $0-2\pi$, there are two possible best fit values of it found in the literature: $306^o$ (NH), $254^o$ (IH) \cite{Gonzalez-Garcia:2014bfa} and $254^o$ (NH), $266^o$ (IH) \cite{Forero:2014bxa}. If neutrinos are Majorana fermions whose masses are generated by conventional seesaw mechanism \cite{Minkowski:1977sc,GellMann:1980vs,Yanagida:1979as,Mohapatra:1979ia,Schechter:1980gr}, then two Majorana phases also appear in the mixing matrix which do not affect neutrino oscillations and hence can not be measured by oscillation experiments. The Majorana phases can however, have interesting implications in lepton number violating process like neutrinoless double beta decay, origin of matter-antimatter asymmetry of the Universe etc. Apart from the mass squared differences and mixing angles, the sum of the absolute neutrino masses are also tightly constrained from cosmology $\sum_i \lvert m_i \rvert < 0.23$ eV \cite{Ade:2013zuv}.

One of the most popular BSM framework to understand the origin of tiny neutrino mass and large leptonic mixing is to identify the possible underlying symmetries. Symmetries can either relate two or more free parameters of the model or make them vanish, making the model more predictive. The widely studied $\mu-\tau$ symmetric neutrino mass matrix giving $\theta_{13}=0$ is one such scenario where discrete flavor symmetries can relate two or more terms in the neutrino mass matrix. Non-zero $\theta_{13}$, as required by latest oscillation data, can be generated by incorporating different possible corrections to leading order $\mu-\tau$ symmetric neutrino mass matrix, as discussed recently in many works including \cite{Borah:2013jia,Borah:2013lva,Borah:2014fga,Borah:2014bda,Kalita:2014mga}. The other possible role symmetries can play is to impose texture zeros in the mass matrices. The symmetry realization of such texture zeros can be found in several earlier as well as recent works 
\cite{Berger:2000zj,Low:2004wx,Low:2005yc,Grimus:2004hf,Dighe:2009xj,Dev:2011jc,Felipe:2014vka}. Recently, a systematic study of texture zeros in lepton mass matrices were done in \cite{Ludl:2014axa}. In the simplest case, one can assume the charged lepton mass matrix to be diagonal and then consider the possible texture zeros in the symmetric Majorana neutrino mass matrix. It turns out that in this simplest case, only certain types of one-zero texture and two-zero textures in the Majorana neutrino mass matrix are consistent with neutrino data.

In this work, we consider all types of texture zeros allowed in the Majorana neutrino mass matrix (in the diagonal charged lepton basis) from neutrino oscillation data and constrain them further from the requirement of producing successful baryon asymmetry through the mechanism of leptogenesis. Some earlier works related to the calculation of lepton asymmetry with texture zero Majorana neutrino mass matrix can be found in \cite{Kaneko:2002yp,Kaneko:2003cy,Dev:2010vy,Bando:2004hi}. Leptogenesis is one of the most widely studied formalism which provides a dynamical origin of the observed baryon asymmetry in the Universe. The asymmetry is created in the leptonic sector first which later gets converted into baryon asymmetry through $B+L$ violating electroweak sphaleron transitions \cite{Kuzmin:1985mm}. As pointed out first by Fukugita and Yanagida \cite{Fukugita:1986hr}, the required lepton asymmetry can be generated by the out of equilibrium CP violating decay of heavy Majorana neutrinos which are present in several BSM frameworks which attempt to explain tiny SM neutrino masses. We consider the framework of type I seesaw mechanism \cite{Minkowski:1977sc,GellMann:1980vs,Yanagida:1979as,Mohapatra:1979ia,Schechter:1980gr} generating tiny SM neutrino masses where right handed neutrinos are present and discriminate between different texture zeros in the neutrino mass matrix from the requirement of producing the correct baryon asymmetry seen by Planck experiment \cite{Ade:2013zuv}
\begin{equation}
Y_B = (8.58 \pm 0.22) \times 10^{-11}
\label{barasym}
\end{equation} 
Usually, the Majorana neutrino mass matrix can be constrained from the neutrino oscillation data on two mass squared differences and three mixing angles. But the most general neutrino mass matrix can still contain those neutrino parameters which are not yet determined experimentally: the lightest neutrino mass, leptonic Dirac CP phase and two Majorana CP phases. All these four free parameters can in general, affect the resulting lepton asymmetry calculated from the lightest right handed neutrino decay. Although it is difficult to make predictions with four free parameters, in case of texture zero Majorana mass matrix, it is possible to reduce the number of free parameters. As we show in details in this work, two of these free neutrino parameters can be determined in terms of the other two in one-zero texture case whereas all four free parameters can be numerically determined in case of two-zero texture mass matrices. To simplify the calculation, we assume equality of Majorana phases in one-zero texture case and write down all the free parameters in the neutrino mass matrix in terms of Dirac CP phase. We then compute the baryon asymmetry as a function of Dirac CP phase. We not only constrain the Dirac CP phase from the requirement of producing the observed baryon asymmetry but also show that some of the texture zeros (allowed from neutrino oscillation data) are disfavored if leptogenesis through the lightest right handed neutrino decay is the only source of baryon asymmetry. Since all the neutrino parameters are fixed in two-zero texture mass matrices, we compute baryon asymmetry for different choices of Dirac neutrino mass matrices. Thus we not only discriminate between different two-zero texture mass matrices, but also constrain the Dirac neutrino mass matrices from the requirement of producing the observed baryon asymmetry.

This paper is organized as follows. In section \ref{sec:texture}, we discuss all the possible texture zeros in the Majorana neutrino mass matrix with diagonal charged lepton basis. In section \ref{sec:lepto}, we briefly outline the mechanism of leptogenesis through right handed neutrino decay. In section \ref{sec:numeric}, we discuss the numerical analysis of all the texture zero models and finally conclude in section \ref{sec:conclude}.

\section{Majorana Texture Zeros}
\label{sec:texture}
A symmetric $3\times 3$ Majorana neutrino mass matrix $M_{\nu}$ can have six independent parameters. If $k$ of them are vanishing then the total number of structurally different Majorana mass matrices with texture zeros is
\begin{equation}
\label{prmtn}
^6C_k=\frac{6!}{k!(6-k)!}
\end{equation}
A symmetric mass matrix with more than 3 texture zeros $k\geq 4$ can not be compatible with lepton masses and mixing. Similarly in the diagonal charged lepton basis, a symmetric Majorana neutrino mass matrix with 3 texture zeros is not compatible with neutrino oscillation data \cite{Xing:2004ik}. Therefore, we are left with either one-zero texture which can be of six different types and two-zero texture which can be of fifteen different types. Different possible Majorana neutrino mass matrices with one-zero texture and one vanishing eigenvalue was studied by the authors of \cite{Xing:2003ic} whereas one-zero texture in the light of recent neutrino oscillation data with non-zero $\theta_{13}$ was discussed in the work \cite{Lashin:2011dn,Deepthi:2011sk}. Implications of one-zero texture for neutrinoless double beta decay can be found in \cite{Merle:2006du}. Two-zero textures in the Majorana neutrino mass matrix have received lots of attention in several works in the last few years, some of which can be found in \cite{Frampton:2002yf,Xing:2002ta,Xing:2002ap,Kageyama:2002zw,Grimus:2004az,Dev:2006qe,Ludl:2011vv,Kumar:2011vf,Fritzsch:2011qv,Meloni:2012sx,Meloni:2014yea,Dev:2014dla}. We briefly discuss these texture zero Majorana neutrino mass matrices in the following subsections  \ref{one} and \ref{two} respectively.
\subsection{One-zero texture}
\label{one}
In case of one-zero texture, the Majorana neutrino mass matrix $M_{\nu}$ contains only one independent zero. There are six possible patterns of such one-zero texture which, following the notations of \cite{Deepthi:2011sk} can be written as
\begin{center}

$ G_1 :\left(\begin{array}{ccc}
0& \times&\times\\
\times& \times&\times \\
\times& \times&\times 
\end{array}\right) , G_2 :\left(\begin{array}{ccc}
\times& 0&\times\\
0& \times&\times \\
\times& \times&\times 
\end{array}\right) , G_3 :\left(\begin{array}{ccc}
\times& \times&0\\
\times& \times&\times \\
0 & \times&\times 
\end{array}\right) ,   G_4 :\left(\begin{array}{ccc}
\times& \times&\times\\
\times & 0 &\times \\
\times& \times&\times 
\end{array}\right)  ,$
 
\end{center}
\begin{equation}
G_5 :\left(\begin{array}{ccc}
\times& \times&\times\\
\times& \times & 0 \\
\times& 0 &\times 
\end{array}\right) , 
 G_6 :\left(\begin{array}{ccc}
\times& \times&\times\\
\times& \times& \times \\
\times& \times & 0
\end{array}\right) 
\end{equation}

Where the crosses ``$\times$'' denote non-zero arbitrary elements of $M_{\nu}$.

\begin{figure}
\centering
\begin{minipage}{.5\textwidth}
  \centering
  \includegraphics[width=1\linewidth]{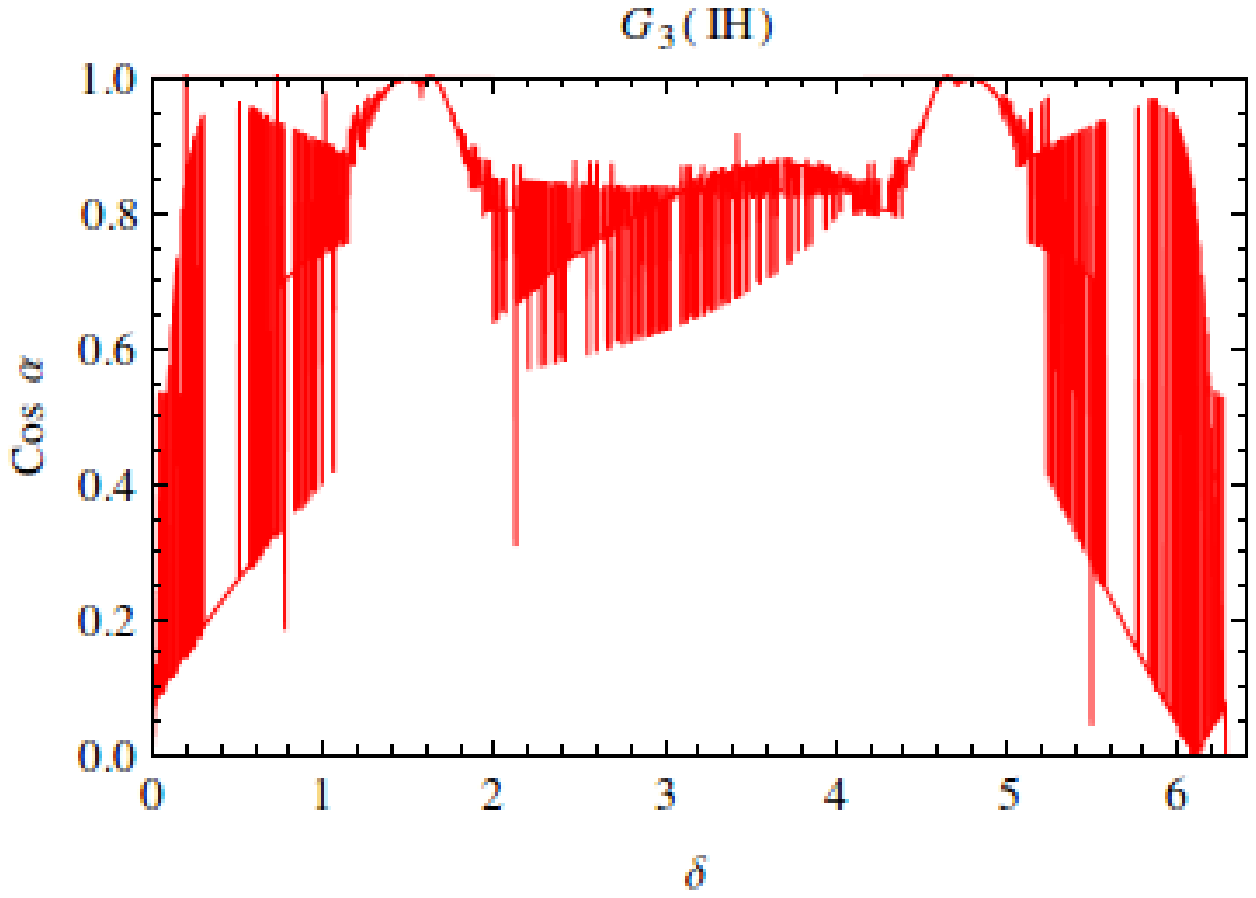}
\end{minipage}%
\begin{minipage}{.5\textwidth}
  \centering
  \includegraphics[width=1\linewidth]{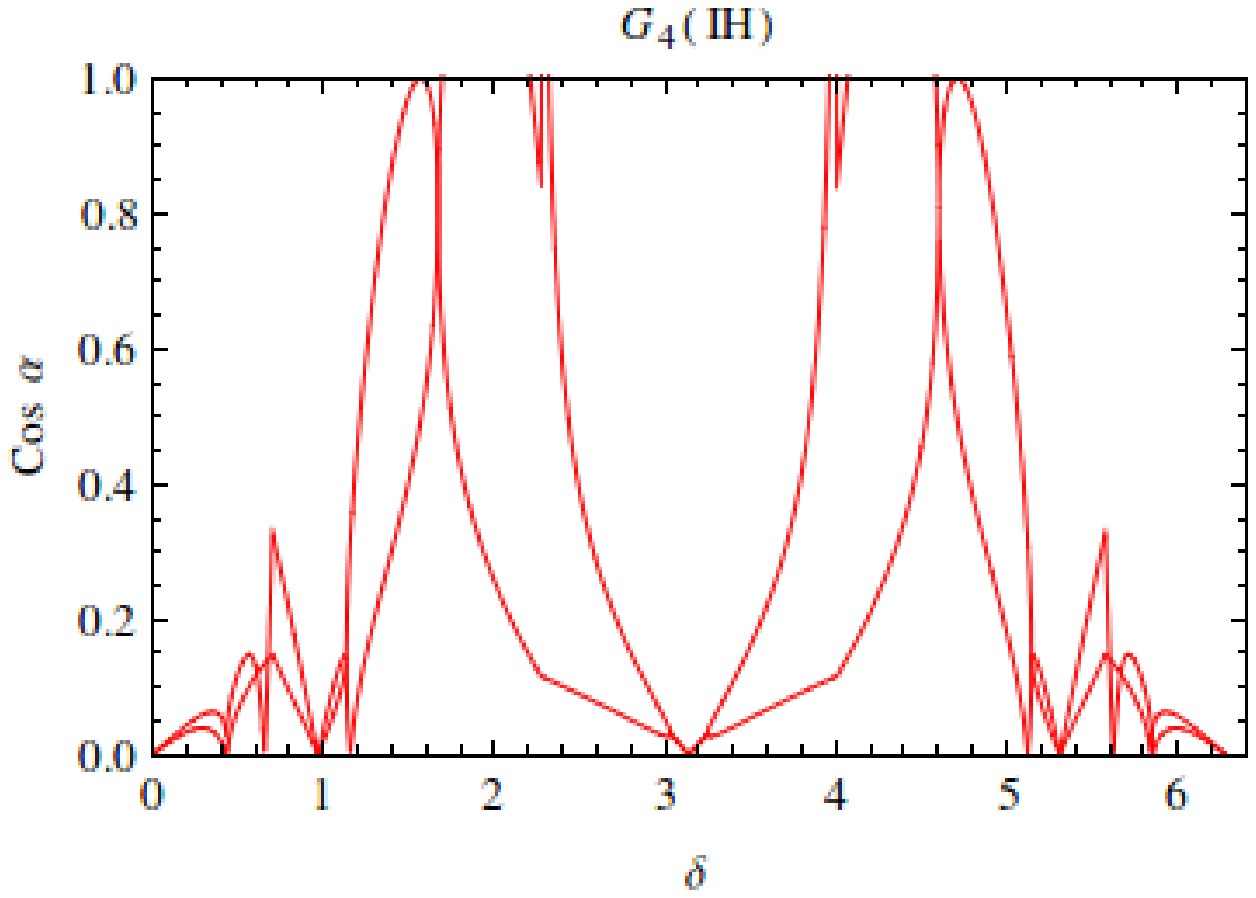}
\end{minipage}
\begin{minipage}{.5\textwidth}
  \centering
  \includegraphics[width=1\linewidth]{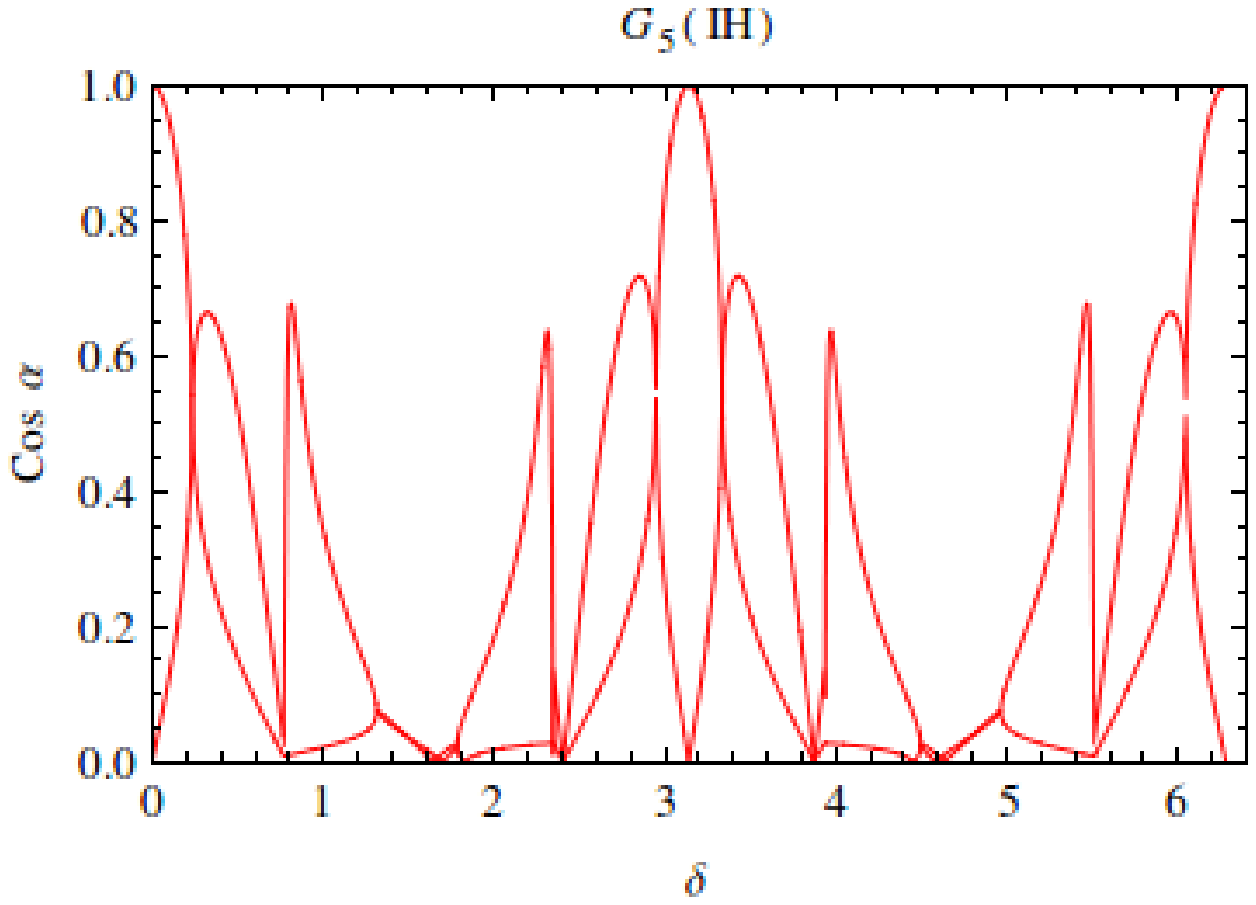}
\end{minipage}%
\begin{minipage}{.5\textwidth}
  \centering
  \includegraphics[width=1\linewidth]{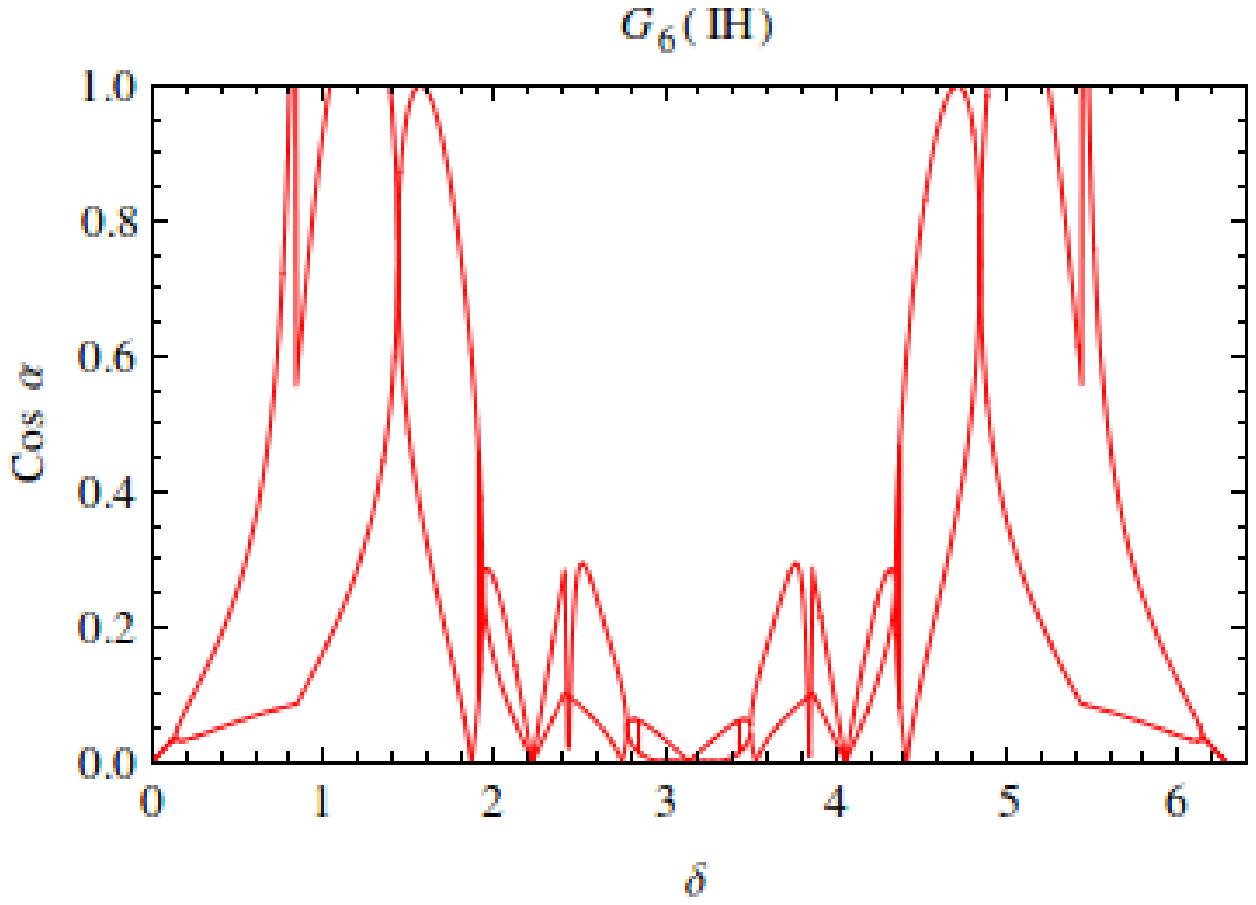}
\end{minipage}
\caption{Variation of $\cos \alpha$ with $\delta$ for one-zero texture with inverted hierarchy.}
\label{fig001}
\end{figure}

\begin{figure}
\centering
\begin{minipage}{.5\textwidth}
  \centering
  \includegraphics[width=1\linewidth]{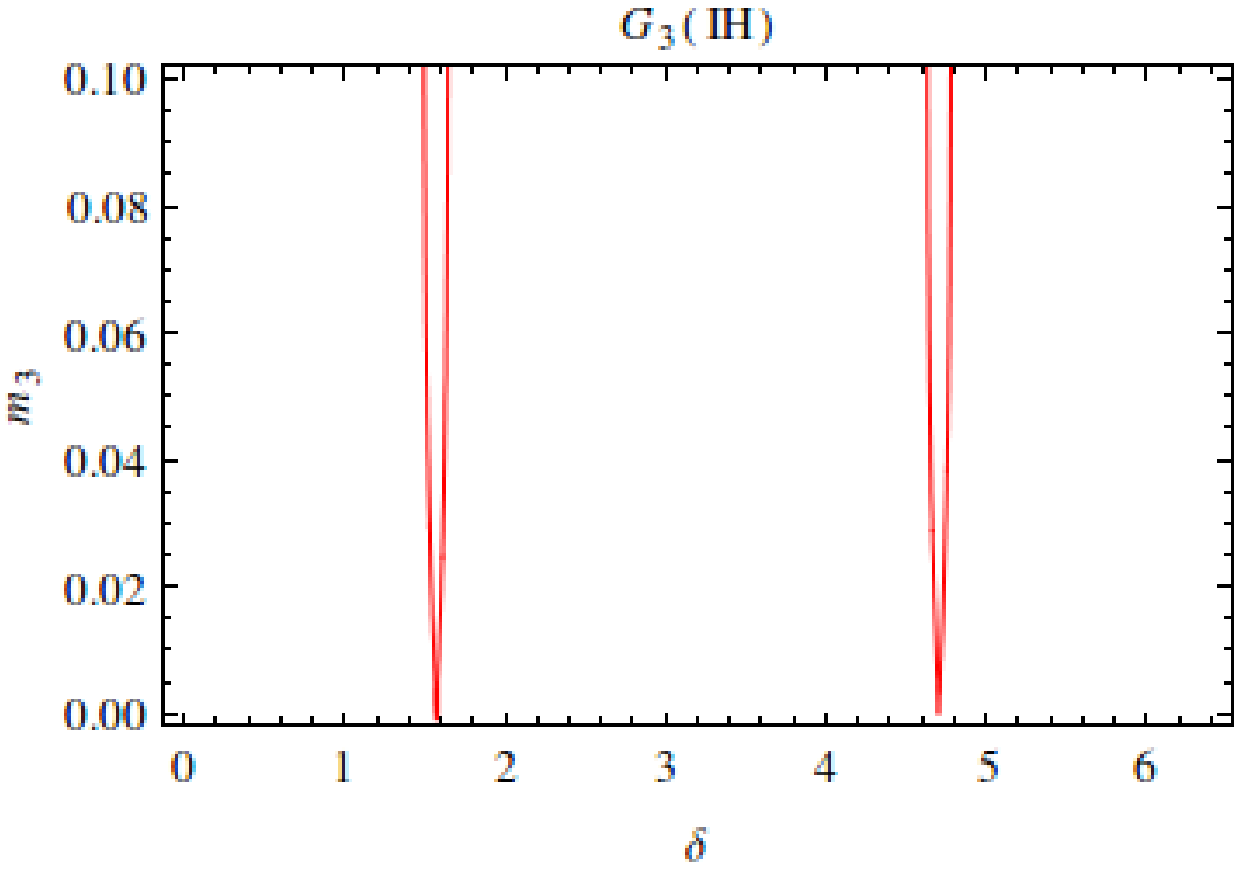}
\end{minipage}%
\begin{minipage}{.5\textwidth}
  \centering
  \includegraphics[width=1\linewidth]{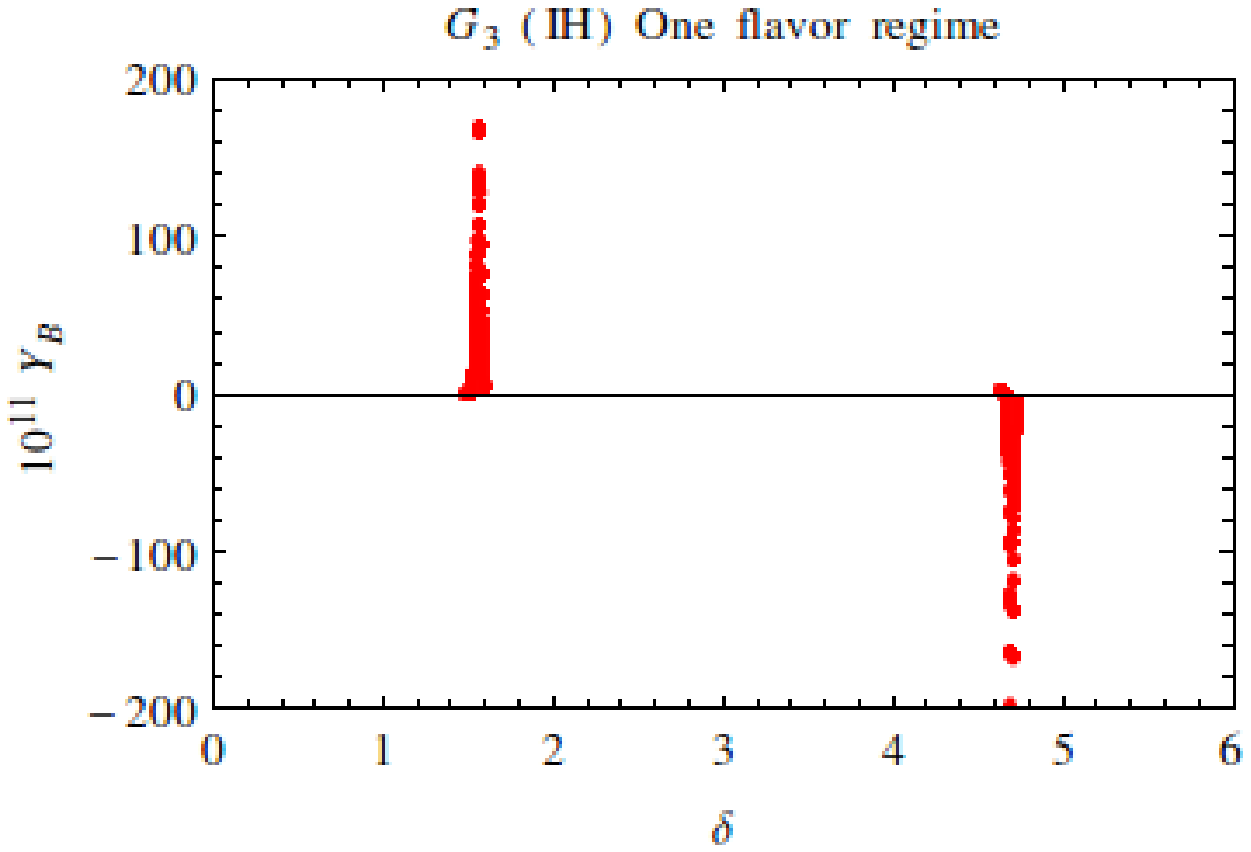}
\end{minipage}
\begin{minipage}{.5\textwidth}
  \centering
  \includegraphics[width=1\linewidth]{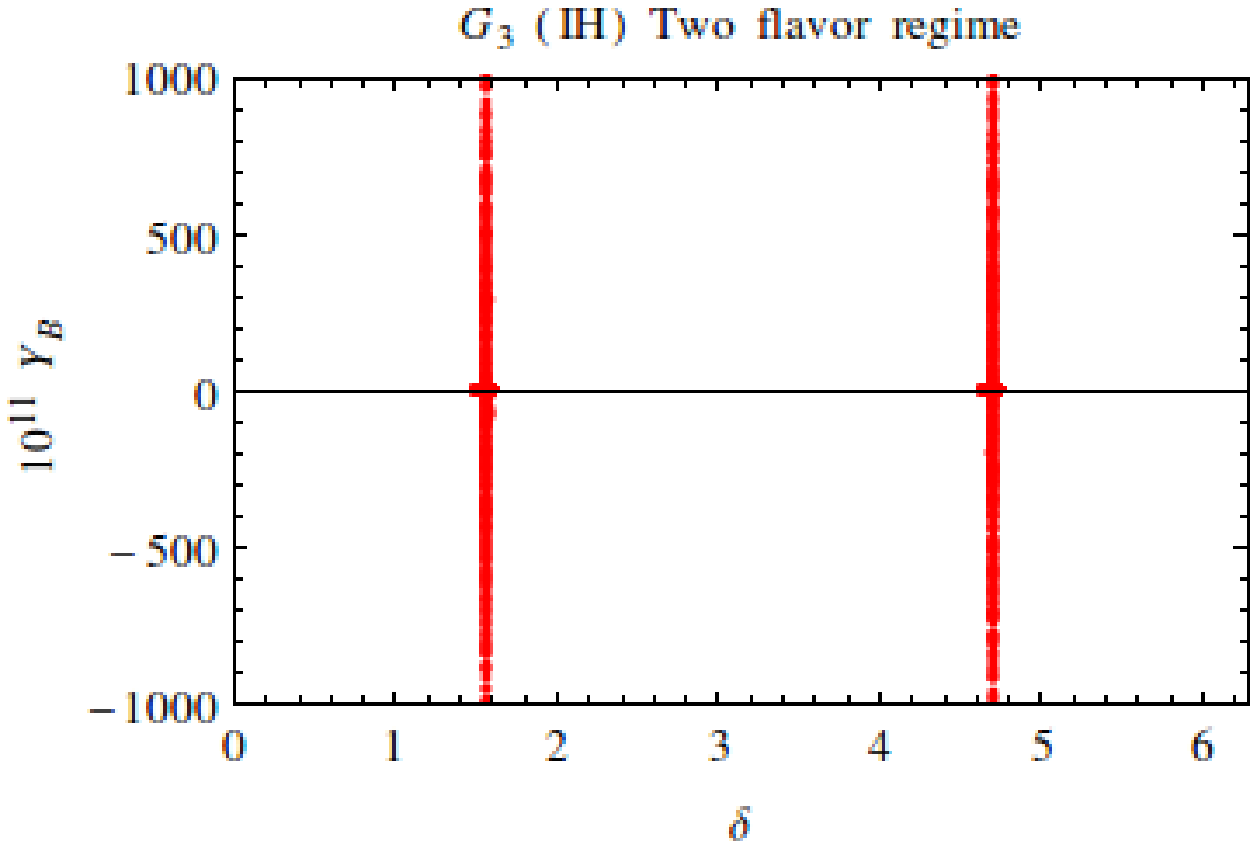}
\end{minipage}%
\begin{minipage}{.5\textwidth}
  \centering
  \includegraphics[width=1\linewidth]{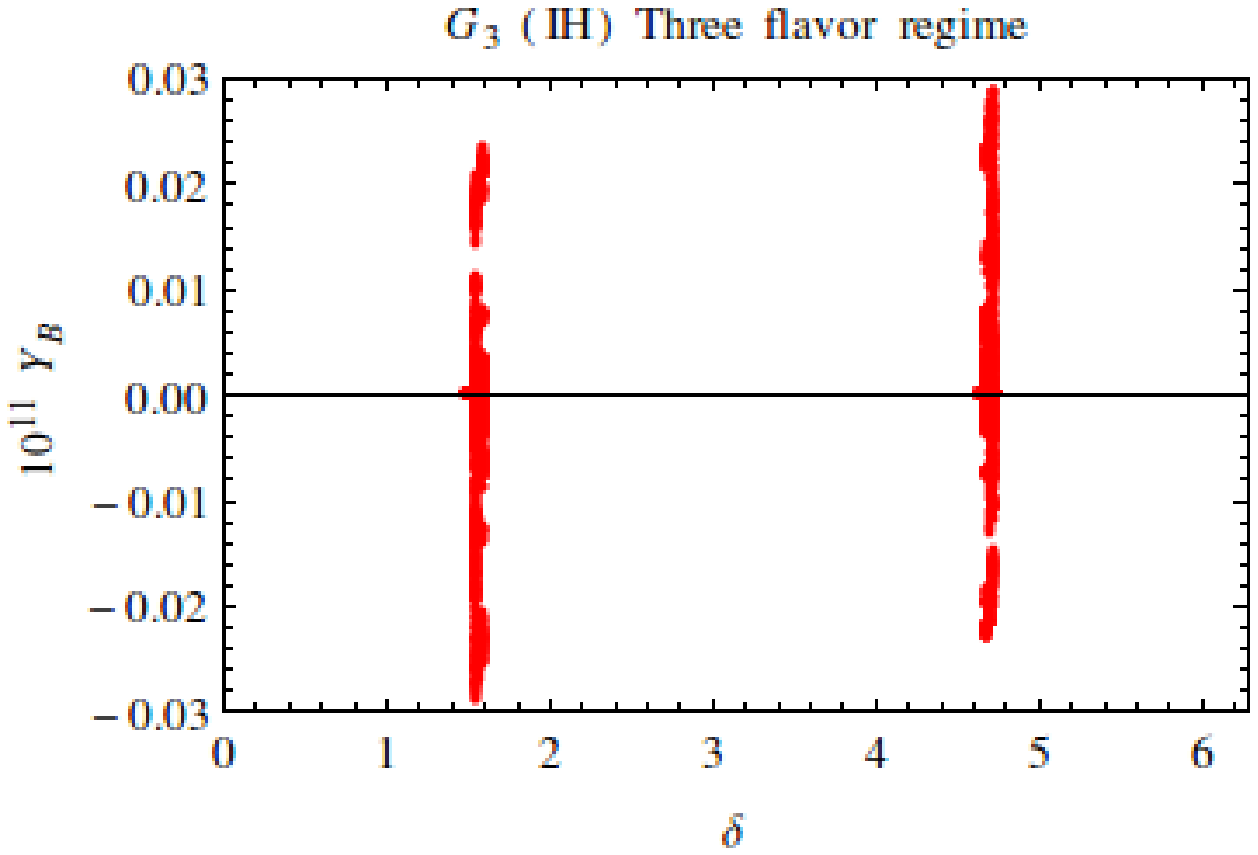}
\end{minipage}
\caption{Variation of lightest neutrino mass $m_3$ and baryon asymmetry with Dirac CP phase $\delta$ for one-zero texture $G_3$ with inverted hierarchy.}
\label{fig1}
\end{figure}

\begin{figure}
\centering
\begin{minipage}{.5\textwidth}
  \centering
  \includegraphics[width=1\linewidth]{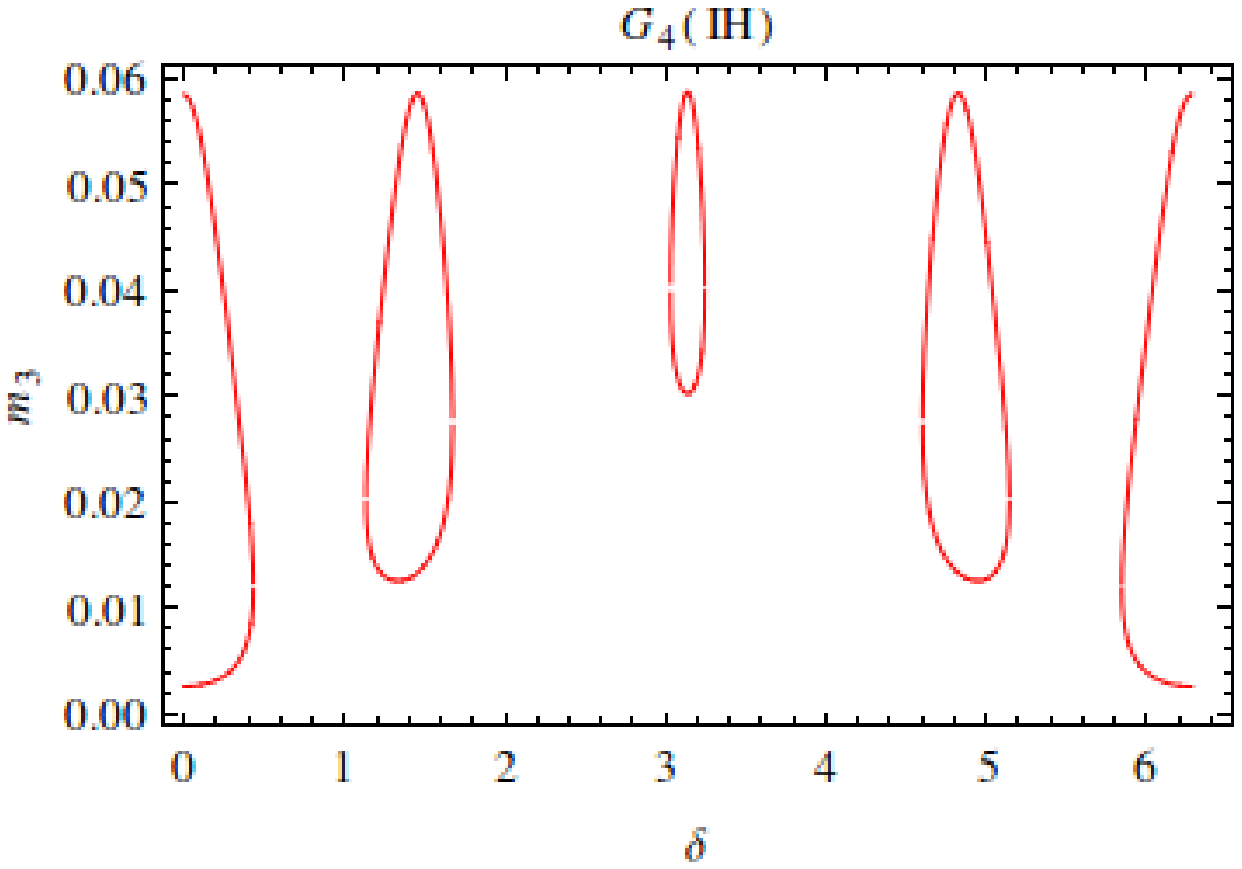}
\end{minipage}%
\begin{minipage}{.5\textwidth}
  \centering
  \includegraphics[width=1\linewidth]{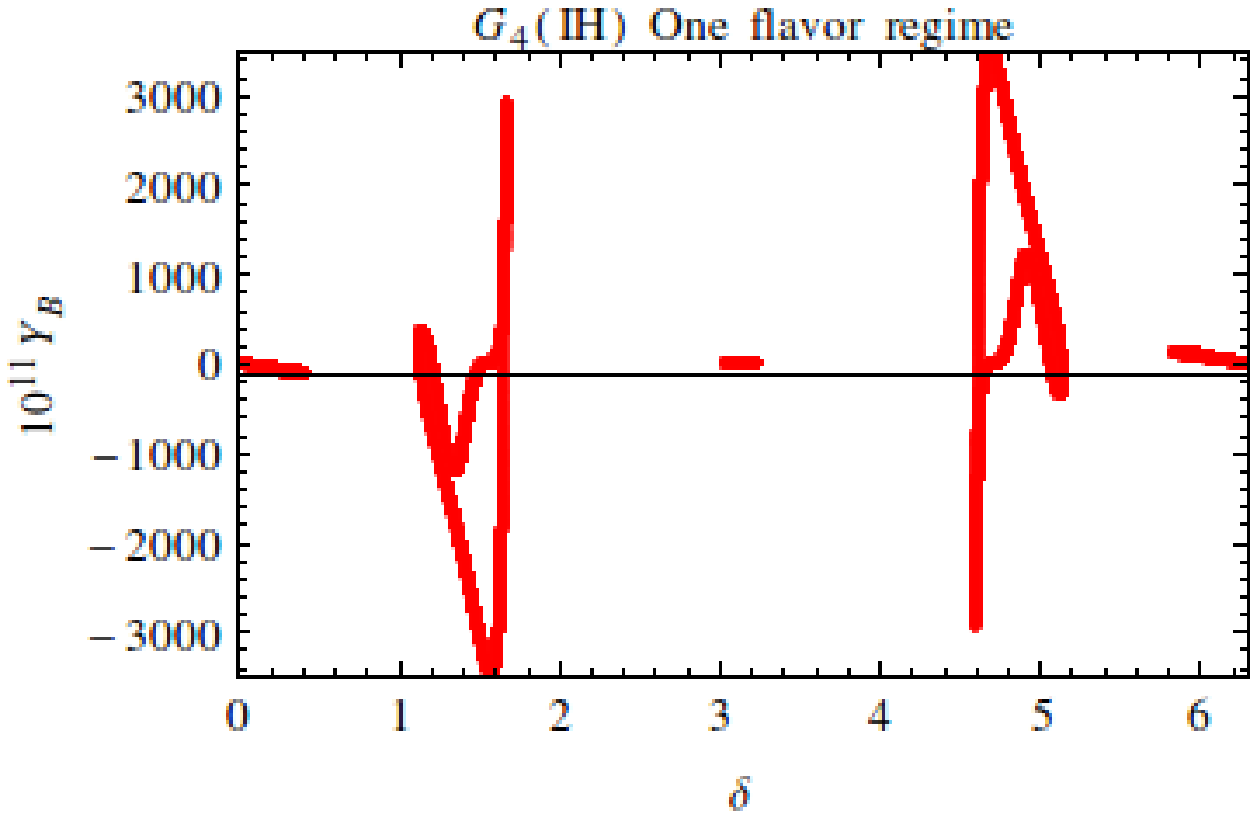}
\end{minipage}
\begin{minipage}{.5\textwidth}
  \centering
  \includegraphics[width=1\linewidth]{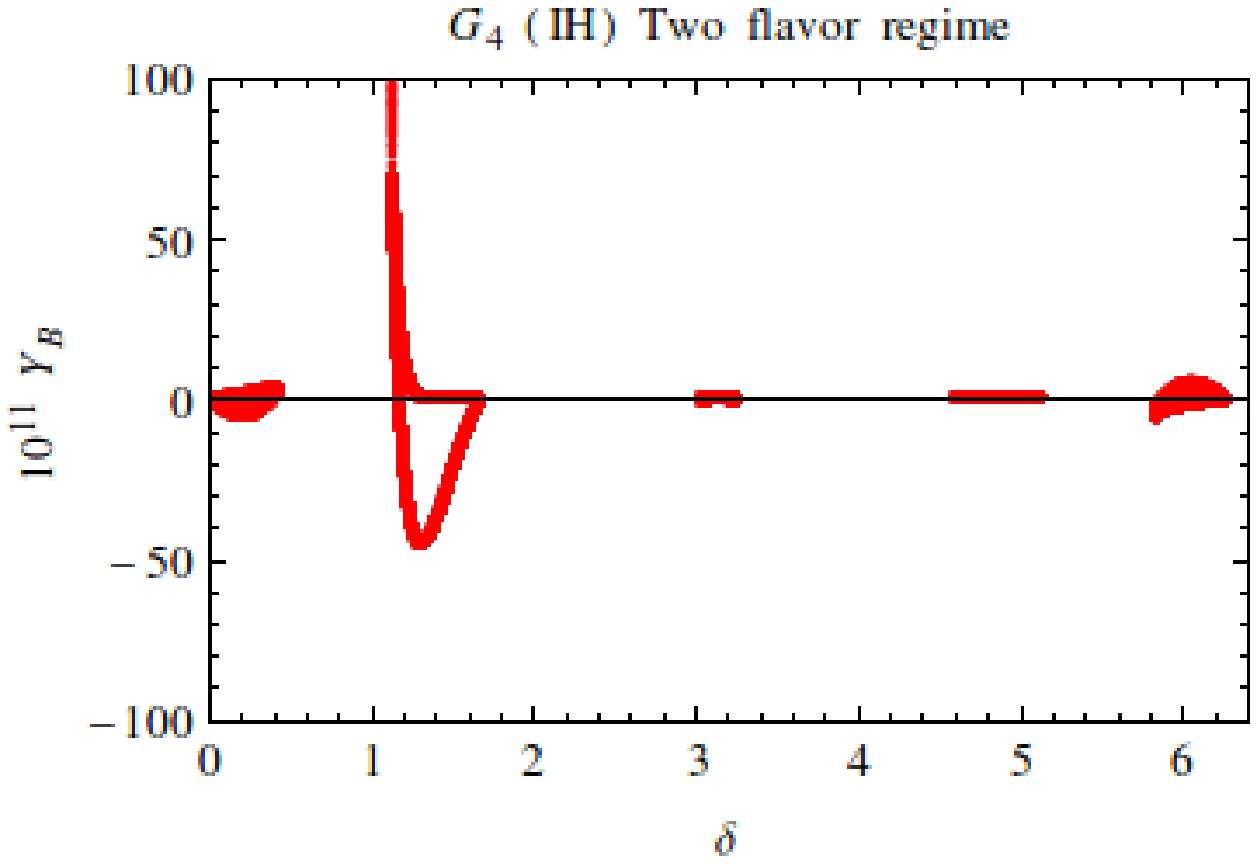}
\end{minipage}%
\begin{minipage}{.5\textwidth}
  \centering
  \includegraphics[width=1\linewidth]{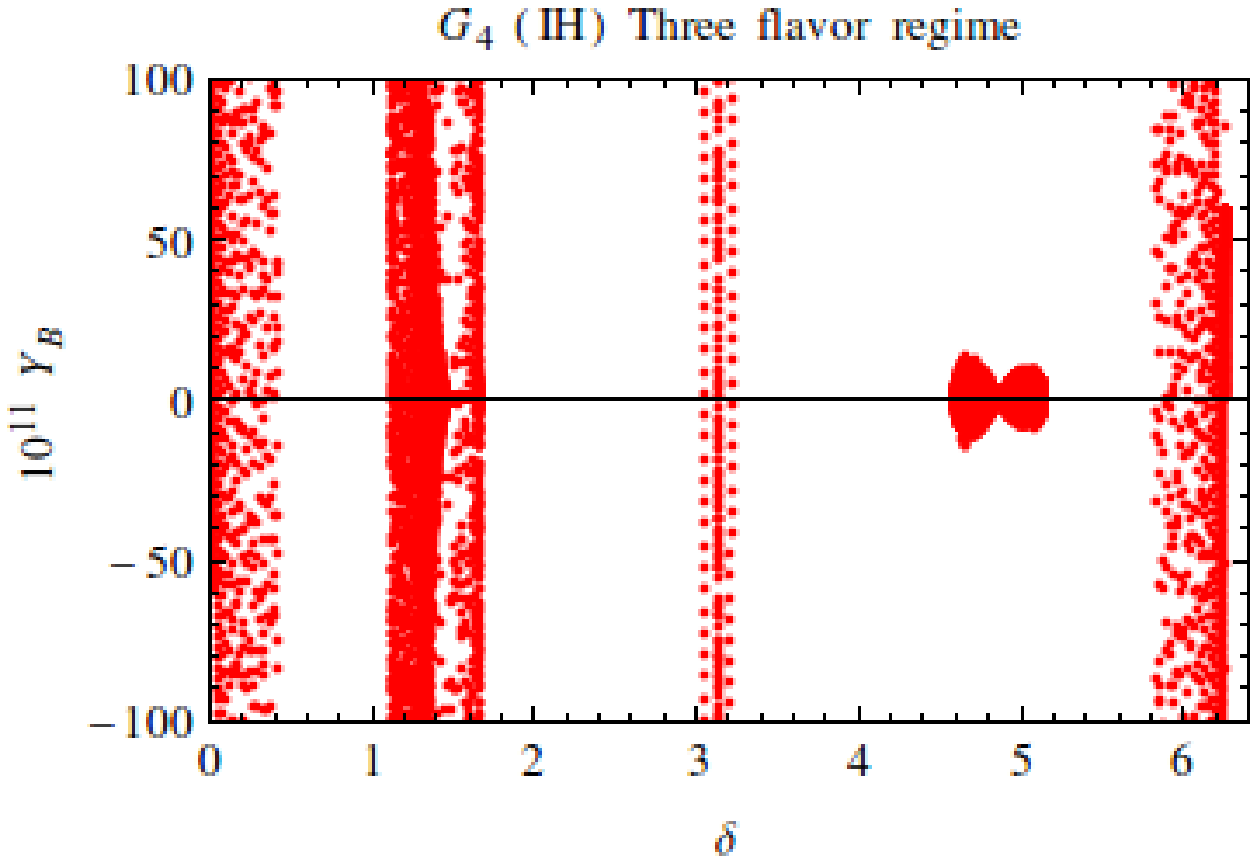}
\end{minipage}
\caption{Variation of lightest neutrino mass $m_3$ and baryon asymmetry with Dirac CP phase $\delta$ for one-zero texture $G_4$ with inverted hierarchy.}
\label{fig2}
\end{figure}

\begin{figure}
\centering
\begin{minipage}{.5\textwidth}
  \centering
  \includegraphics[width=1\linewidth]{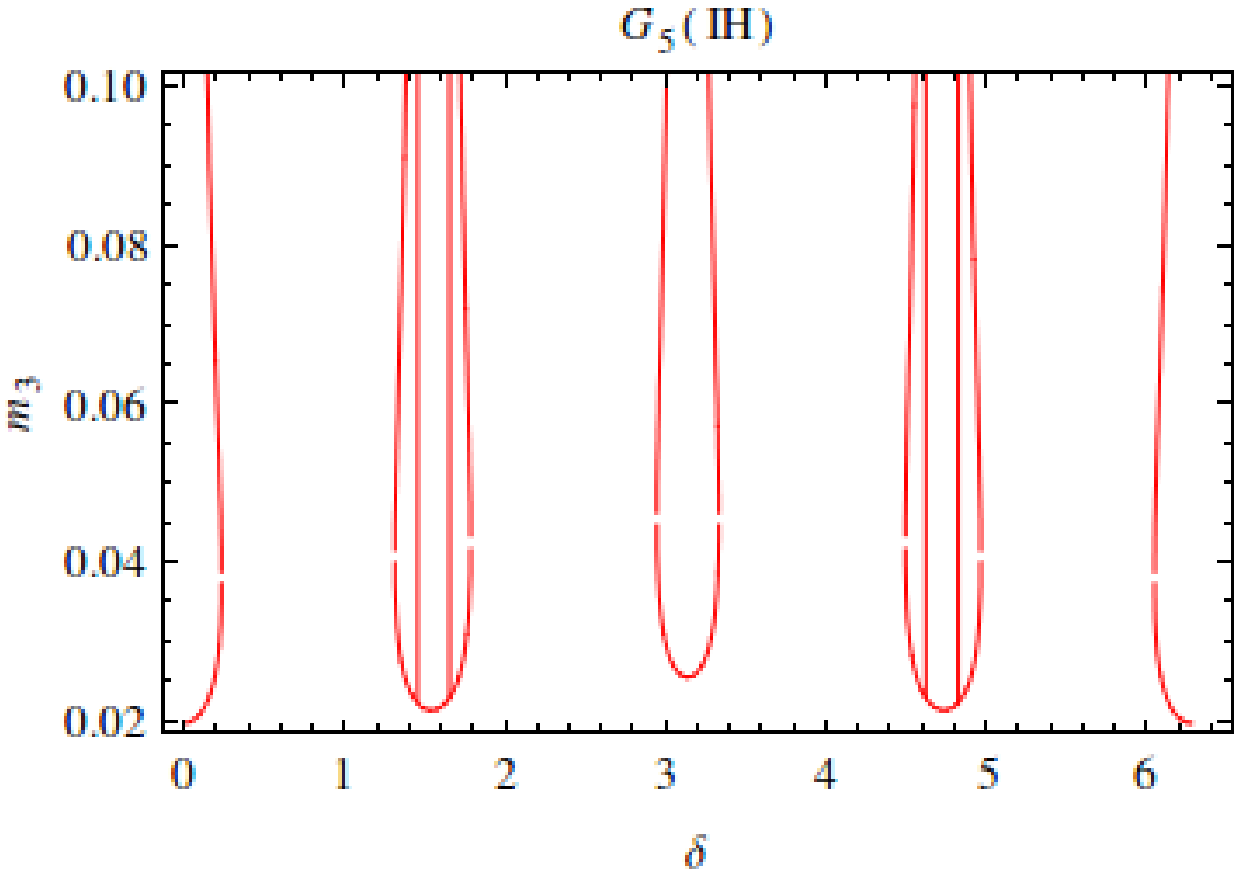}
\end{minipage}%
\begin{minipage}{.5\textwidth}
  \centering
  \includegraphics[width=1\linewidth]{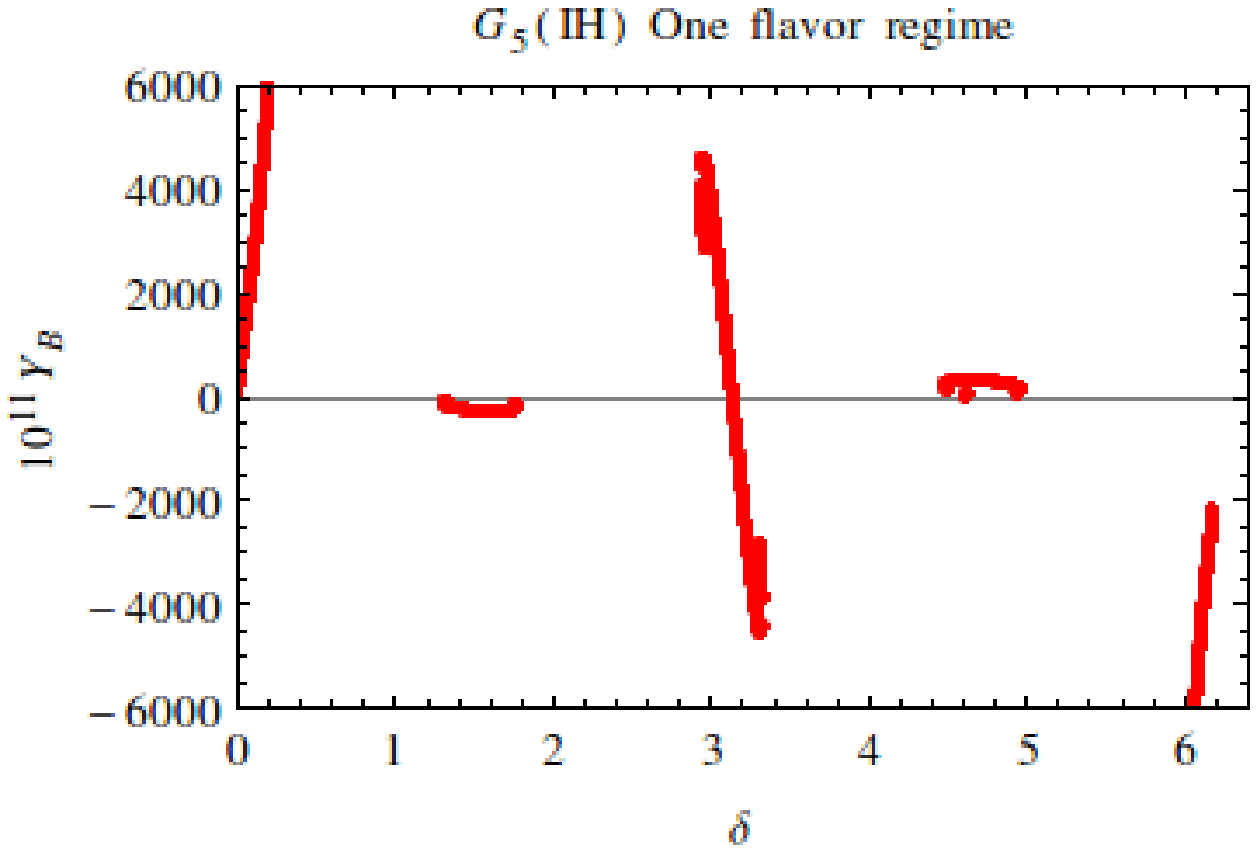}
\end{minipage}
\begin{minipage}{.5\textwidth}
  \centering
  \includegraphics[width=1\linewidth]{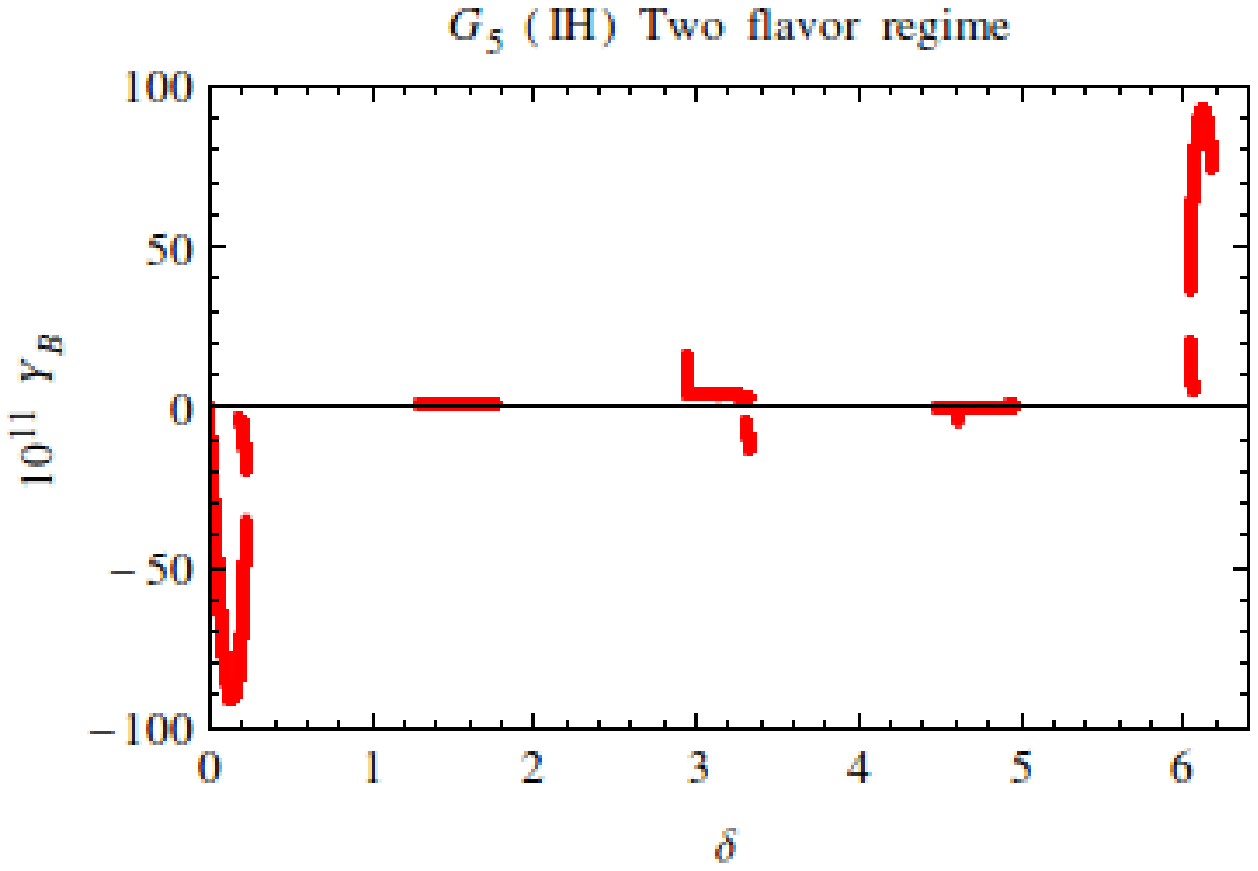}
\end{minipage}%
\begin{minipage}{.5\textwidth}
  \centering
  \includegraphics[width=1\linewidth]{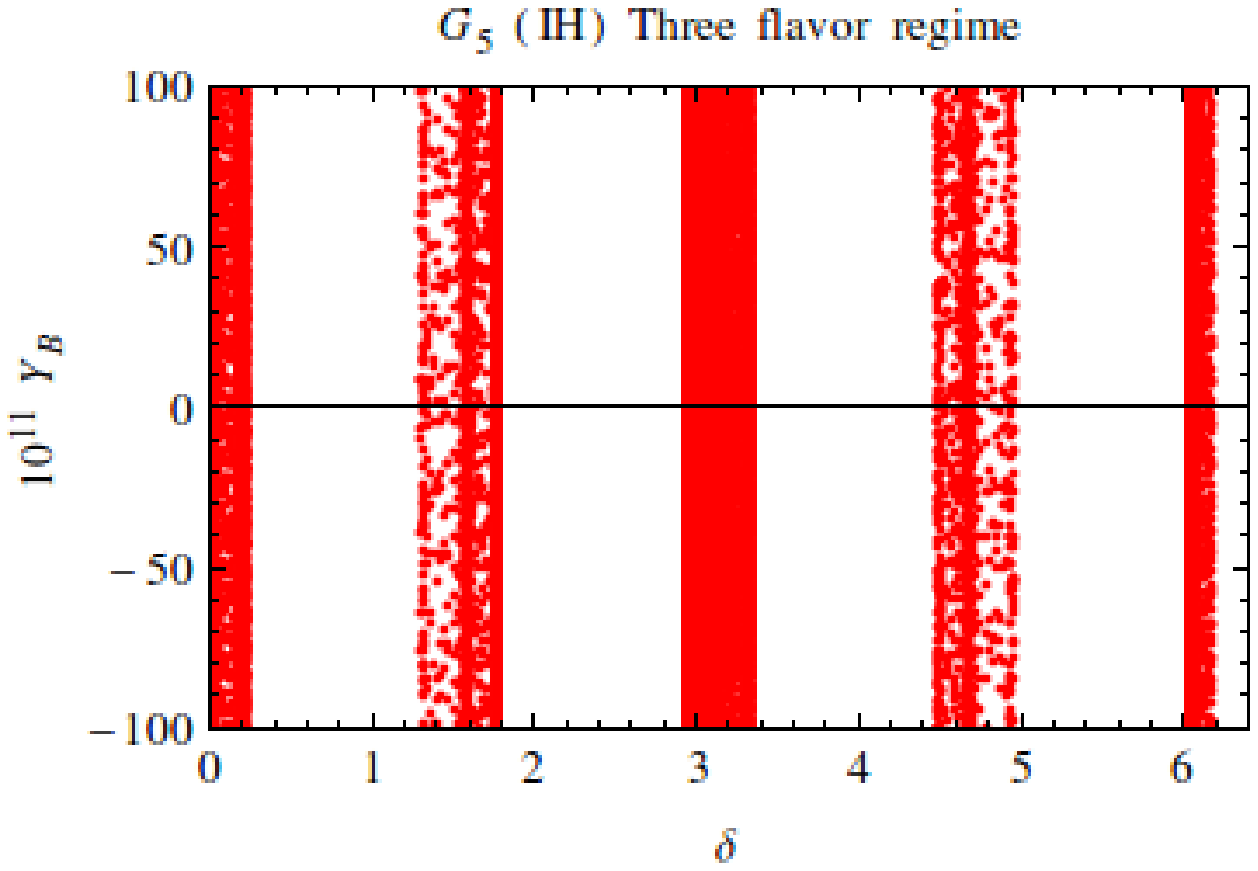}
\end{minipage}
\caption{Variation of lightest neutrino mass $m_3$ and baryon asymmetry with Dirac CP phase $\delta$ for one-zero texture $G_5$ with inverted hierarchy.}
\label{fig3}
\end{figure}

\begin{figure}
\centering
\begin{minipage}{.5\textwidth}
  \centering
  \includegraphics[width=1\linewidth]{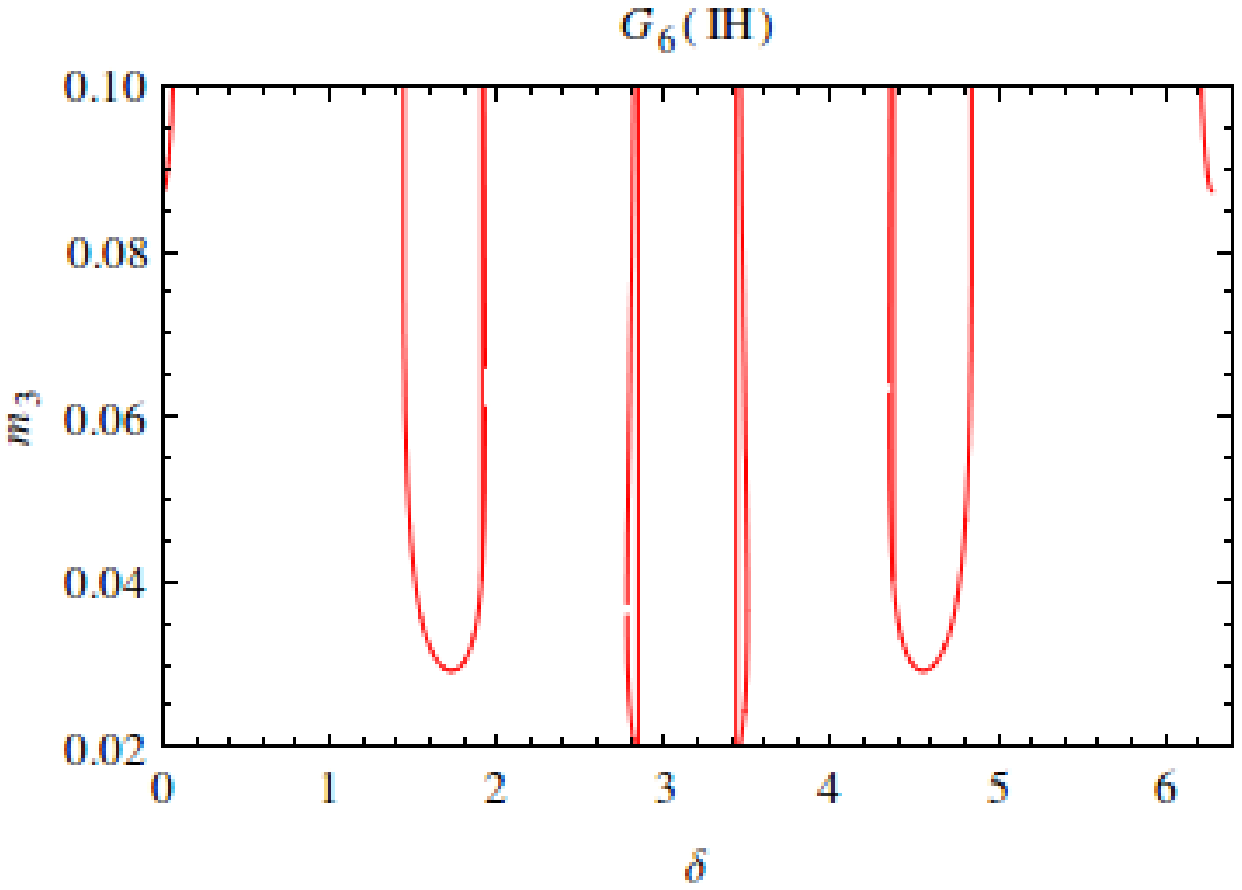}
\end{minipage}%
\begin{minipage}{.5\textwidth}
  \centering
  \includegraphics[width=1\linewidth]{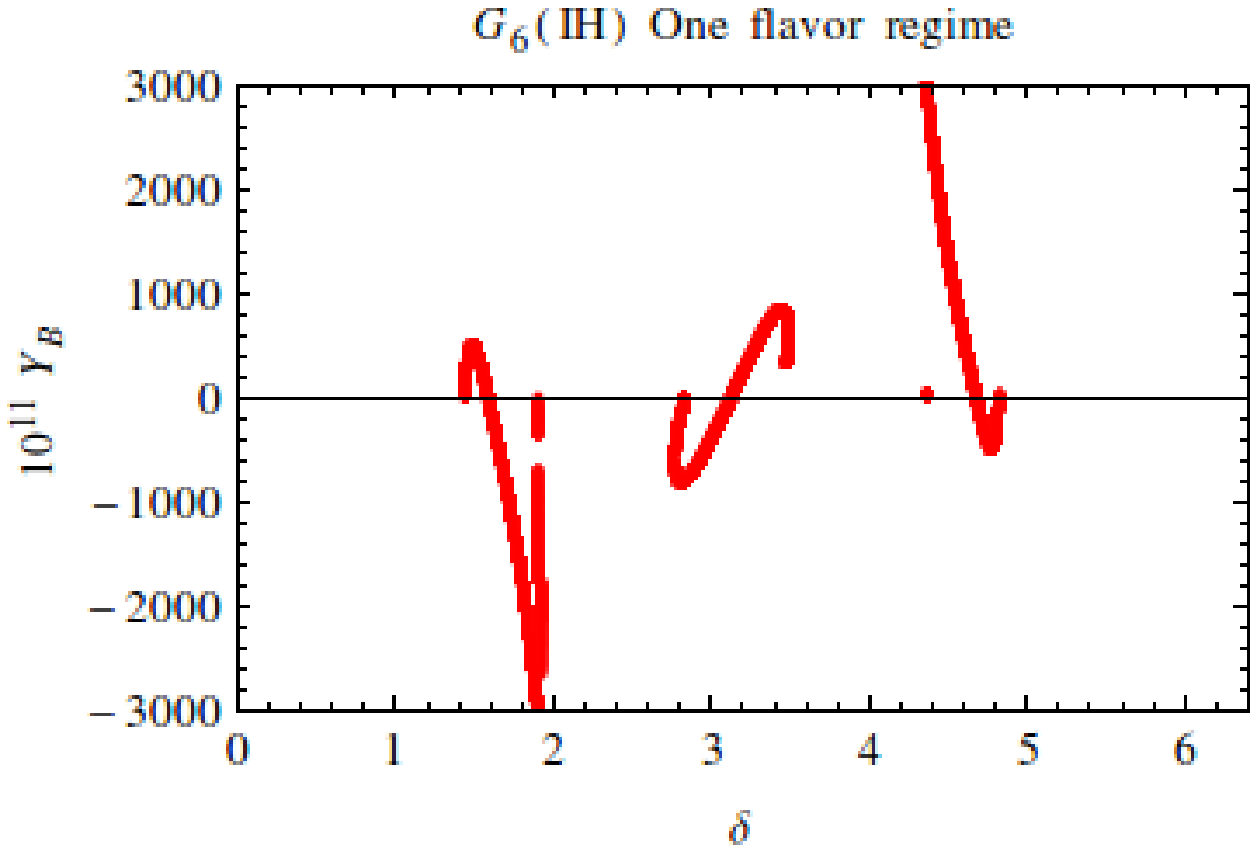}
\end{minipage}
\begin{minipage}{.5\textwidth}
  \centering
  \includegraphics[width=1\linewidth]{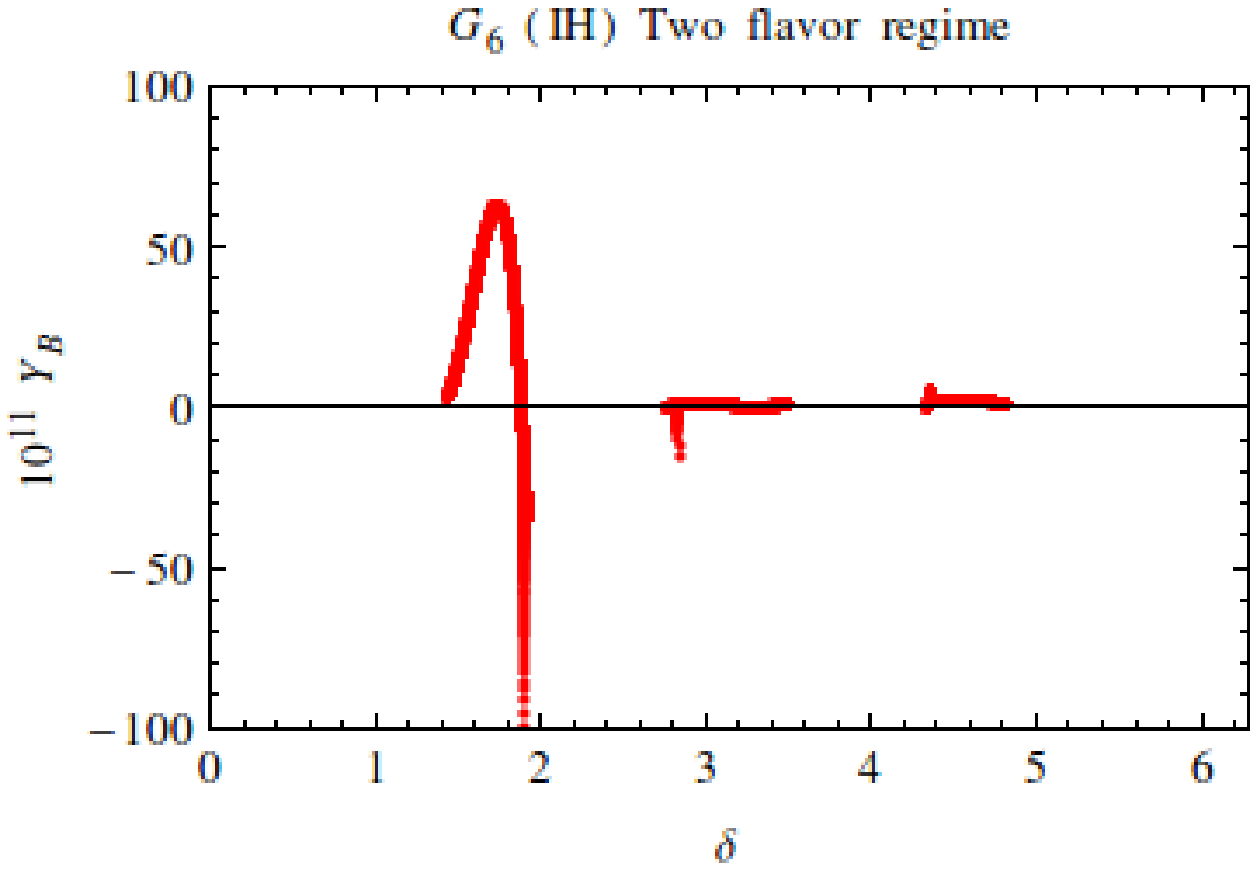}
\end{minipage}%
\begin{minipage}{.5\textwidth}
  \centering
  \includegraphics[width=1\linewidth]{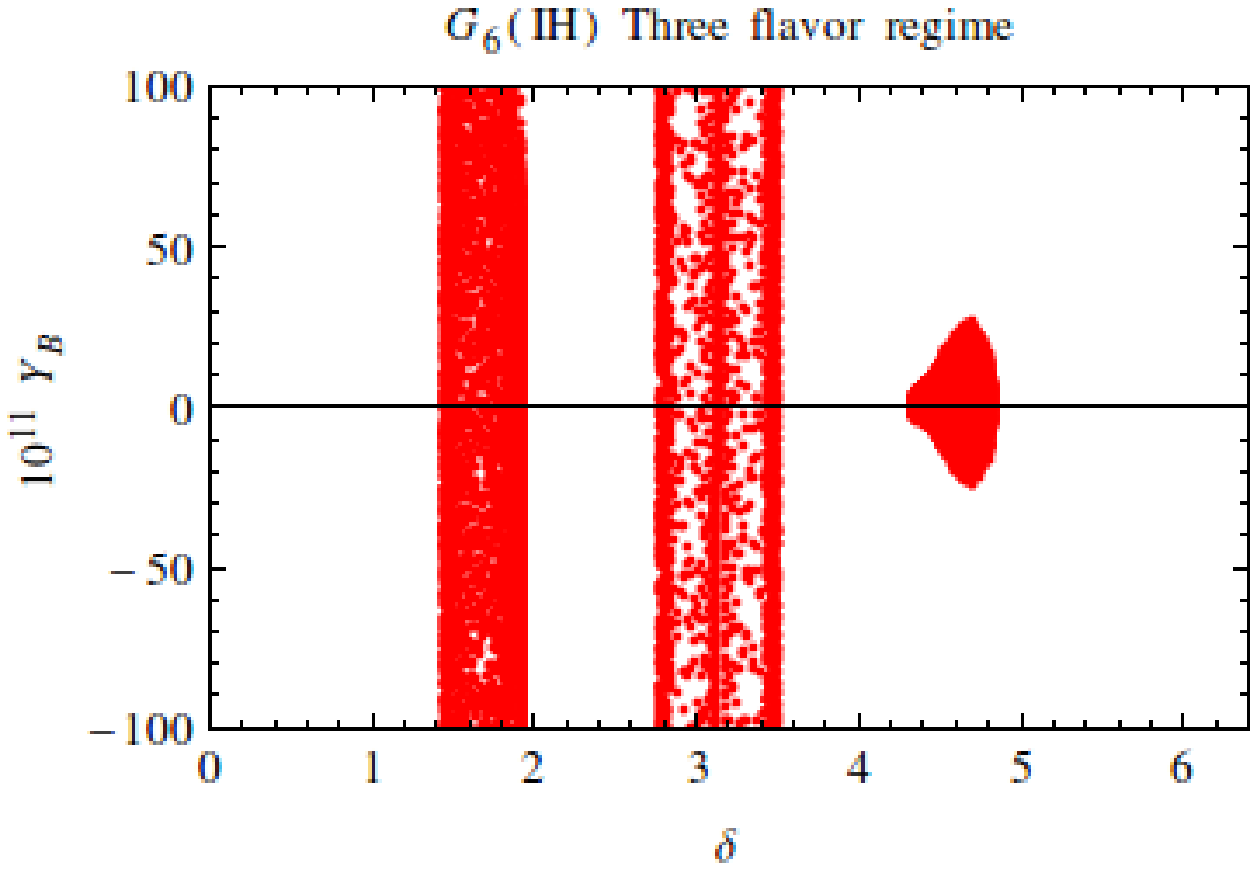}
\end{minipage}
\caption{Variation of lightest neutrino mass $m_3$ and baryon asymmetry with Dirac CP phase $\delta$ for one-zero texture $G_6$ with inverted hierarchy.}
\label{fig4}
\end{figure}

\begin{figure}
\centering
\begin{minipage}{.5\textwidth}
  \centering
  \includegraphics[width=1\linewidth]{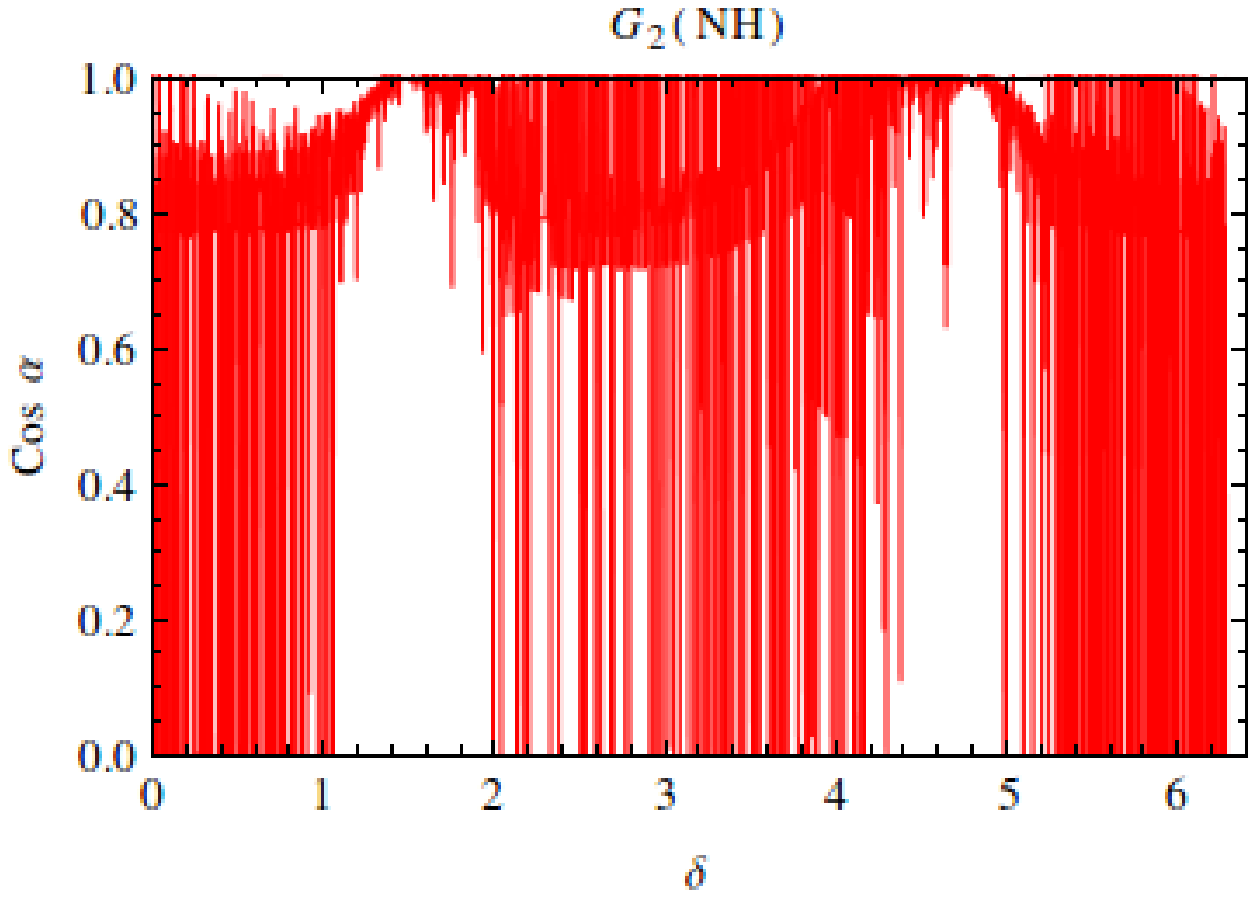}
\end{minipage}%
\begin{minipage}{.5\textwidth}
  \centering
  \includegraphics[width=1\linewidth]{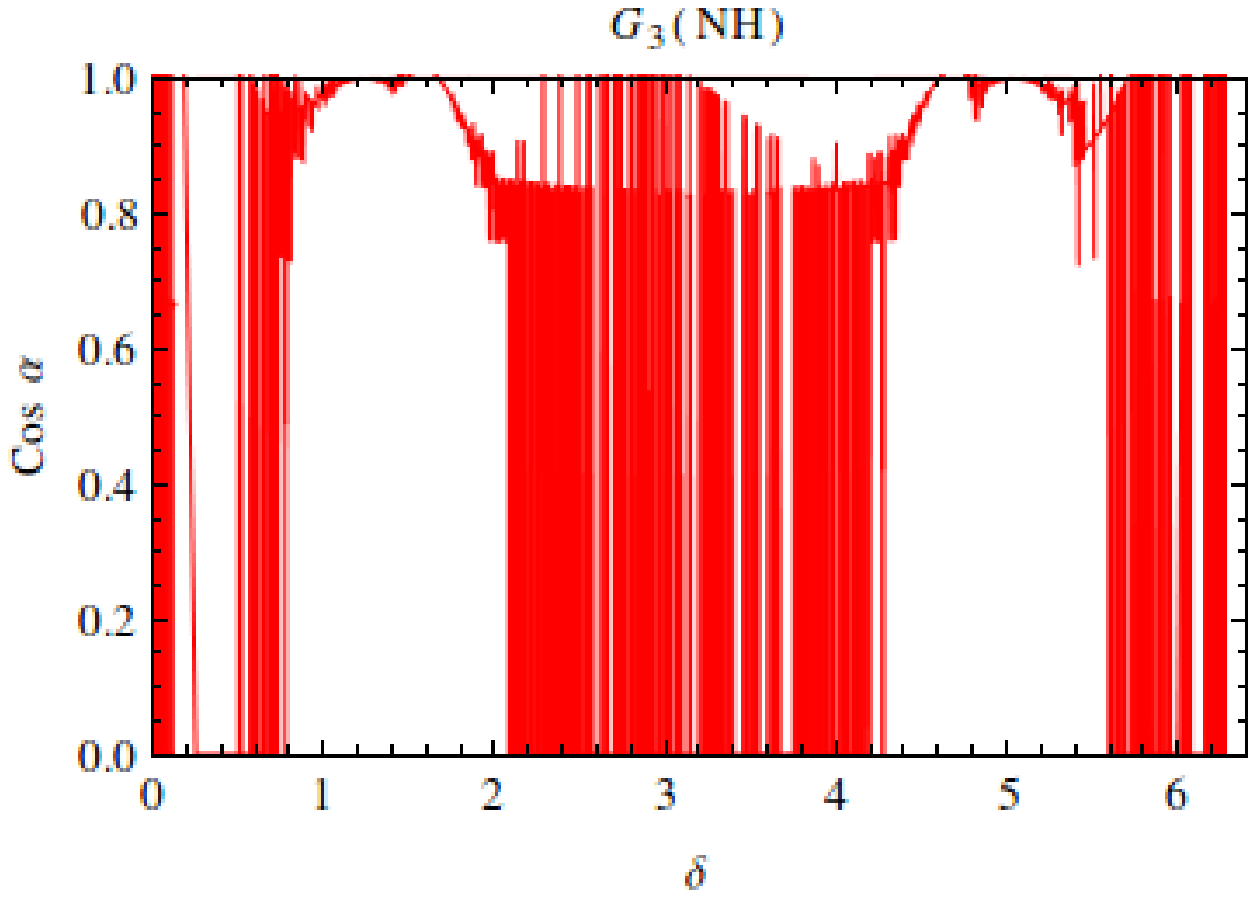}
\end{minipage}
\caption{Variation of $\cos \alpha$ with $\delta$ for one-zero texture with normal hierarchy.}
\label{fig01}
\end{figure}

\begin{figure}
\centering
\begin{minipage}{.5\textwidth}
  \centering
  \includegraphics[width=1\linewidth]{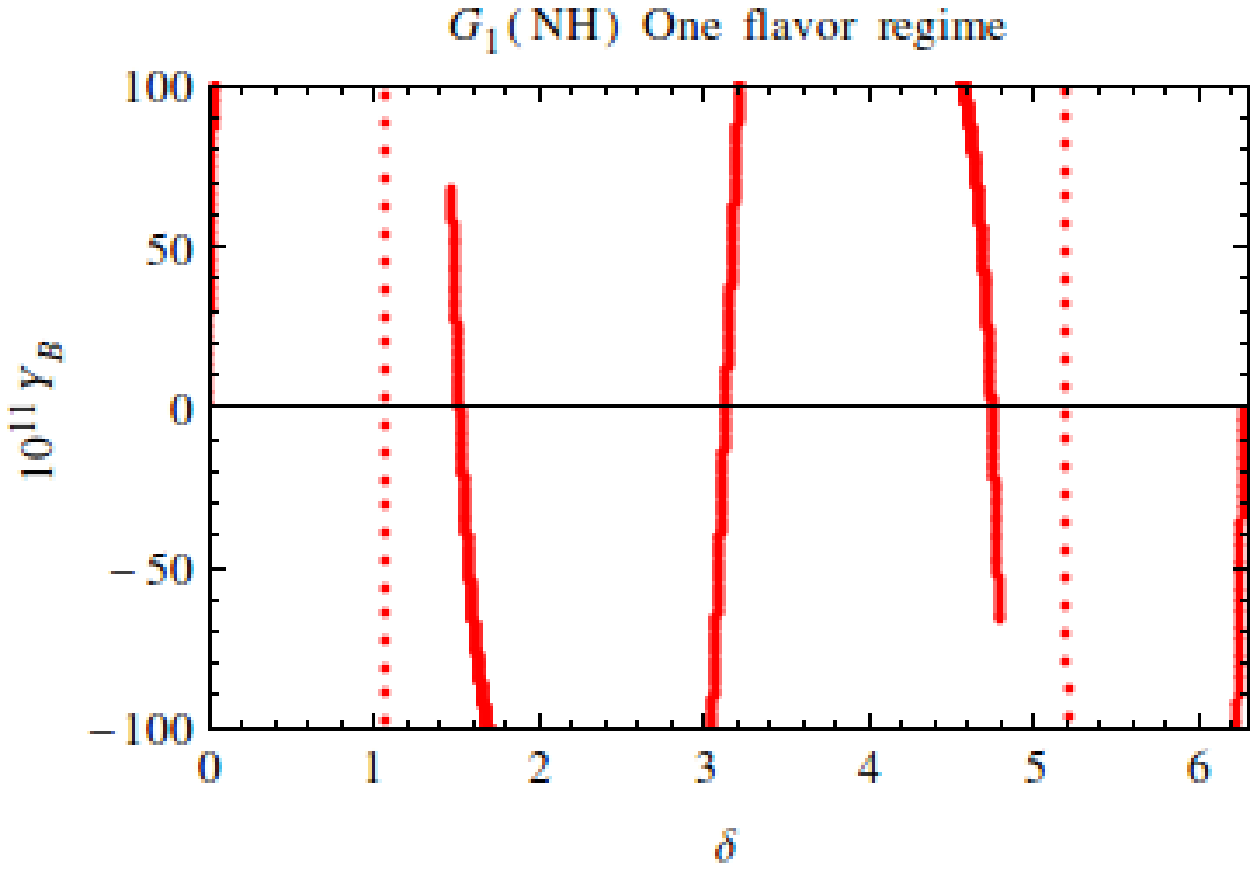}
\end{minipage}
\begin{minipage}{.5\textwidth}
  \centering
  \includegraphics[width=1\linewidth]{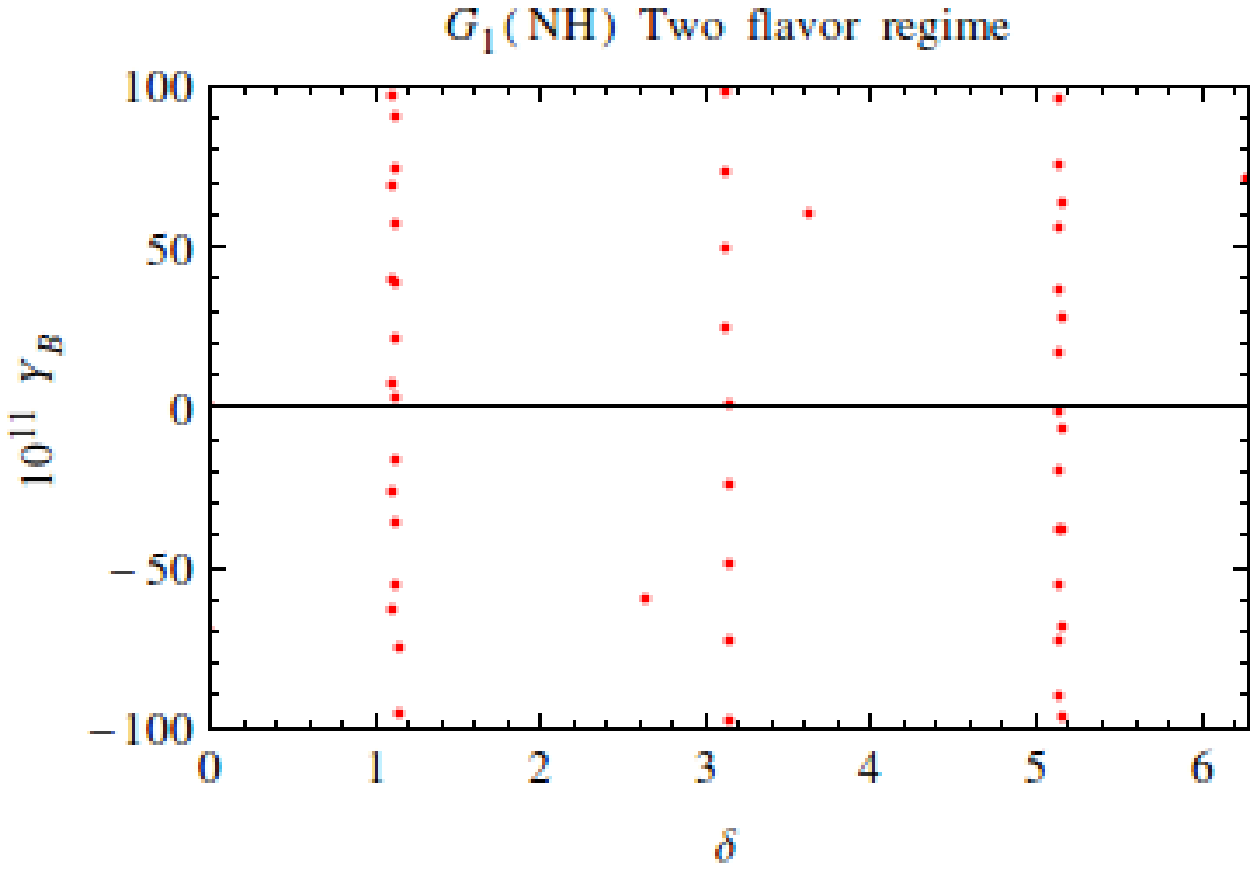}
\end{minipage}%
\begin{minipage}{.5\textwidth}
  \centering
  \includegraphics[width=1\linewidth]{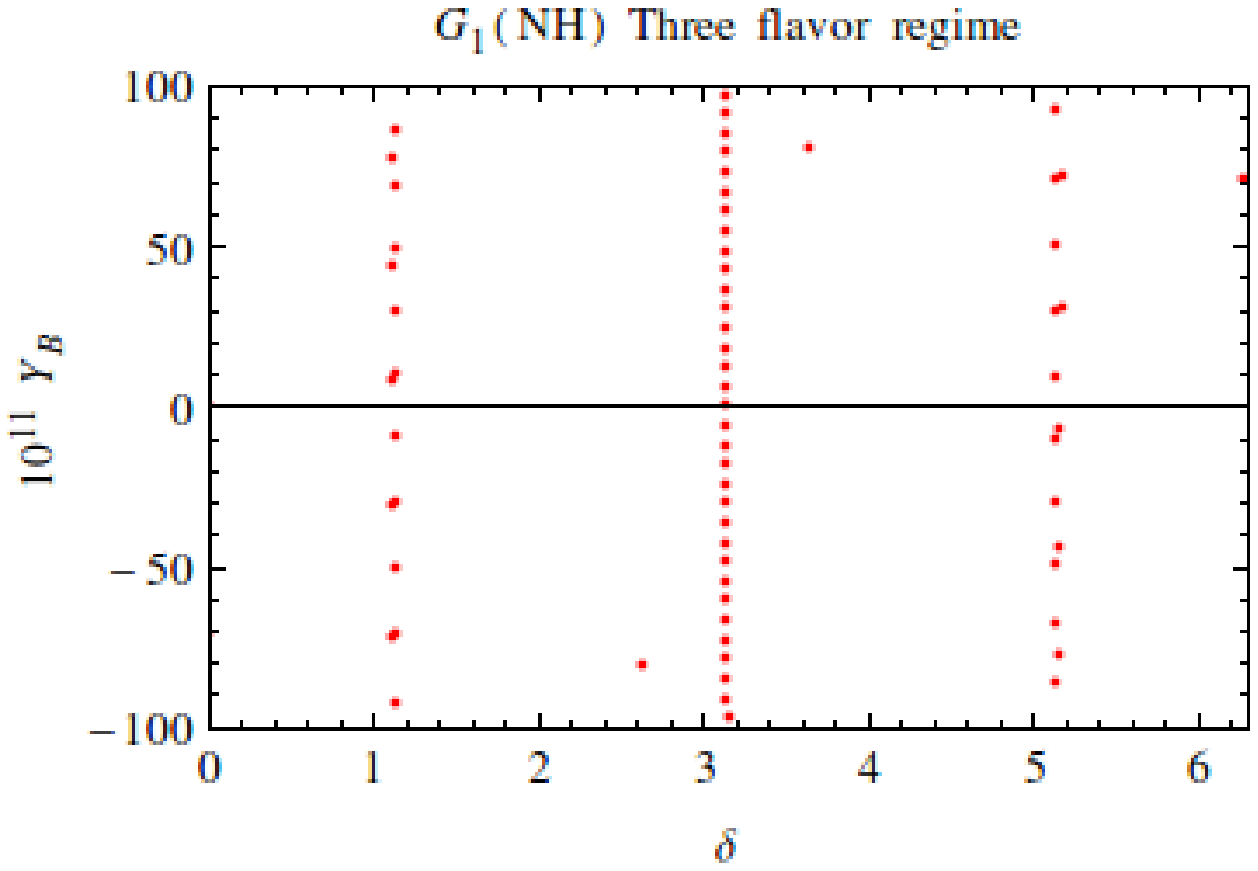}
\end{minipage}%
\caption{Variation of baryon asymmetry with Dirac CP phase $\delta$ for one-zero texture $G_1$ with normal hierarchy.}
\label{fig5}
\end{figure}

\begin{figure}
\centering
\begin{minipage}{.5\textwidth}
  \centering
  \includegraphics[width=1\linewidth]{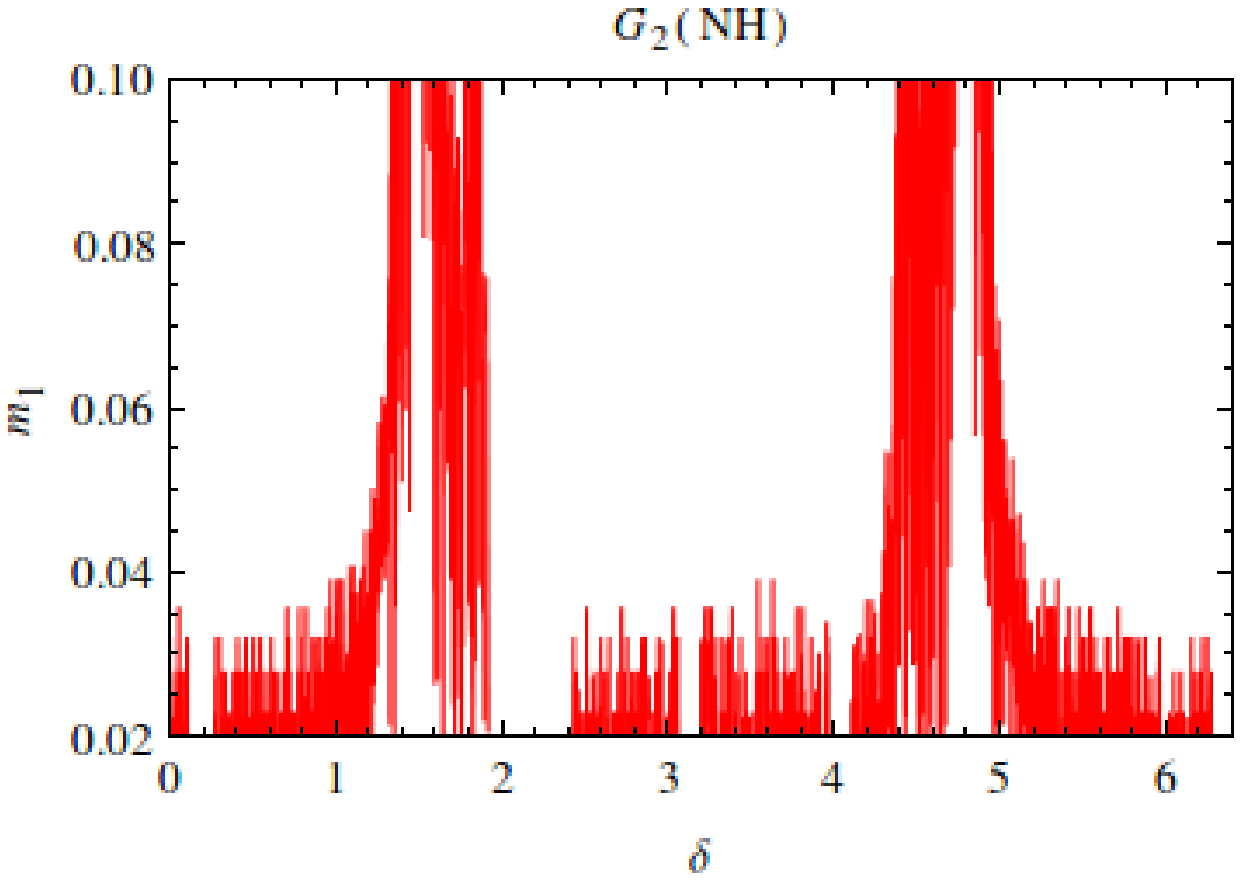}
\end{minipage}%
\begin{minipage}{.5\textwidth}
  \centering
  \includegraphics[width=1\linewidth]{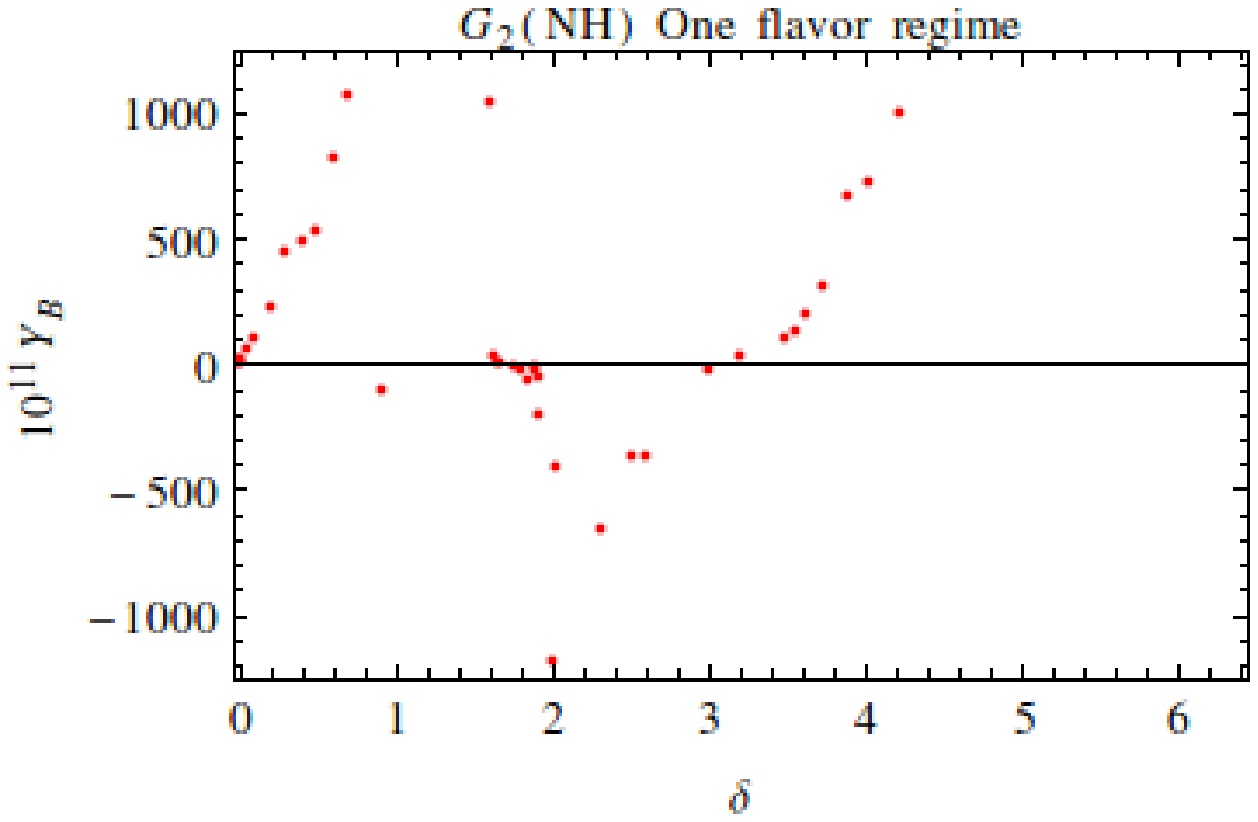}
\end{minipage}
\begin{minipage}{.5\textwidth}
  \centering
  \includegraphics[width=1\linewidth]{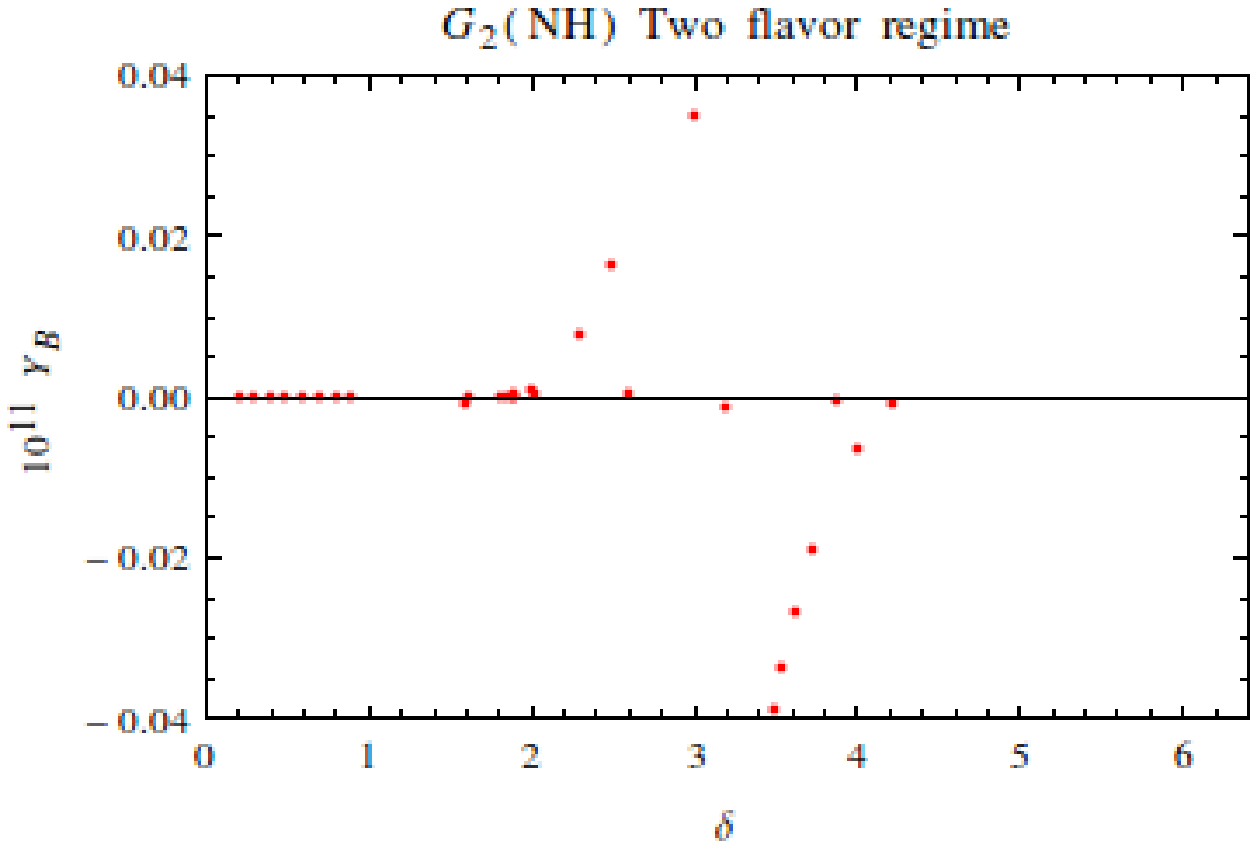}
\end{minipage}%
\begin{minipage}{.5\textwidth}
  \centering
  \includegraphics[width=1\linewidth]{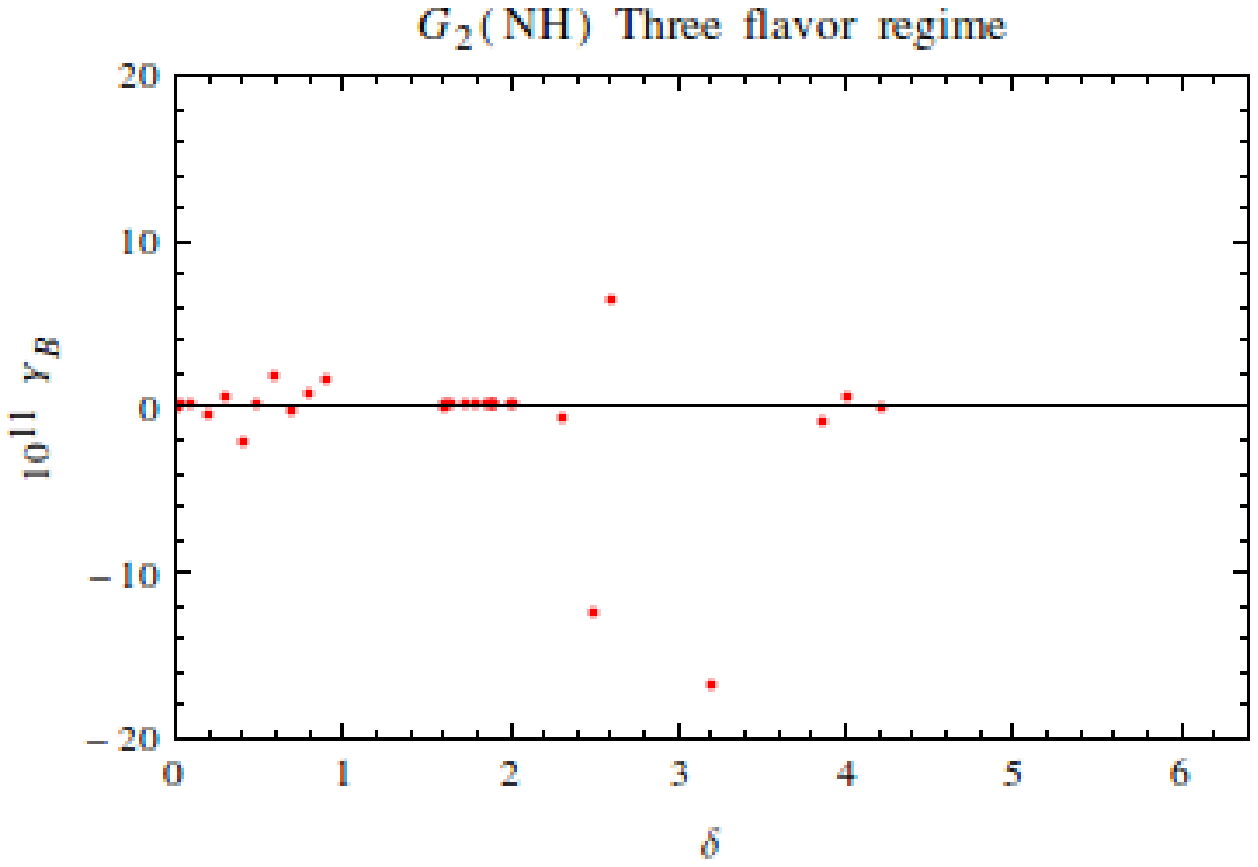}
\end{minipage}
\caption{Variation of lightest neutrino mass $m_1$ and baryon asymmetry with Dirac CP phase $\delta$ for one-zero texture $G_2$ with normal hierarchy.}
\label{fig6}
\end{figure}

\begin{figure}
\centering
\begin{minipage}{.5\textwidth}
  \centering
  \includegraphics[width=1\linewidth]{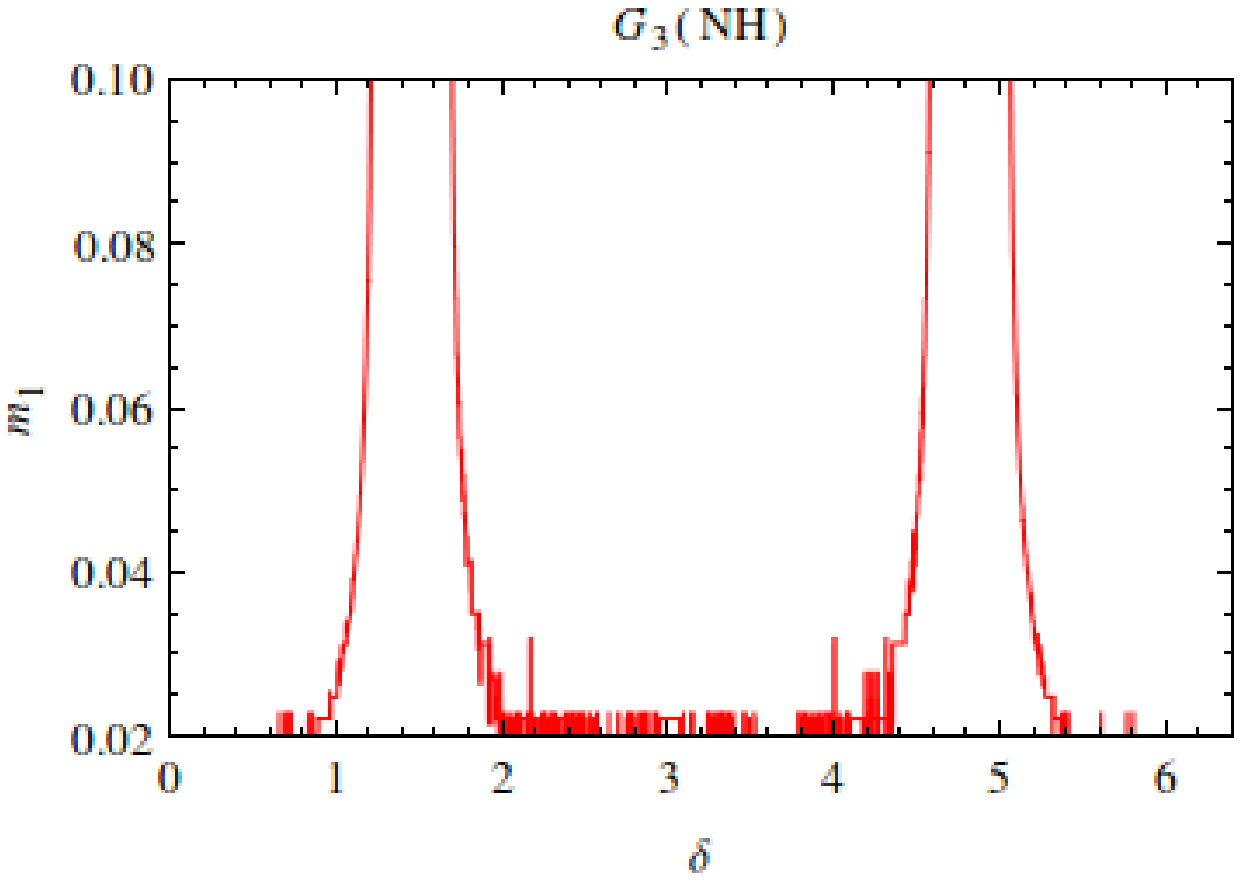}
\end{minipage}%
\begin{minipage}{.5\textwidth}
  \centering
  \includegraphics[width=1\linewidth]{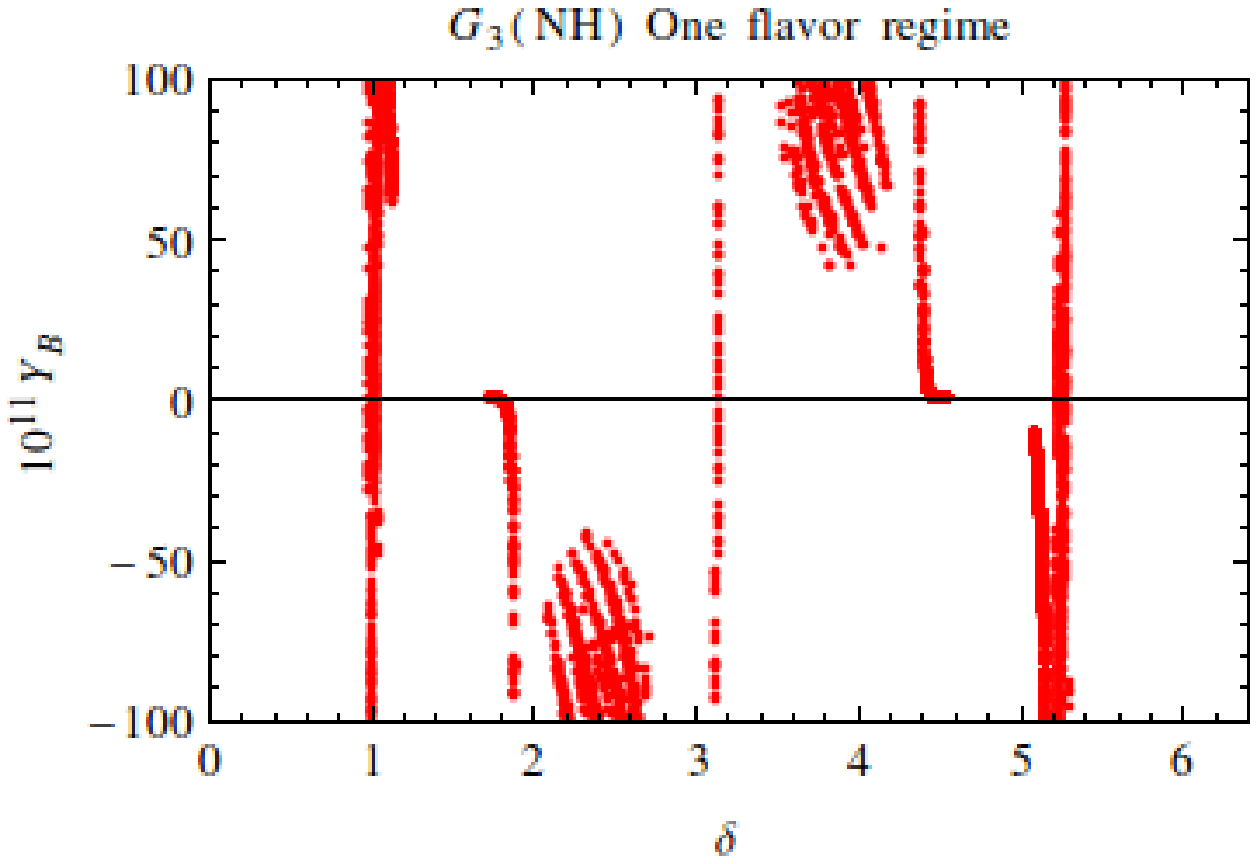}
\end{minipage}
\begin{minipage}{.5\textwidth}
  \centering
  \includegraphics[width=1\linewidth]{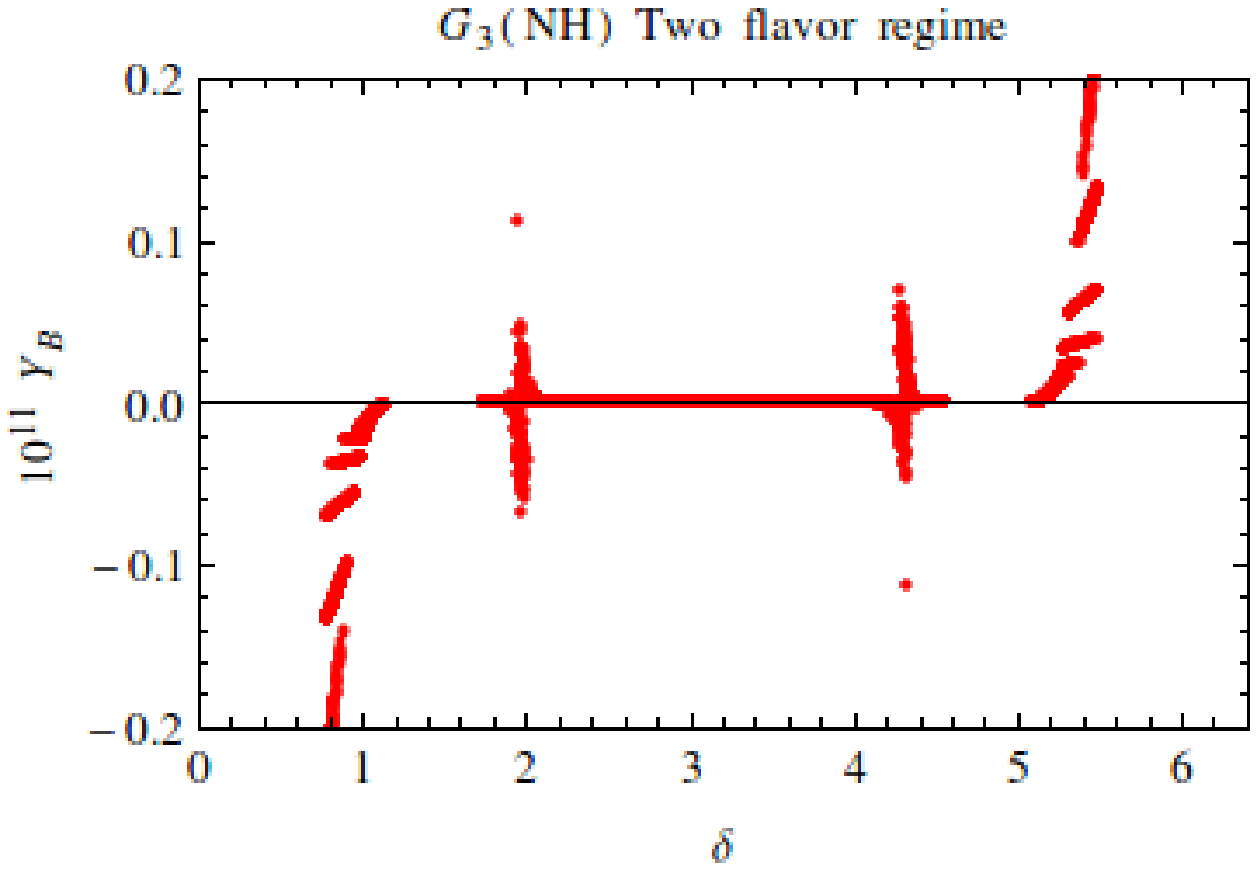}
\end{minipage}%
\begin{minipage}{.5\textwidth}
  \centering
  \includegraphics[width=1\linewidth]{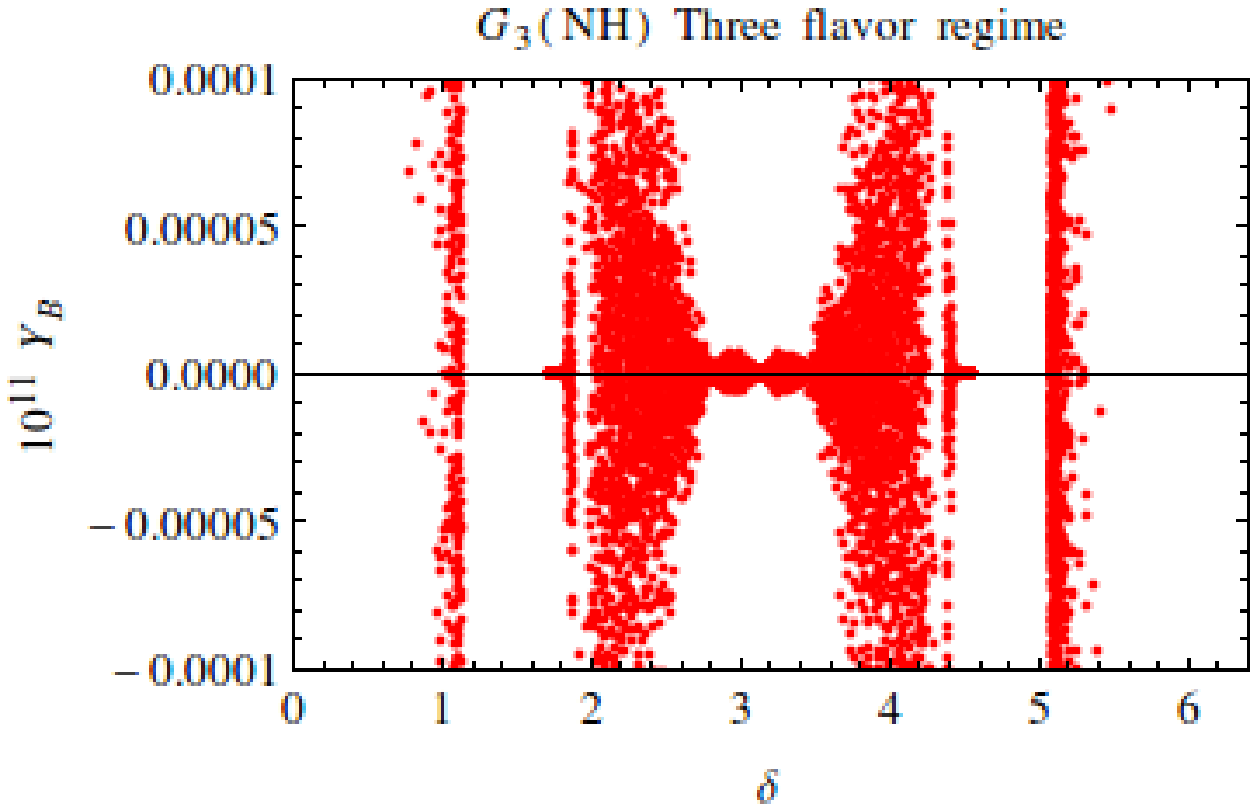}
\end{minipage}
\caption{Variation of lightest neutrino mass $m_1$ and baryon asymmetry with Dirac CP phase $\delta$ for one-zero texture $G_3$ with normal hierarchy.}
\label{fig7}
\end{figure}

\begin{table}[!h]
\begin{tabular}{ccccc}
 \hline

Patterns     & $m_3$ & $\delta$&$\alpha$ &$\beta$\\
        \hline \hline
\mbox{$A1$}     & 0.00019    &  0.0059      & 2.99   & 0.99\\
\mbox{$A2$}     & 0.00038    &  0.34      &  2.99   &0.87\\ 
\mbox{$B1$}     & 0.049   &   1.57     & 3.10   &3.11\\
\mbox{$B2$}     &  0.0048   &  1.62      & 0.93   &3.88\\
\mbox{$B3$}     &  0.00055   &  0.055      & 4.99   &6.098\\
\mbox{$B4$}     &  0.0052   &  0.37      & 2.00   &0.73\\
        \hline
\end{tabular}
 
\caption{Values of $m_3$, $\delta$, $\alpha$ and $\beta$ for two-zero texture with inverted hierarchy.}
\label{table-2zero1}
\end{table}

\begin{table}[!h]
\begin{tabular}{ccccc}
 \hline

Patterns     & $m_1$ & $\delta$&$\alpha$ &$\beta$\\
        \hline \hline
\mbox{$A1$}     & 0.005 & 4.36  & 1.73 &4.21\\
\mbox{$A2$}     & 0.0069 & 0.039 &1.57  & 1.55\\ 
\mbox{$B1$}     &  0.068 &1.59  &0.041  &3.14\\ 
\mbox{$B2$}     & 0.022 & 0.84 &2.72  & 3.31\\  
\mbox{$B3$}     &0.07  & 1.55 & 3.11 &0.0017\\  
\mbox{$B4$}     & 0.07 & 4.79 &3.15  &6.19\\  
        \hline
\end{tabular}
 
\caption{Values of $m_1$, $\delta$, $\alpha$ and $\beta$ for two-zero texture with normal hierarchy.}
\label{table-2zero2}
\end{table}
\begin{figure}
\centering
\begin{minipage}{.5\textwidth}
  \centering
  \includegraphics[width=1\linewidth]{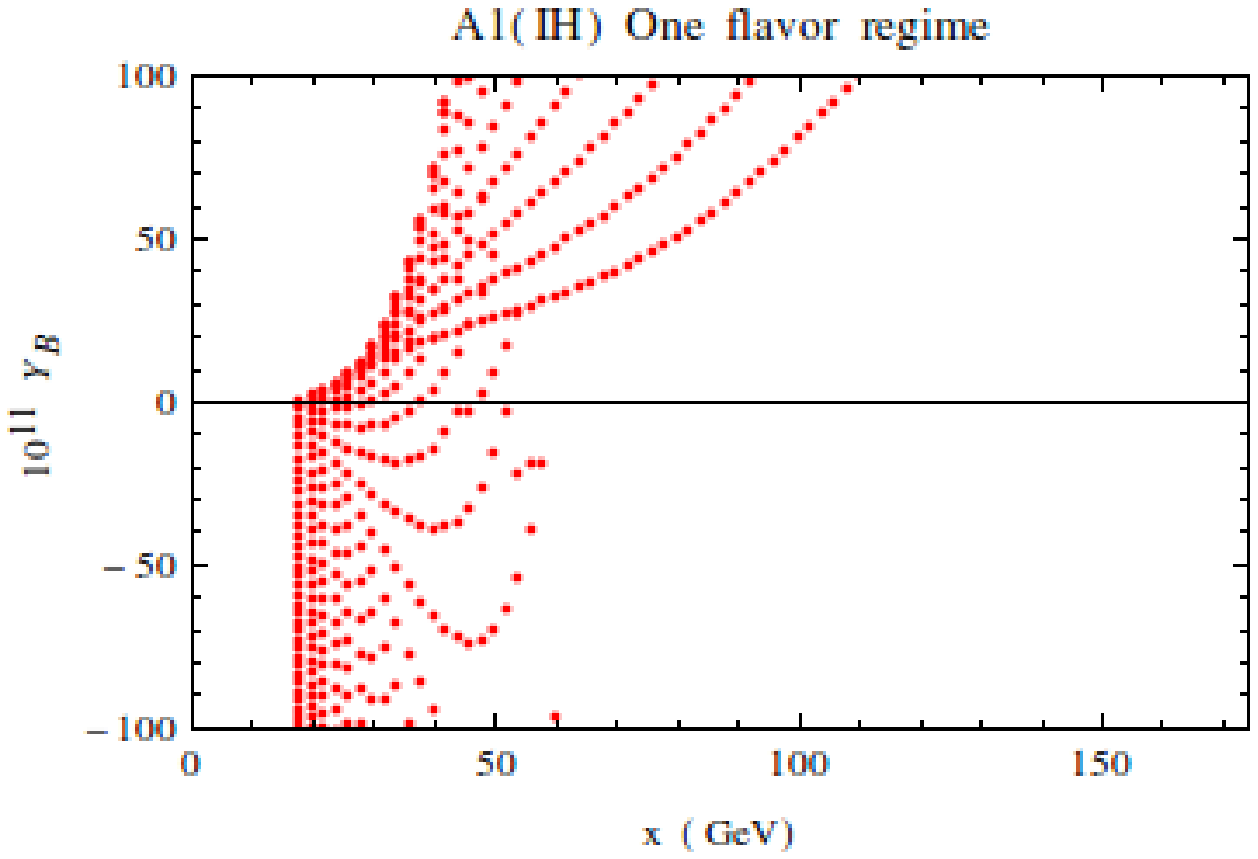}
\end{minipage}%
\begin{minipage}{.5\textwidth}
  \centering
  \includegraphics[width=1\linewidth]{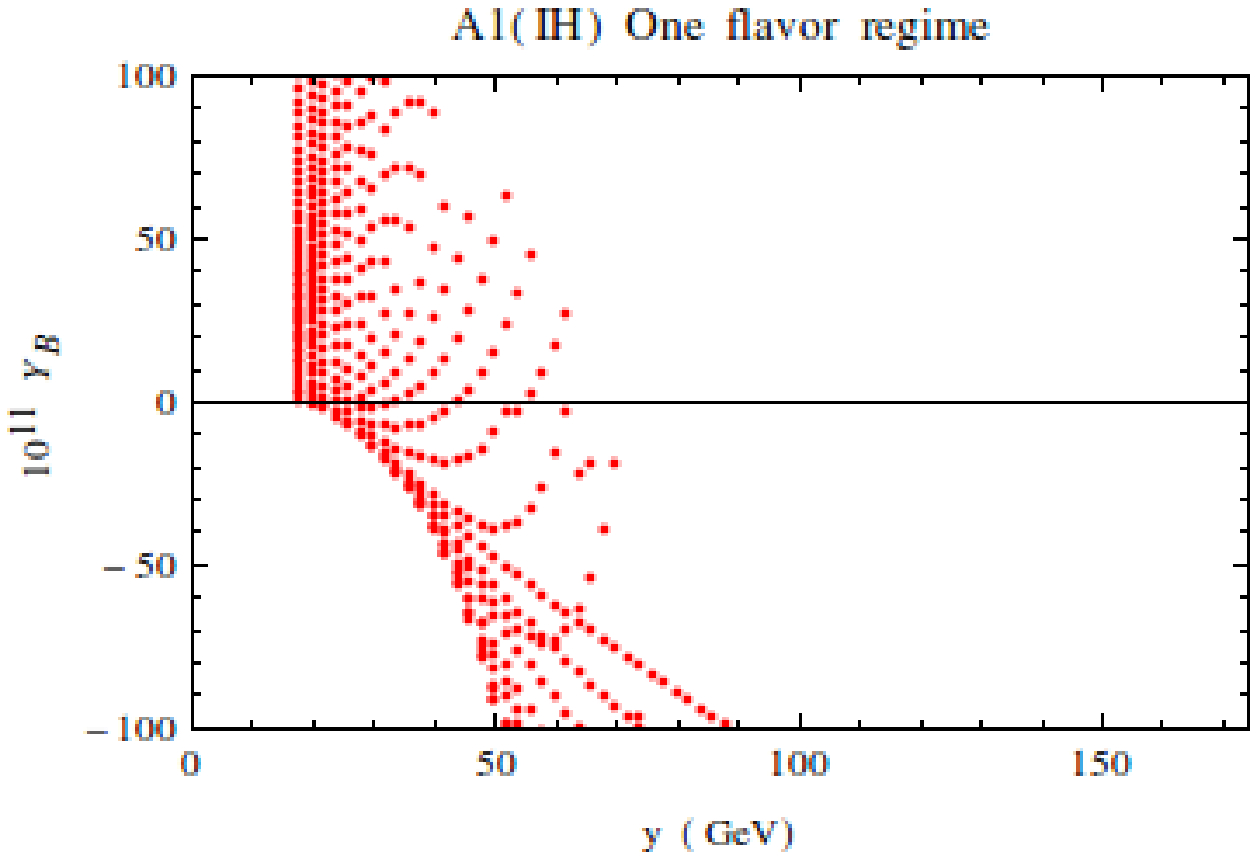}
\end{minipage}
\begin{minipage}{.5\textwidth}
  \centering
  \includegraphics[width=1\linewidth]{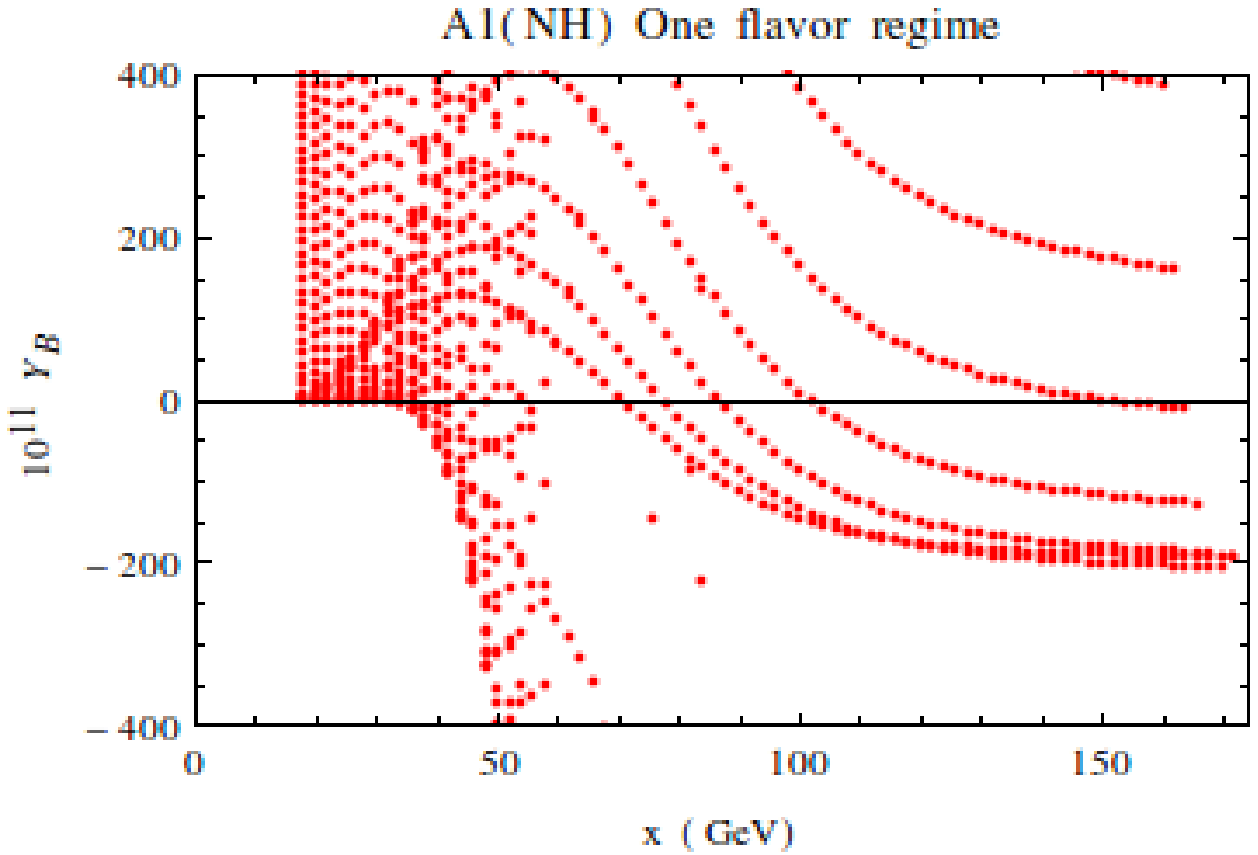}
\end{minipage}%
\begin{minipage}{.5\textwidth}
  \centering
  \includegraphics[width=1\linewidth]{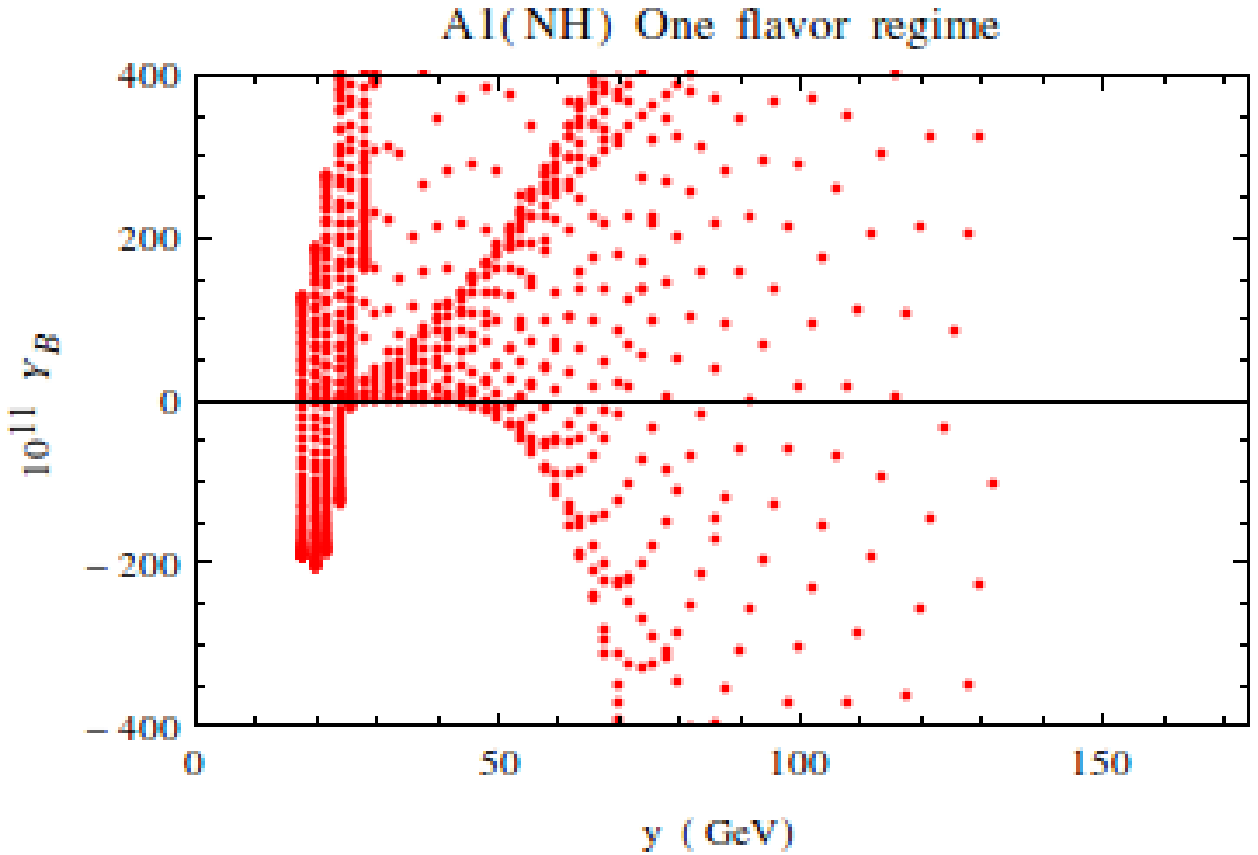}
\end{minipage}
\caption{Variation of baryon asymmetry in one flavor regime with Dirac neutrino masses for two-zero texture $A_1$.}
\label{fig8}
\end{figure}

\begin{figure}
\centering
\begin{minipage}{.5\textwidth}
  \centering
  \includegraphics[width=1\linewidth]{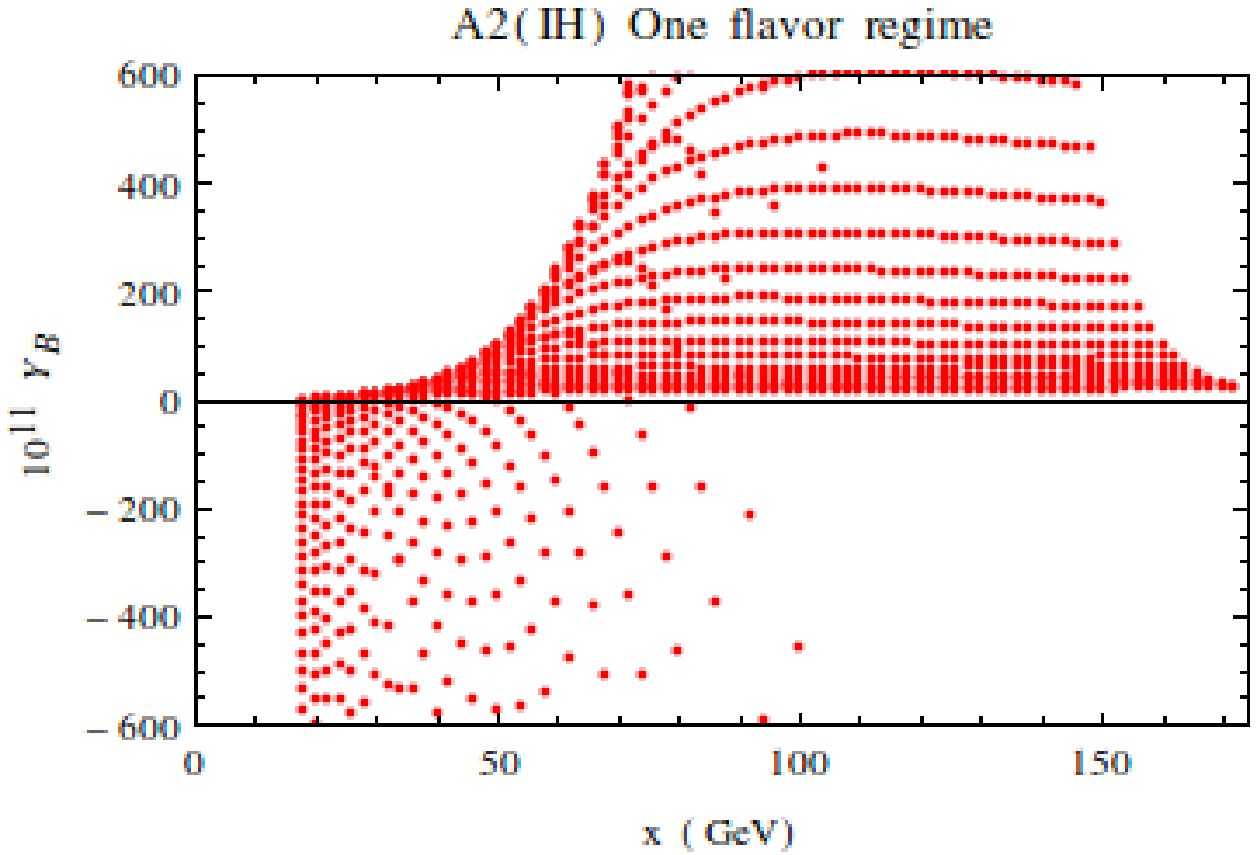}
\end{minipage}%
\begin{minipage}{.5\textwidth}
  \centering
  \includegraphics[width=1\linewidth]{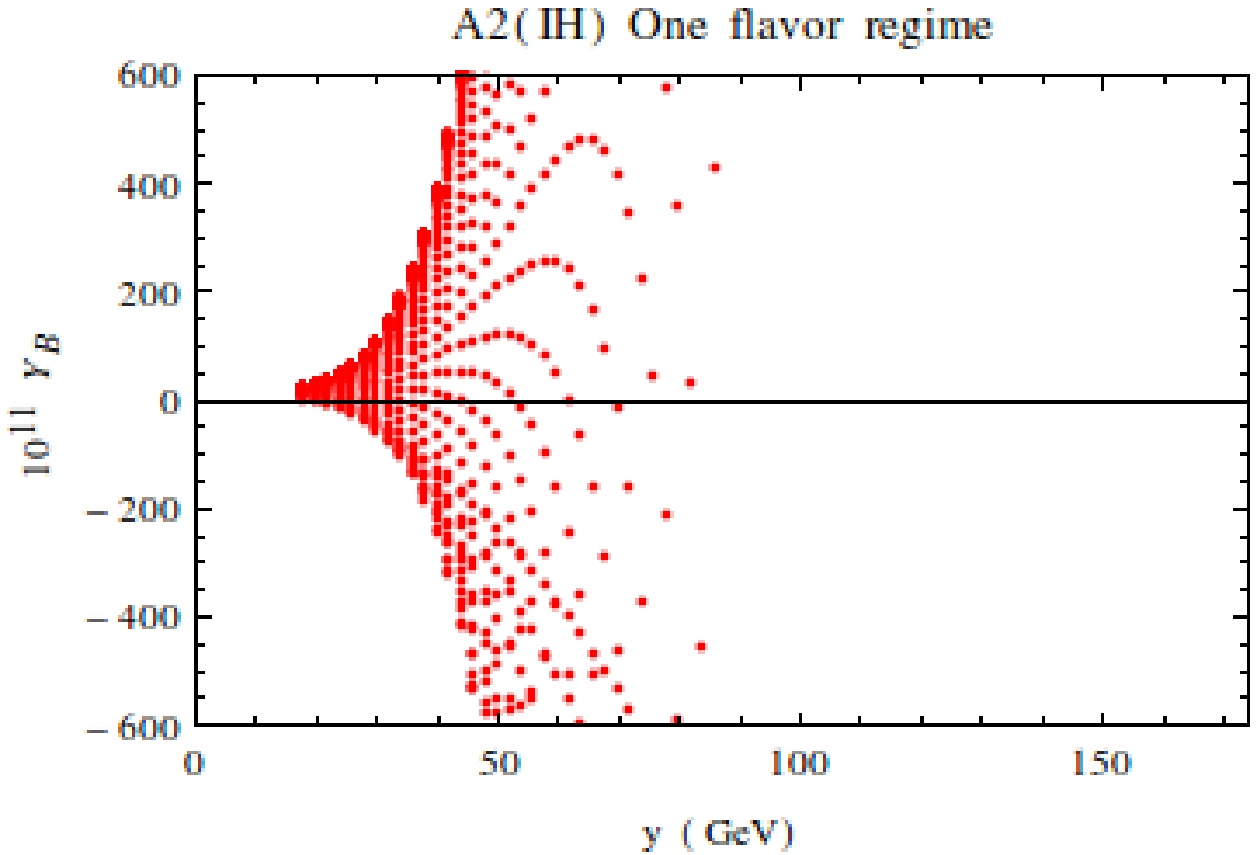}
\end{minipage}
\begin{minipage}{.5\textwidth}
  \centering
  \includegraphics[width=1\linewidth]{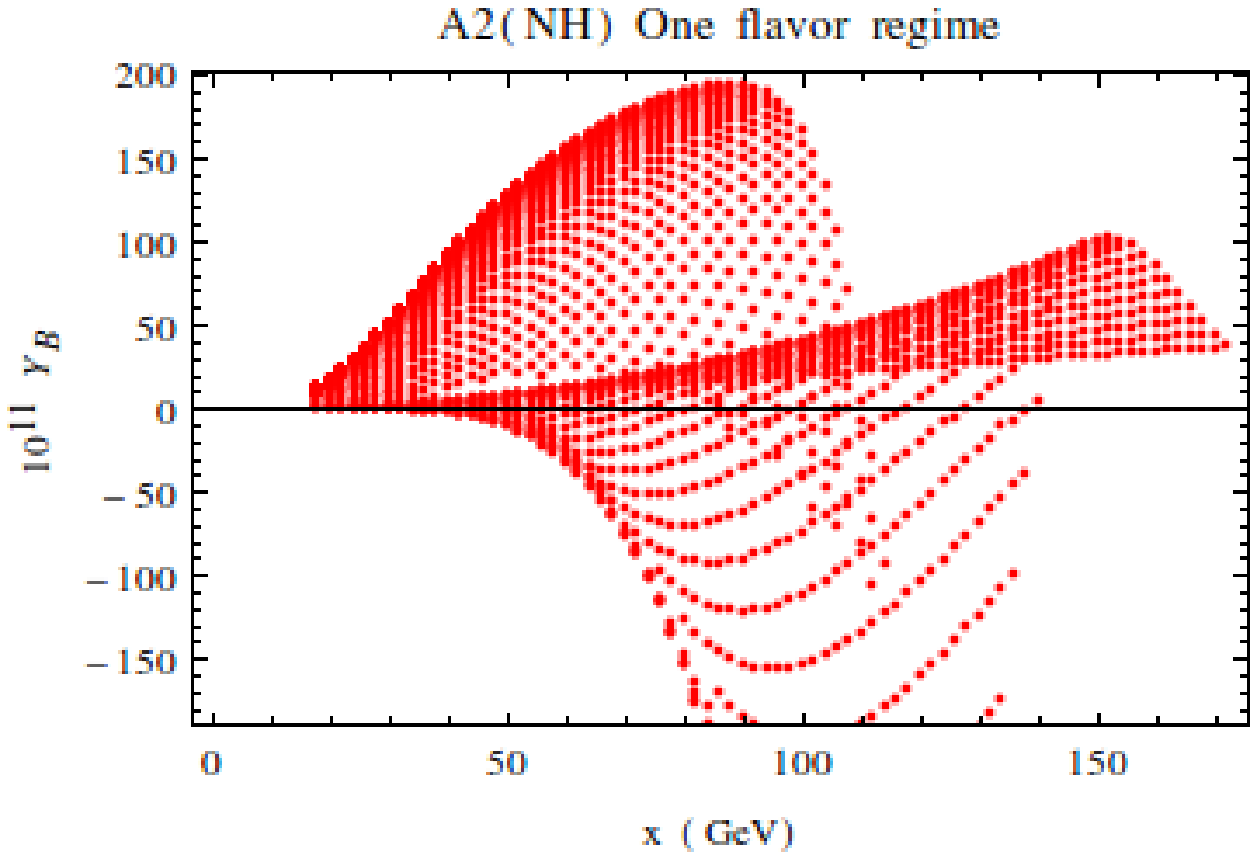}
\end{minipage}%
\begin{minipage}{.5\textwidth}
  \centering
  \includegraphics[width=1\linewidth]{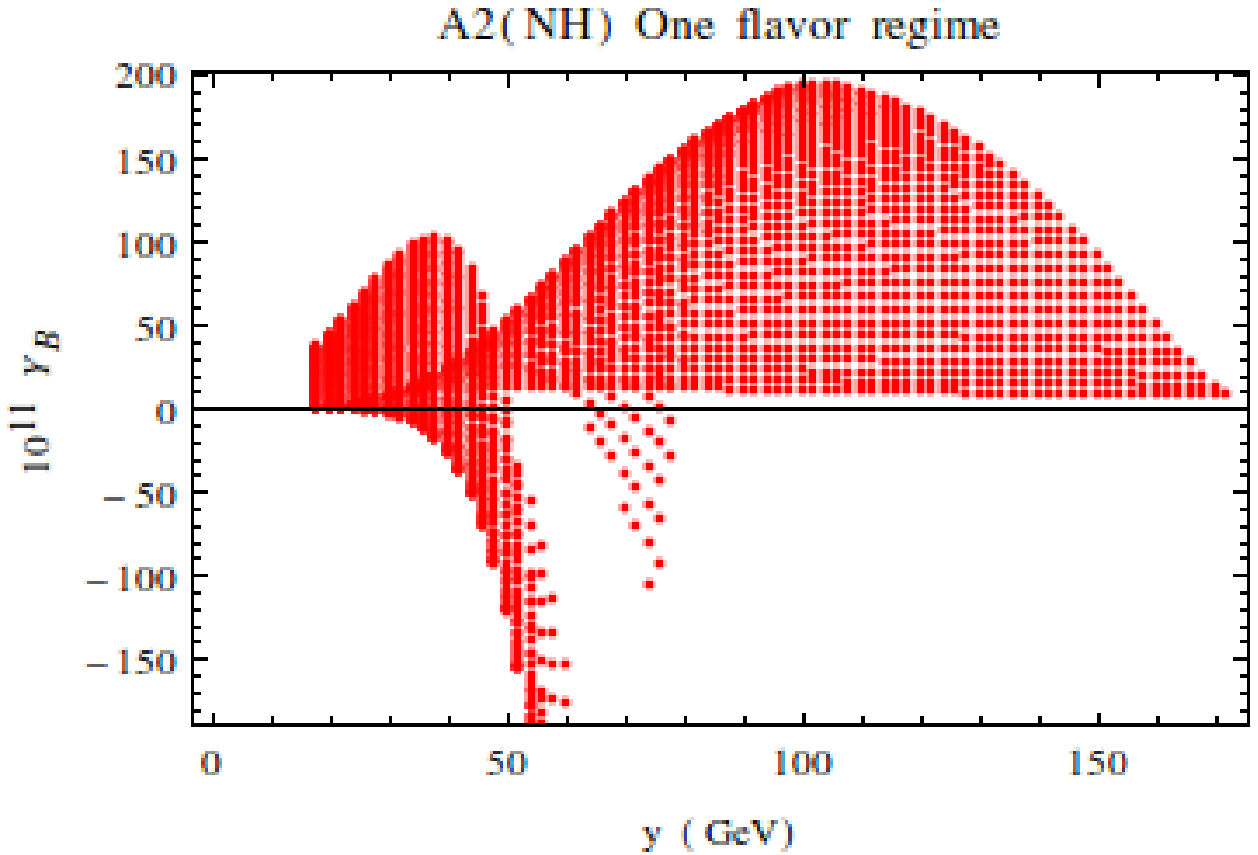}
\end{minipage}
\caption{Variation of baryon asymmetry in one flavor regime with Dirac neutrino masses for two-zero texture $A_2$.}
\label{fig9}
\end{figure}

\begin{figure}
\centering
\begin{minipage}{.5\textwidth}
  \centering
  \includegraphics[width=1\linewidth]{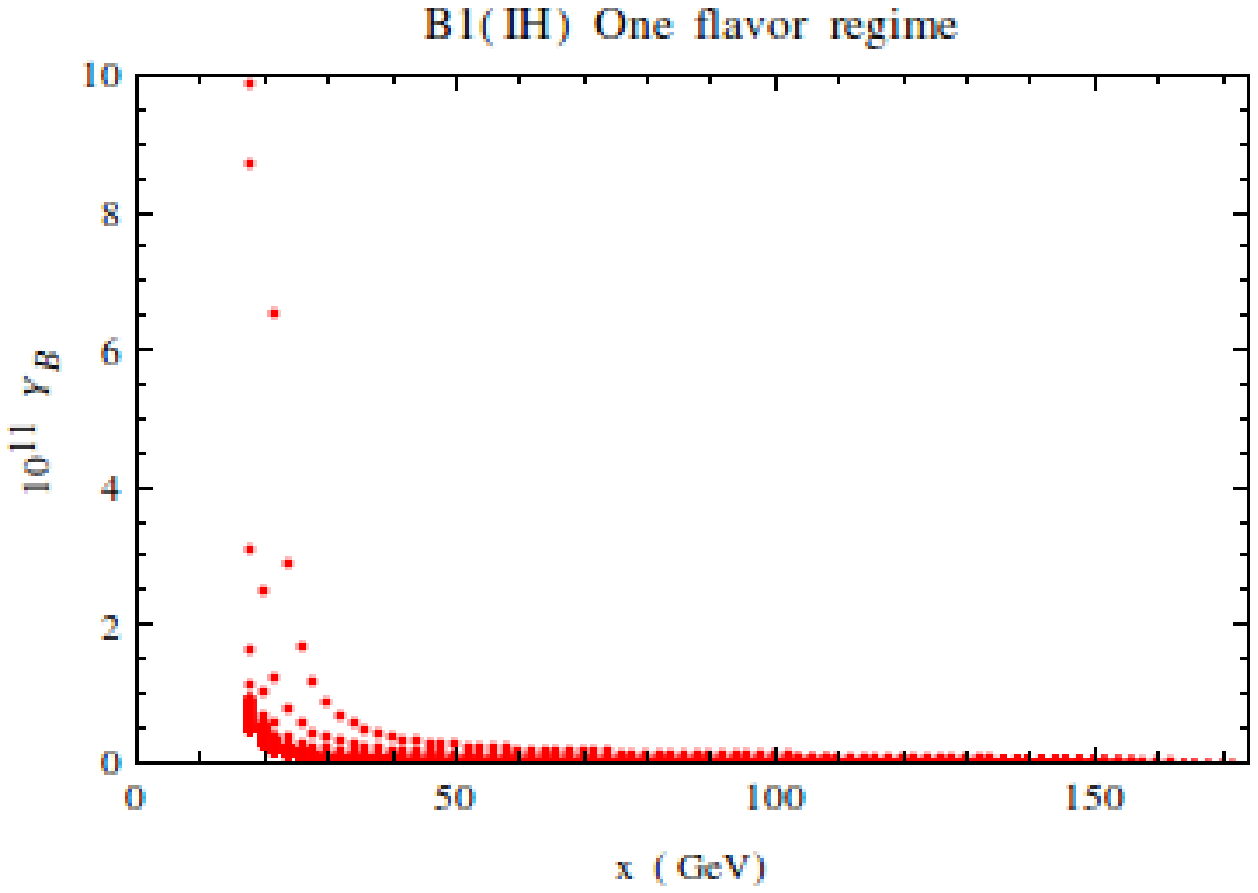}
\end{minipage}%
\begin{minipage}{.5\textwidth}
  \centering
  \includegraphics[width=1\linewidth]{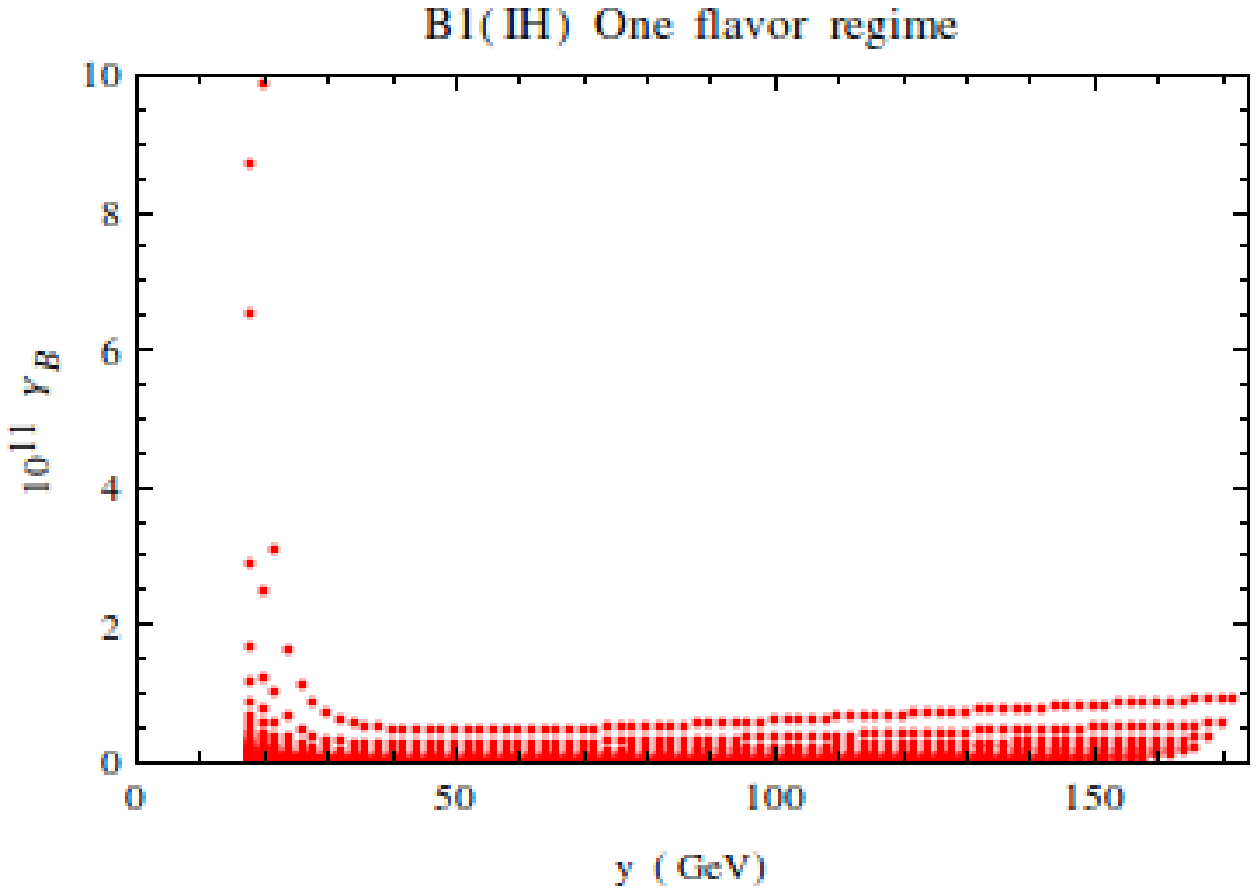}
\end{minipage}
\begin{minipage}{.5\textwidth}
  \centering
  \includegraphics[width=1\linewidth]{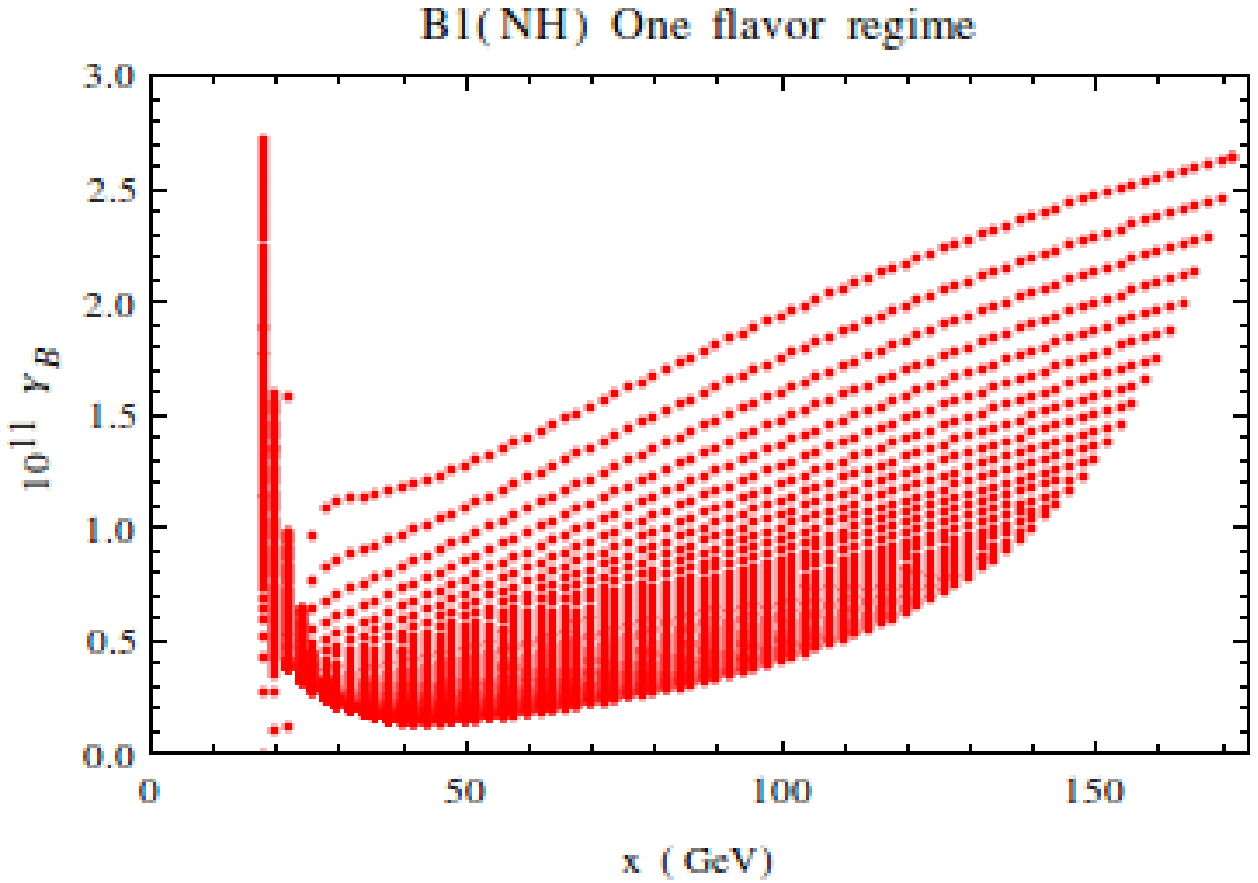}
\end{minipage}%
\begin{minipage}{.5\textwidth}
  \centering
  \includegraphics[width=1\linewidth]{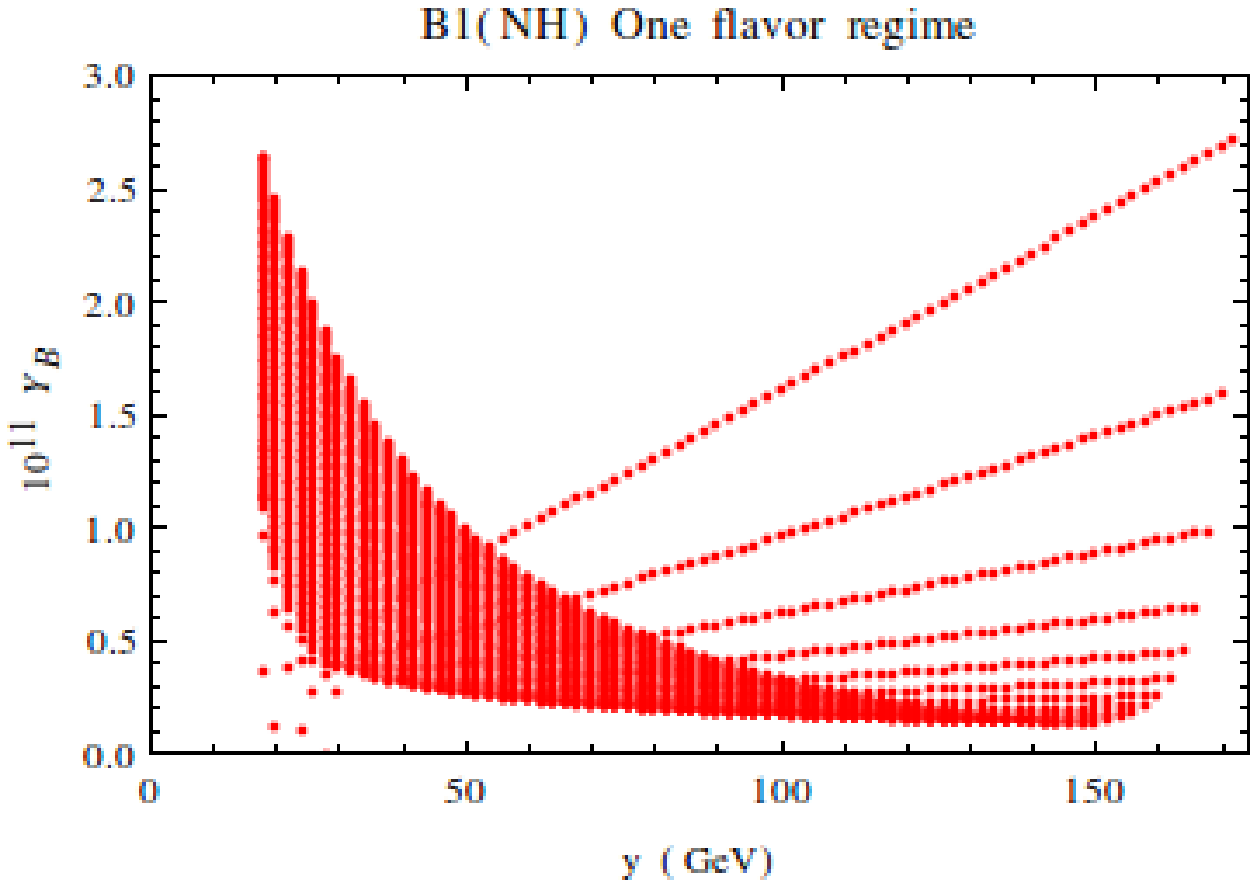}
\end{minipage}
\caption{Variation of baryon asymmetry in one flavor regime with Dirac neutrino masses for two-zero texture $B_1$.}
\label{fig10}
\end{figure}

\begin{figure}
\centering
\begin{minipage}{.5\textwidth}
  \centering
  \includegraphics[width=1\linewidth]{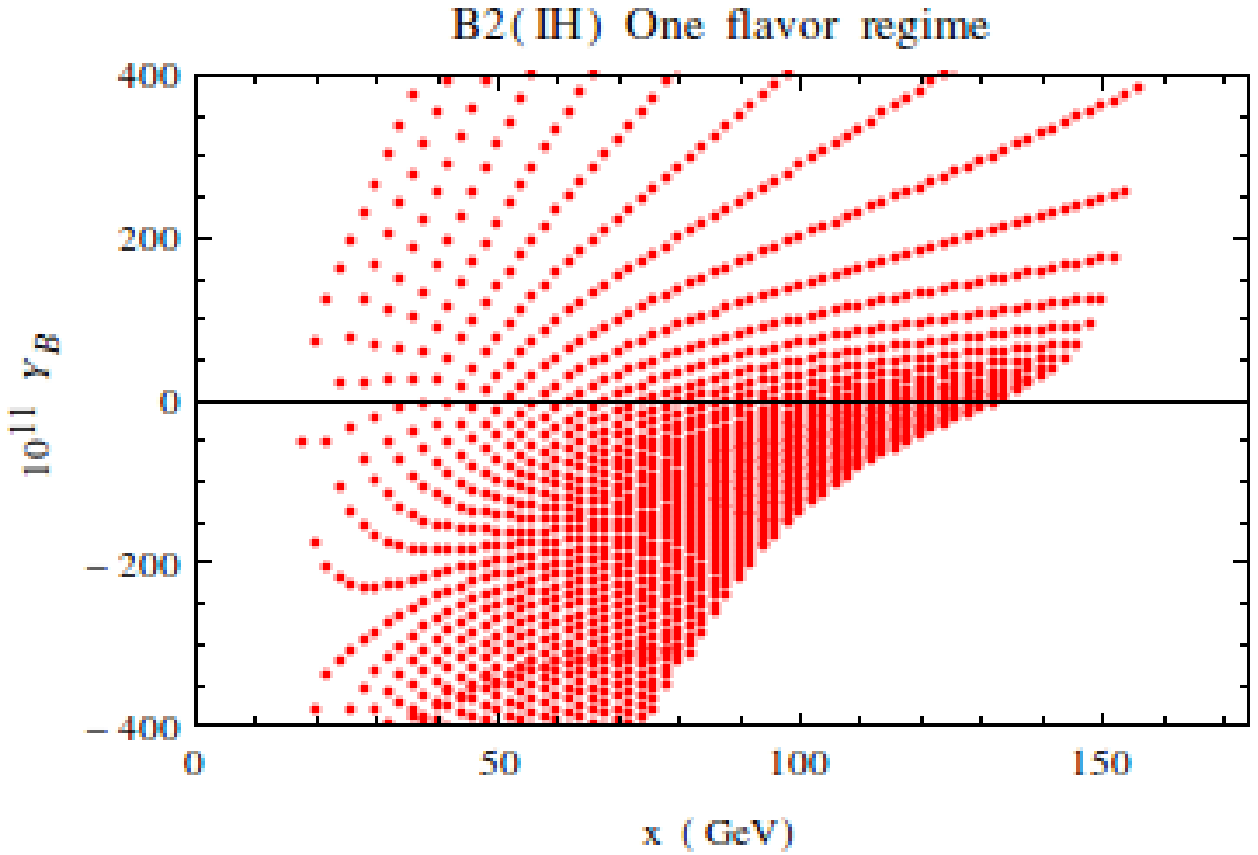}
\end{minipage}%
\begin{minipage}{.5\textwidth}
  \centering
  \includegraphics[width=1\linewidth]{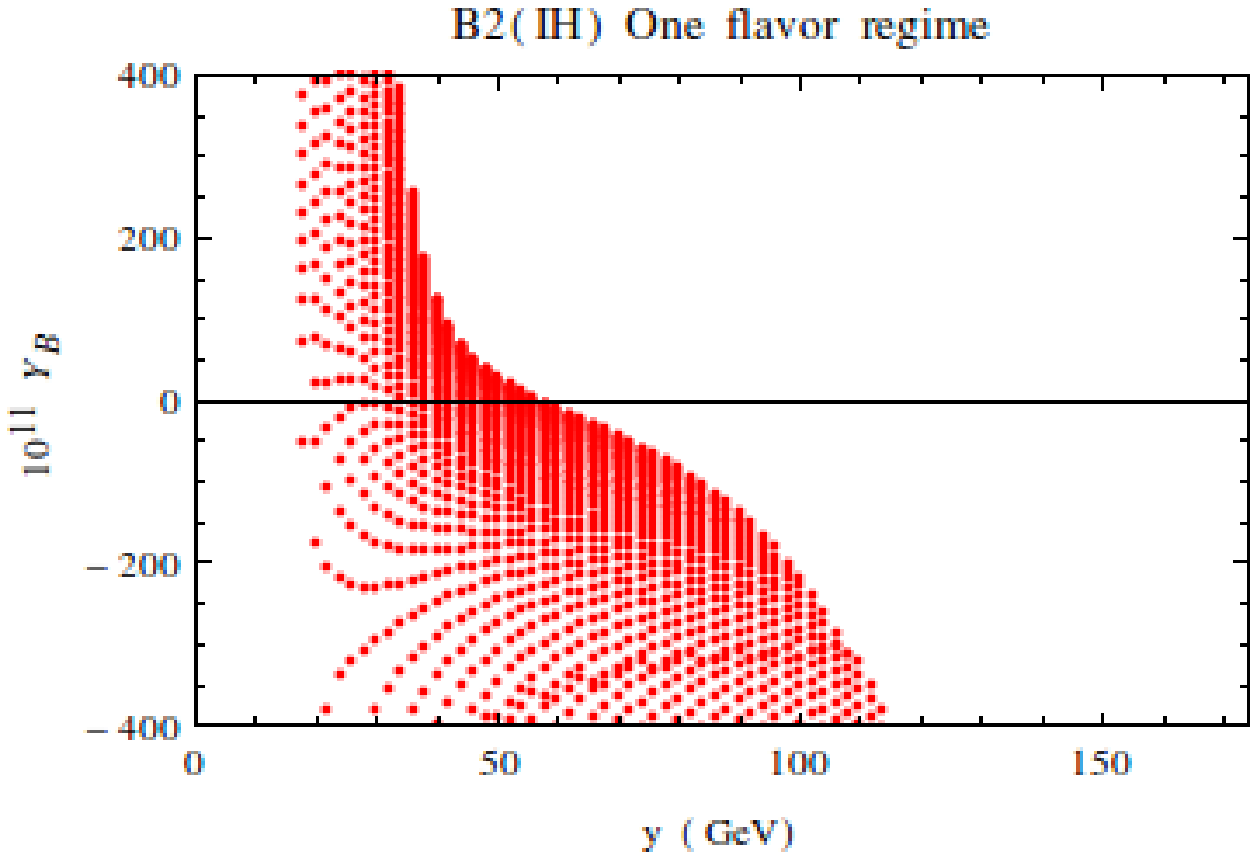}
\end{minipage}
\begin{minipage}{.5\textwidth}
  \centering
  \includegraphics[width=1\linewidth]{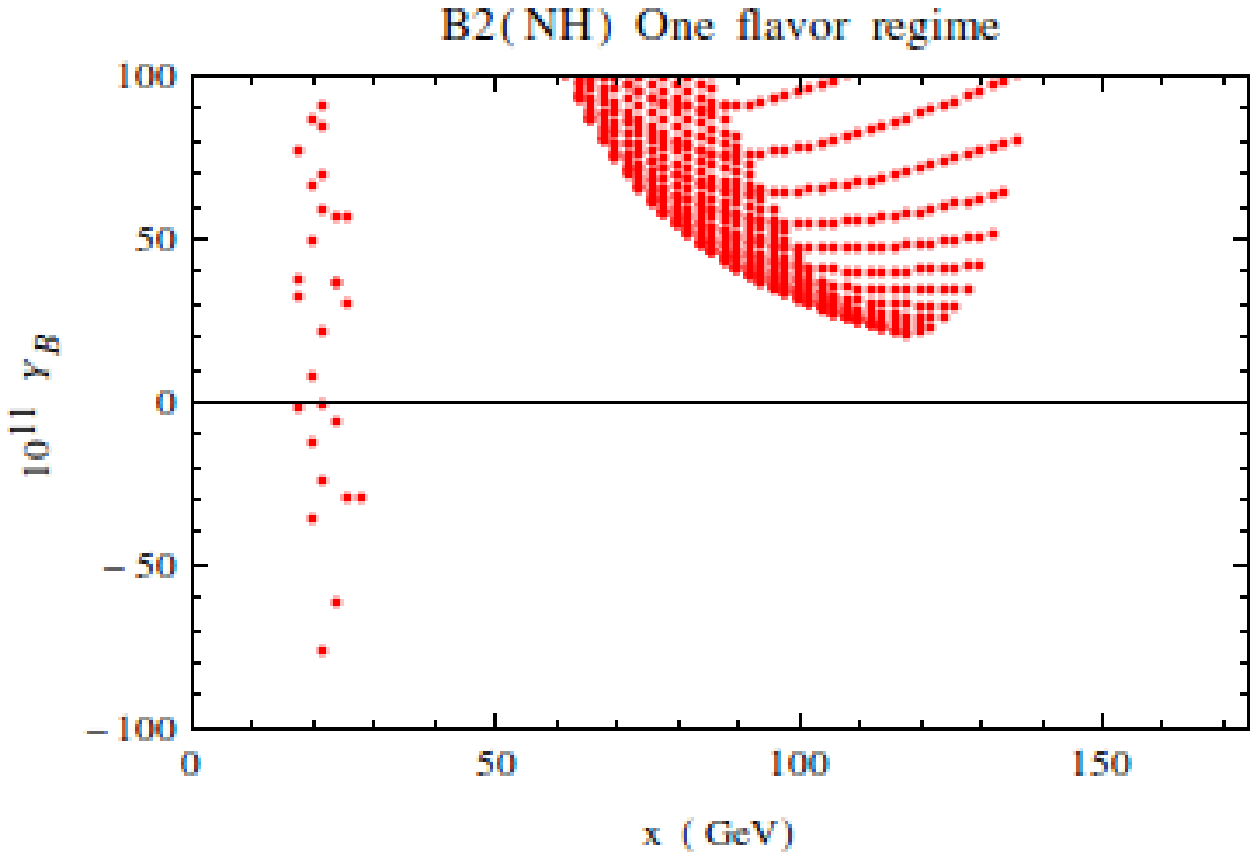}
\end{minipage}%
\begin{minipage}{.5\textwidth}
  \centering
  \includegraphics[width=1\linewidth]{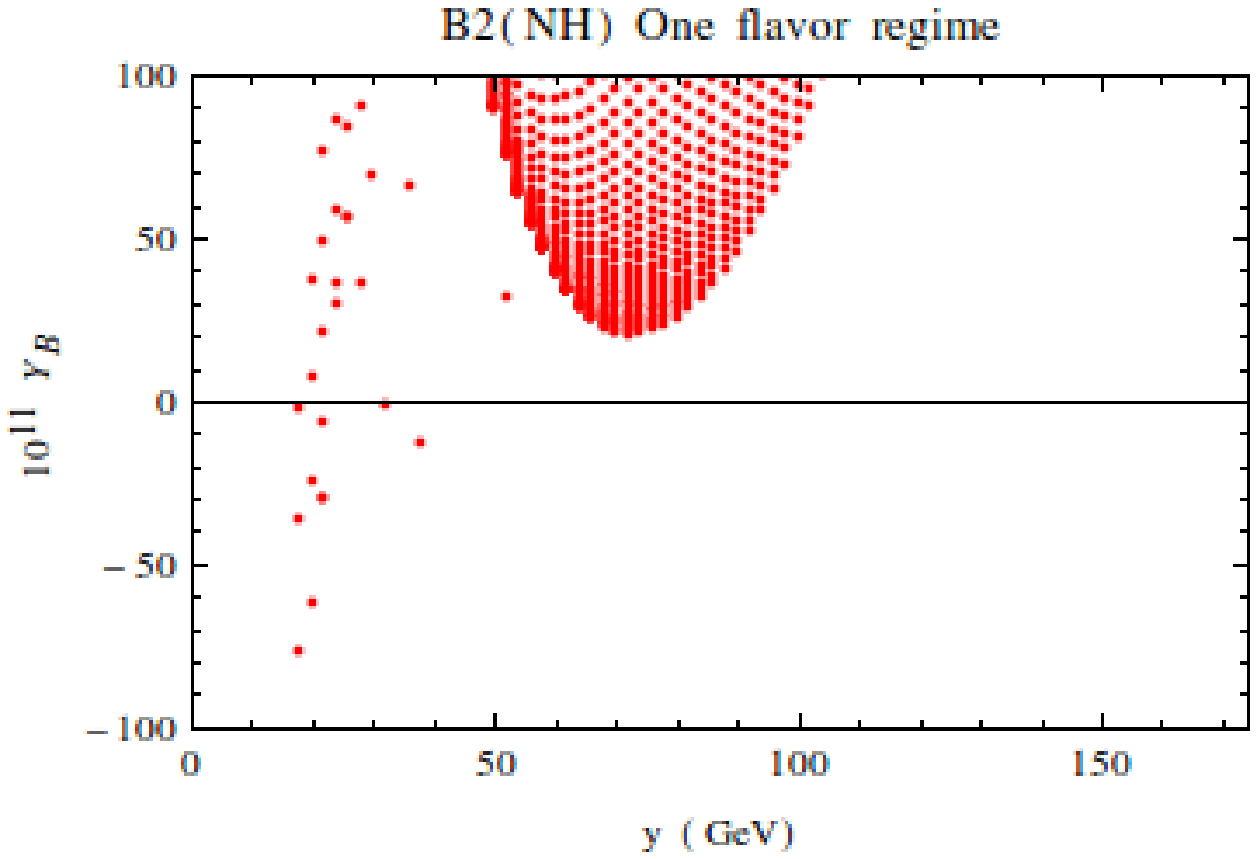}
\end{minipage}
\caption{Variation of baryon asymmetry in one flavor regime with Dirac neutrino masses for two-zero texture $B_2$.}
\label{fig11}
\end{figure}

\begin{figure}
\centering
\begin{minipage}{.5\textwidth}
  \centering
  \includegraphics[width=1\linewidth]{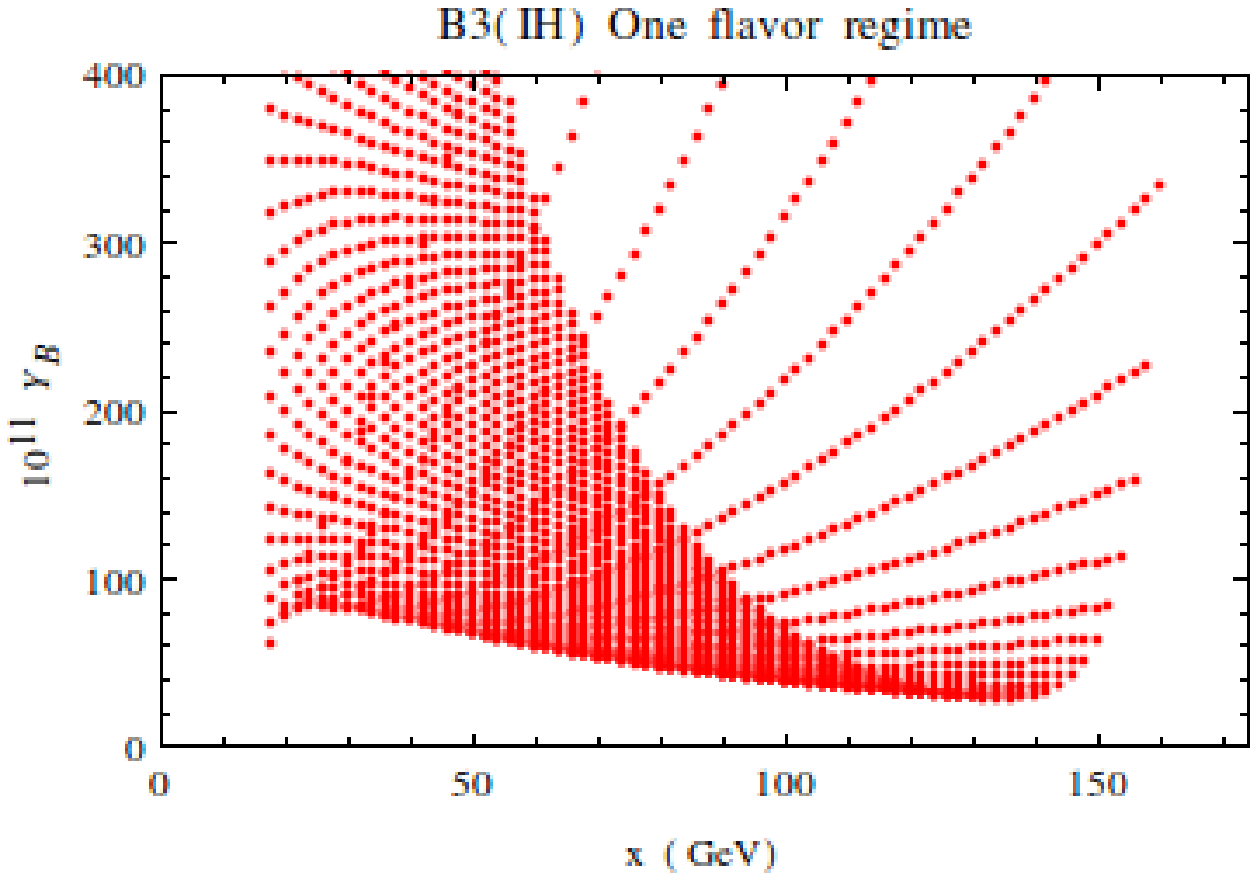}
\end{minipage}%
\begin{minipage}{.5\textwidth}
  \centering
  \includegraphics[width=1\linewidth]{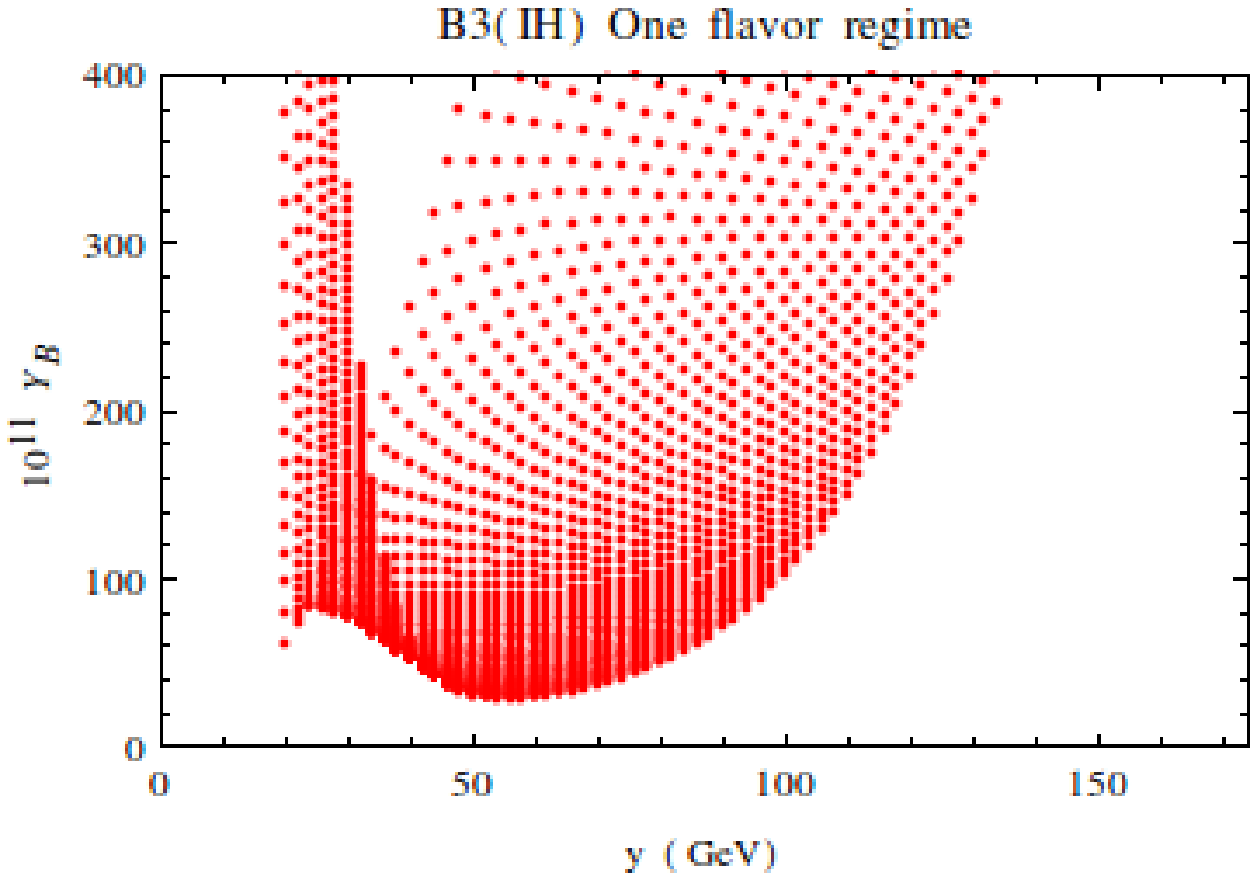}
\end{minipage}
\begin{minipage}{.5\textwidth}
  \centering
  \includegraphics[width=1\linewidth]{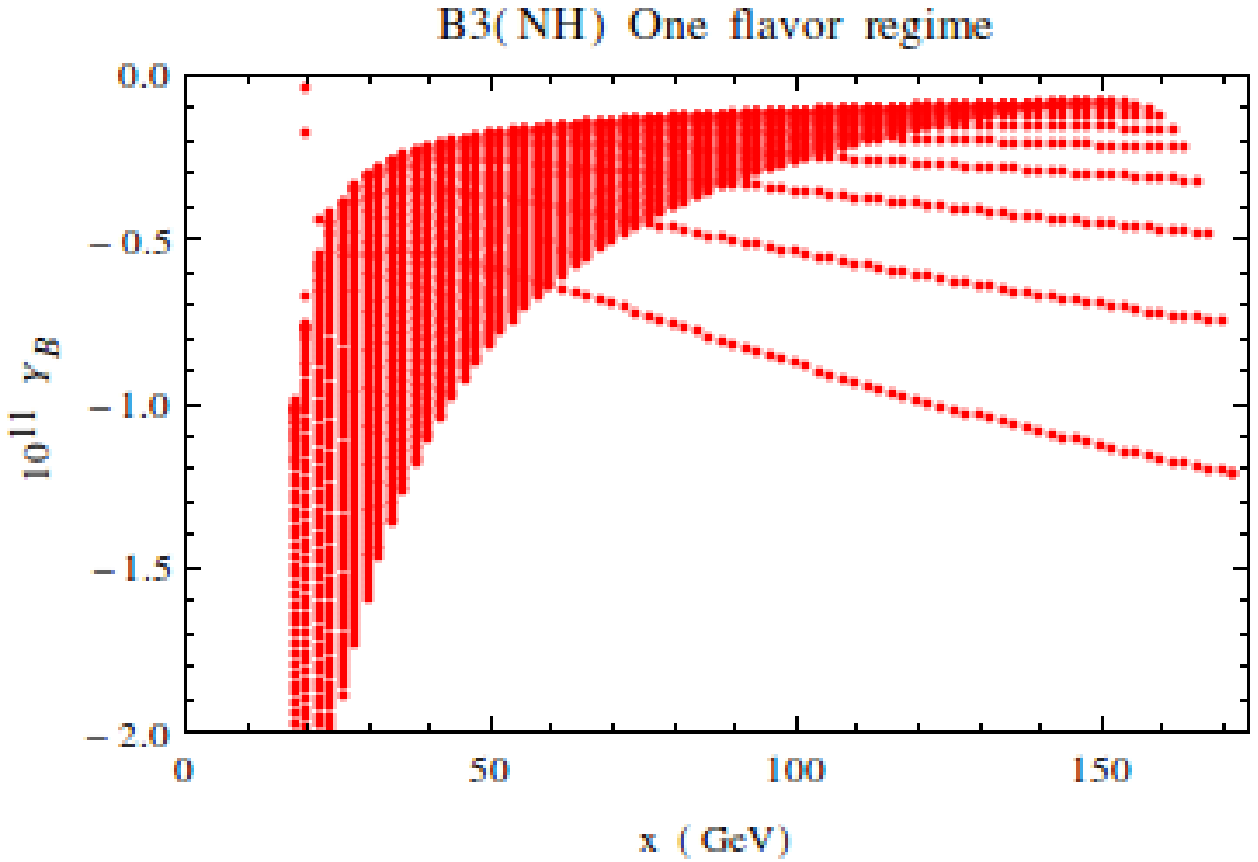}
\end{minipage}%
\begin{minipage}{.5\textwidth}
  \centering
  \includegraphics[width=1\linewidth]{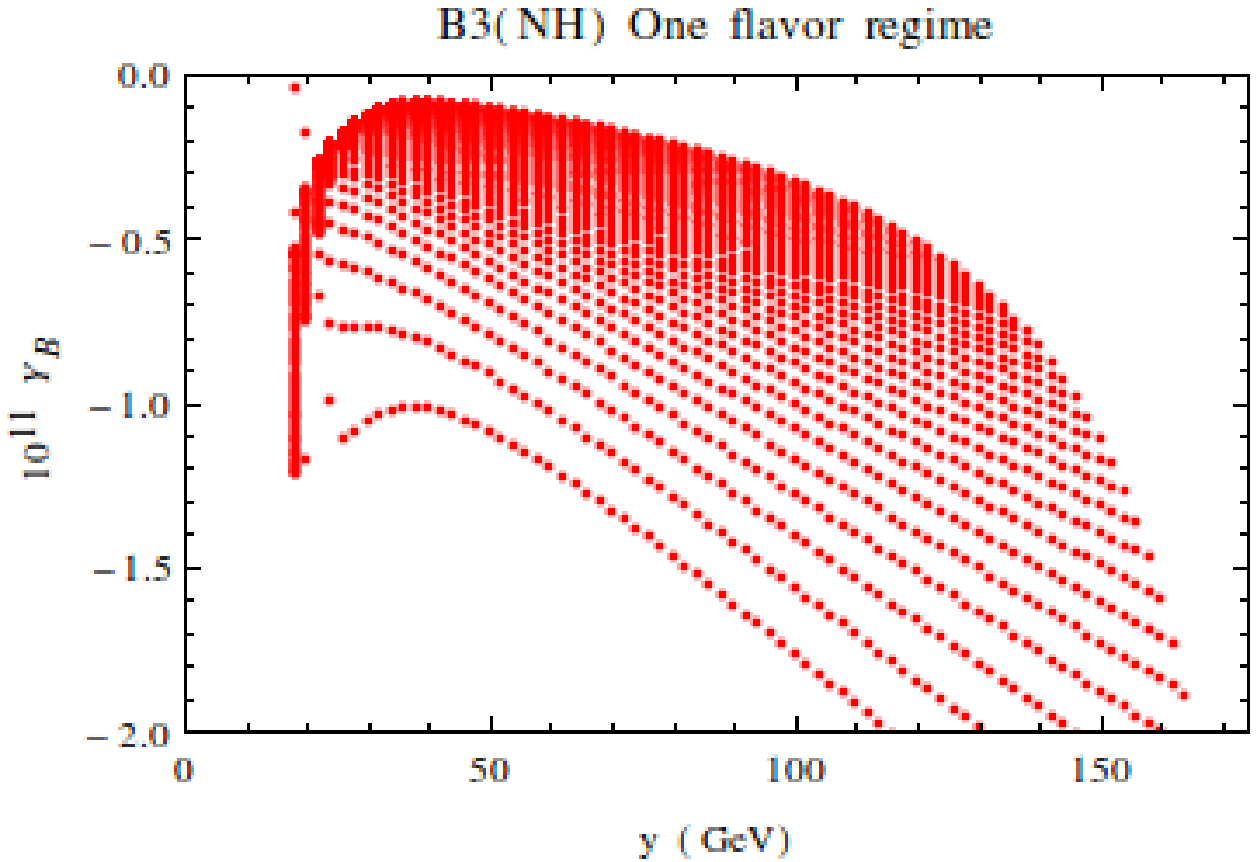}
\end{minipage}
\caption{Variation of baryon asymmetry in one flavor regime with Dirac neutrino masses for two-zero texture $B_3$.}
\label{fig12}
\end{figure}

\begin{figure}
\centering
\begin{minipage}{.5\textwidth}
  \centering
  \includegraphics[width=1\linewidth]{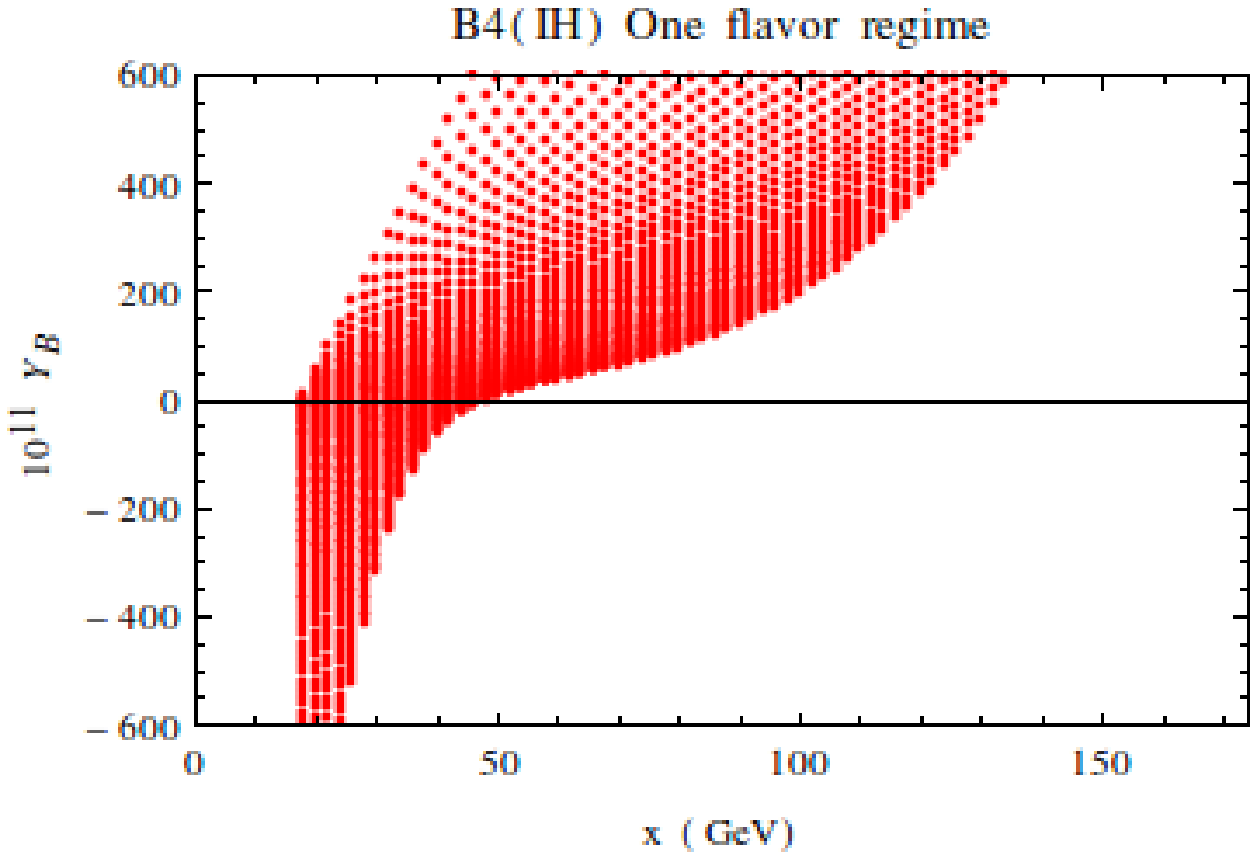}
\end{minipage}%
\begin{minipage}{.5\textwidth}
  \centering
  \includegraphics[width=1\linewidth]{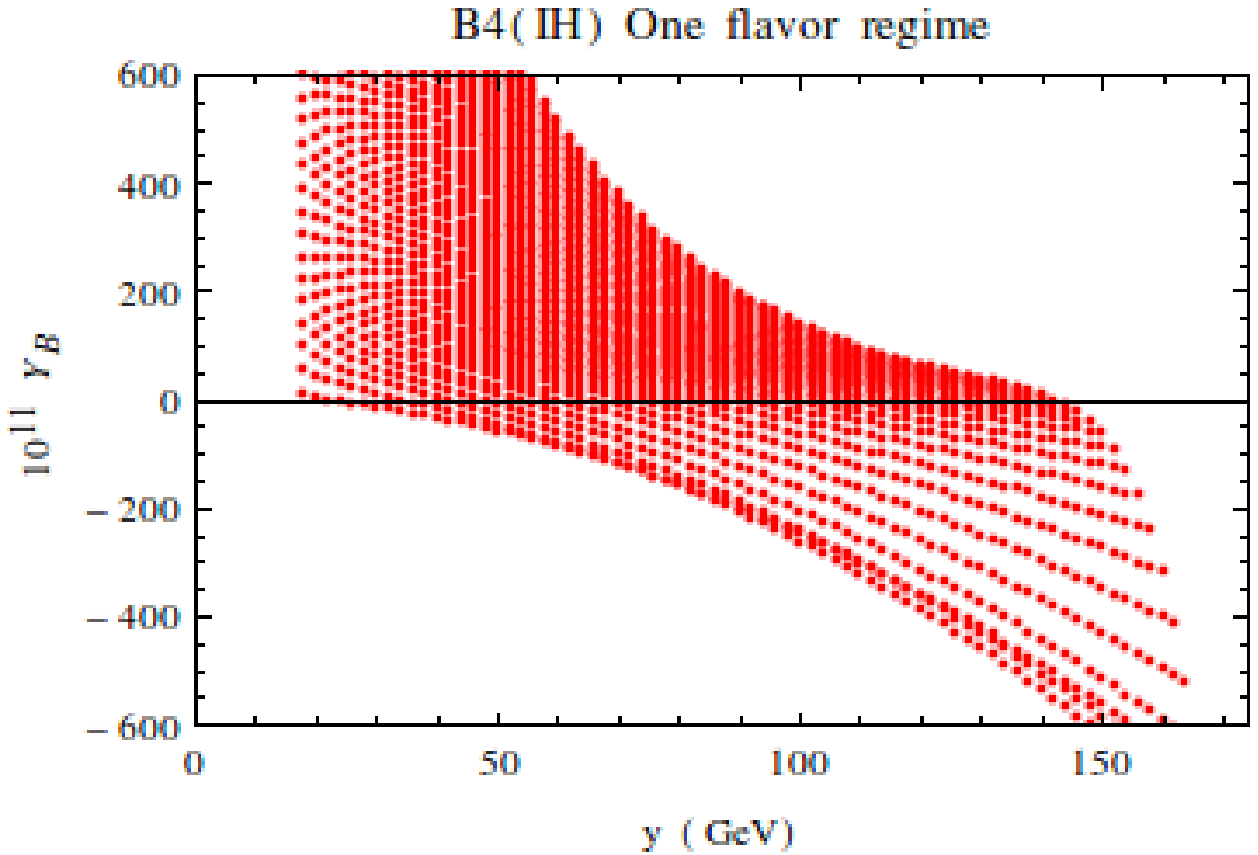}
\end{minipage}
\begin{minipage}{.5\textwidth}
  \centering
  \includegraphics[width=1\linewidth]{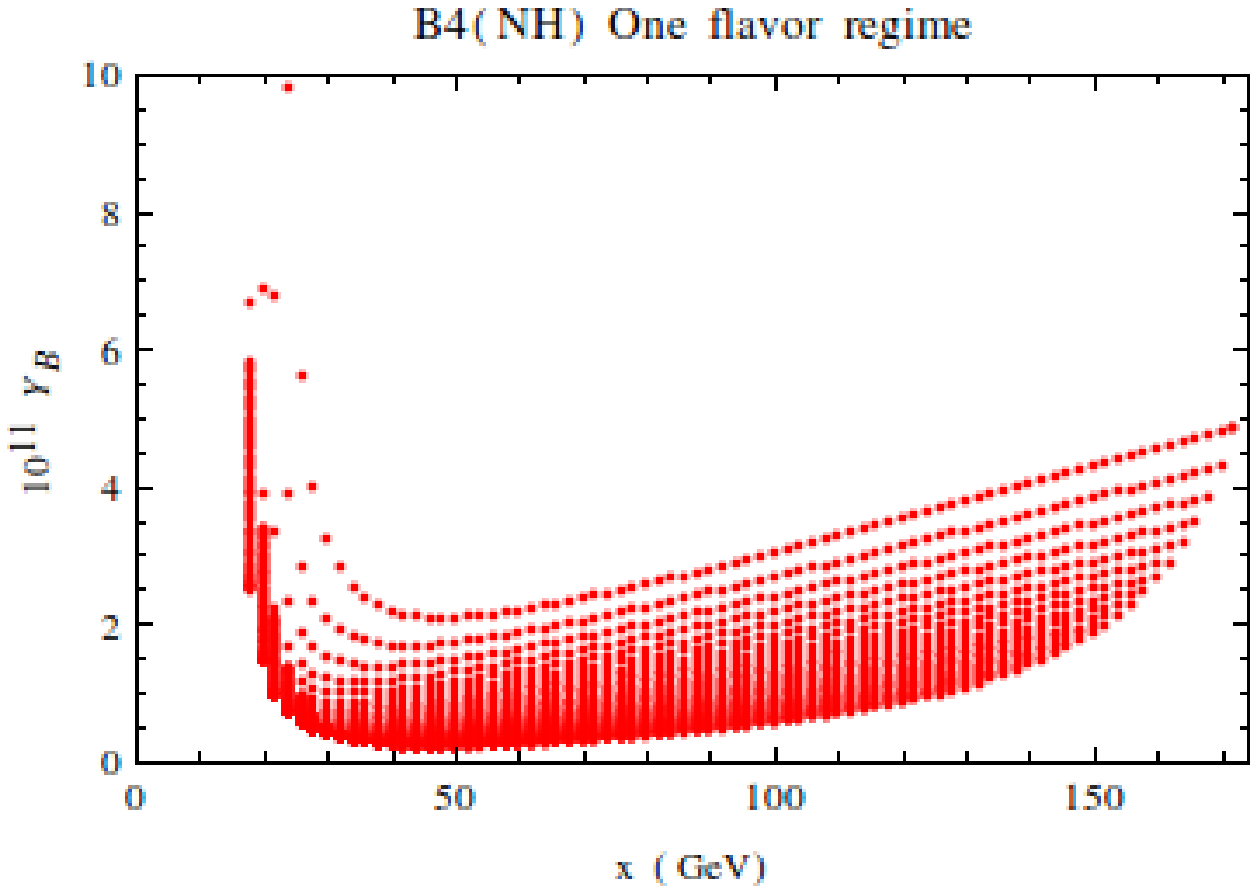}
\end{minipage}%
\begin{minipage}{.5\textwidth}
  \centering
  \includegraphics[width=1\linewidth]{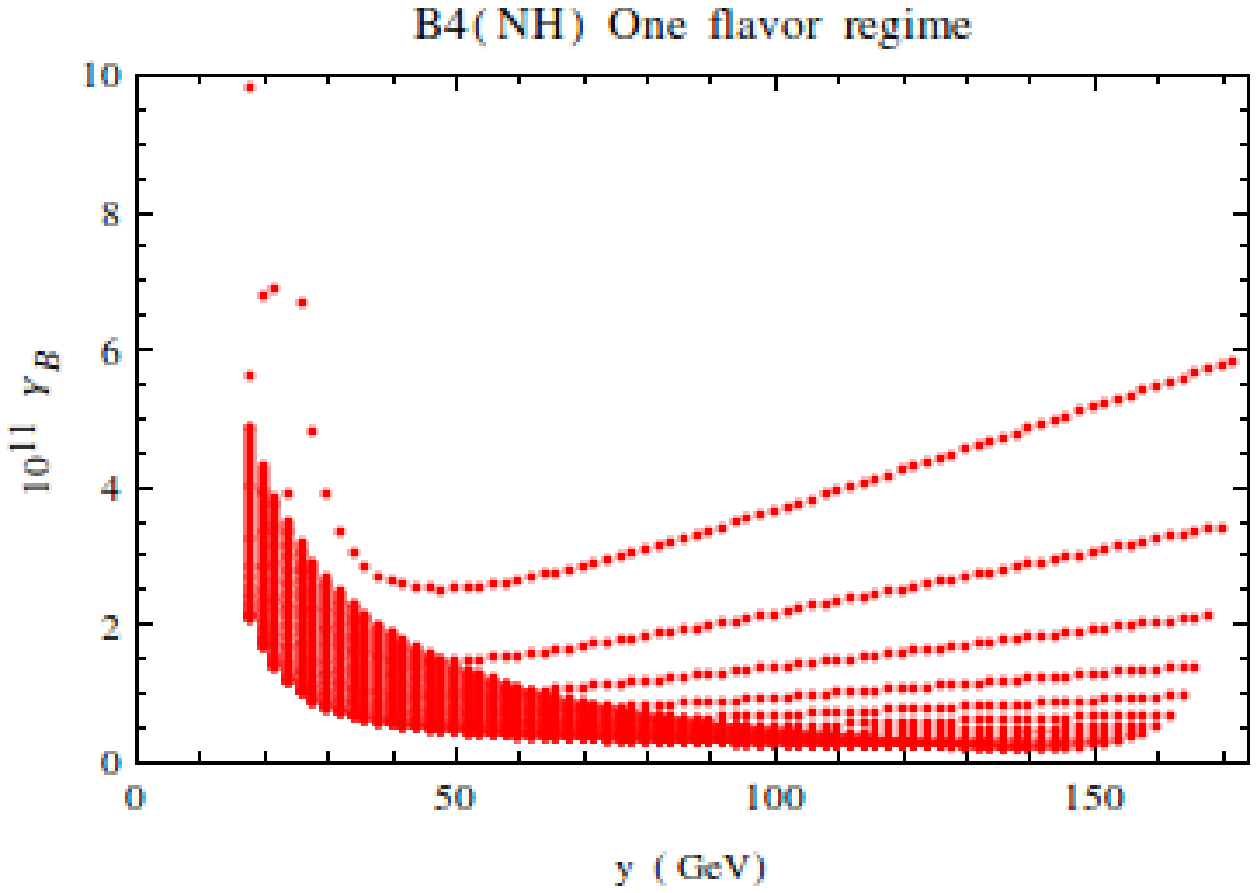}
\end{minipage}
\caption{Variation of baryon asymmetry in one flavor regime with Dirac neutrino masses for two-zero texture $B_4$.}
\label{fig13}
\end{figure}

\subsection{Two-zero texture}
\label{two}
There are fifteen possible two-zero textures of the Majorana neutrino mass matrix $M_{\nu}$ . Using the notations of \cite{Fritzsch:2011qv}, these fifteen two-zero textures of $M_{\nu}$ can be classified into six categories given below:

\begin{equation}
A_1 :\left(\begin{array}{ccc}
0& 0&\times\\
0& \times&\times \\
\times& \times&\times 
\end{array}\right) , 
 A_2 :\left(\begin{array}{ccc}
0& \times&0\\
\times& \times&\times \\
0& \times&\times 
\end{array}\right);
\end{equation}

\begin{equation}
B_1 :\left(\begin{array}{ccc}
\times& \times&0\\
\times& 0&\times \\
0& \times&\times 
\end{array}\right) , 
 B_2 :\left(\begin{array}{ccc}
\times& 0 &\times\\
0& \times&\times \\
\times& \times&0 
\end{array}\right),
 B_3 :\left(\begin{array}{ccc}
\times& 0&\times\\
0& 0&\times \\
\times& \times&\times 
\end{array}\right), 
B_4 :\left(\begin{array}{ccc}
\times& \times&0\\
\times& \times&\times \\
0& \times&0 
\end{array}\right);
\end{equation}

\begin{equation}
C :\left(\begin{array}{ccc}
\times& \times&\times\\
\times& 0&\times \\
\times& \times&0
\end{array}\right);
\end{equation}

\begin{equation}
D_1 :\left(\begin{array}{ccc}
\times& \times&\times\\
\times& 0&0 \\
\times& 0&\times 
\end{array}\right) , 
 D_2 :\left(\begin{array}{ccc}
\times& \times&\times\\
\times& \times&0 \\
\times& 0&0 
\end{array}\right);
\end{equation}

\begin{equation}
E_1 :\left(\begin{array}{ccc}
0& \times&\times\\
\times& 0&\times \\
\times& \times&\times 
\end{array}\right) , 
 E_2 :\left(\begin{array}{ccc}
0& \times &\times\\
\times& \times&\times \\
\times& \times&0 
\end{array}\right),
 E_3 :\left(\begin{array}{ccc}
0& \times&\times\\
\times& \times&0 \\
\times& 0&\times 
\end{array}\right);
\end{equation}

\begin{equation}
F_1 :\left(\begin{array}{ccc}
\times& 0&0\\
0& \times&\times \\
0& \times&\times 
\end{array}\right) , 
 F_2 :\left(\begin{array}{ccc}
\times& 0 &\times\\
0& \times&0 \\
\times& 0&\times 
\end{array}\right),
 F_3 :\left(\begin{array}{ccc}
\times& \times&0\\
\times& \times&0 \\
0& 0&\times 
\end{array}\right), 
\end{equation}
Where the crosses ``$\times$'' imply non-zero arbitrary elements of $M_{\nu}$. In the light of recent oscillation as well as cosmology data, only six different two-zero textures namely, $A_{1,2}$ and $B_{1,2,3,4}$ are favorable as discussed by the authors of \cite{Fritzsch:2011qv,Meloni:2014yea}. We therefore, consider only these six possible two-zero textures for our analysis.

\begin{figure}
\centering
\begin{minipage}{.5\textwidth}
  \centering
  \includegraphics[width=1\linewidth]{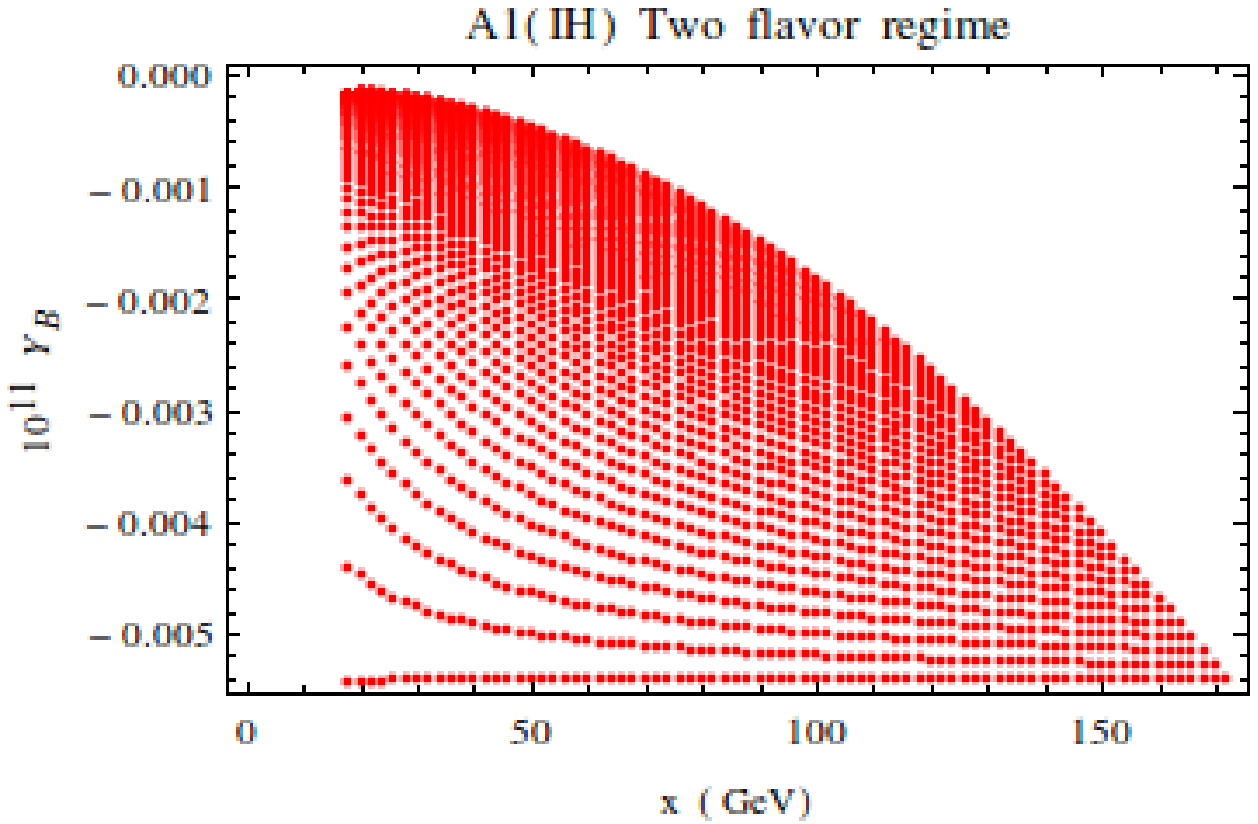}
\end{minipage}%
\begin{minipage}{.5\textwidth}
  \centering
  \includegraphics[width=1\linewidth]{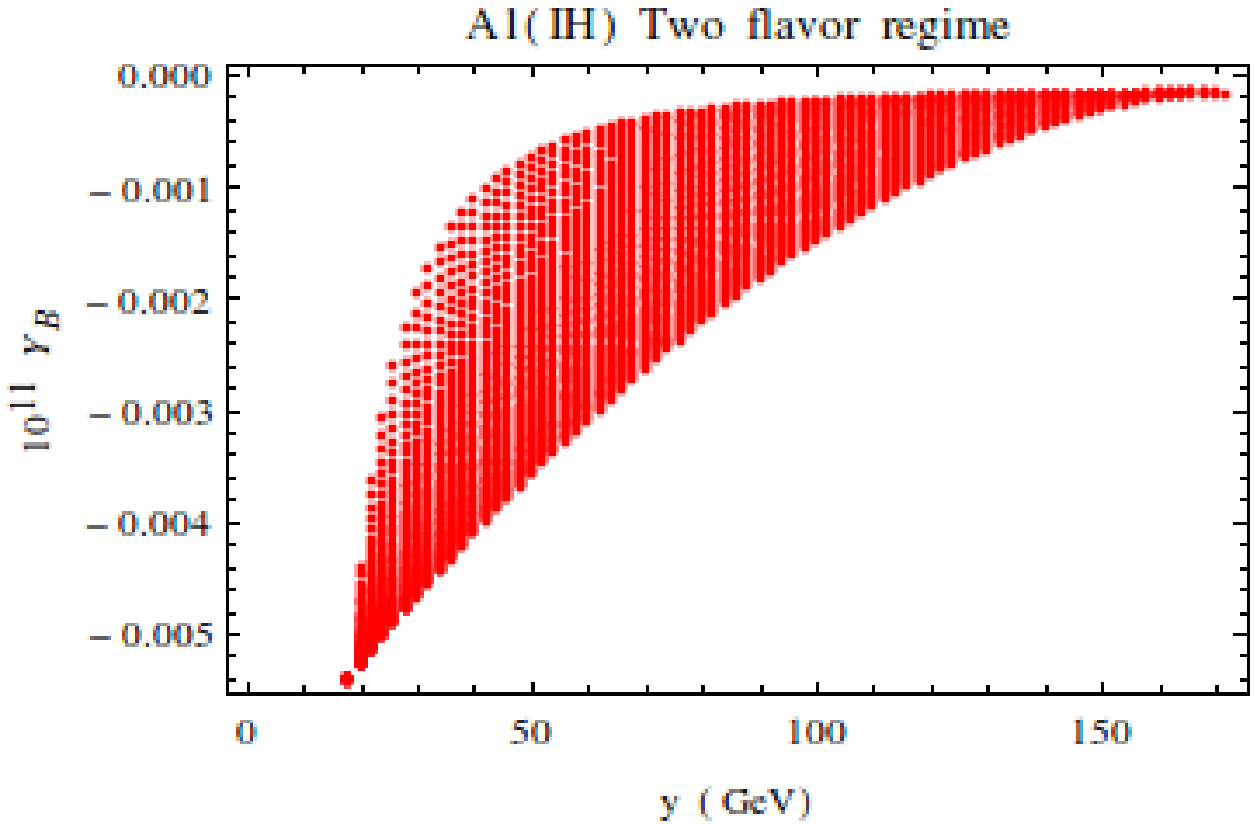}
\end{minipage}
\begin{minipage}{.5\textwidth}
  \centering
  \includegraphics[width=1\linewidth]{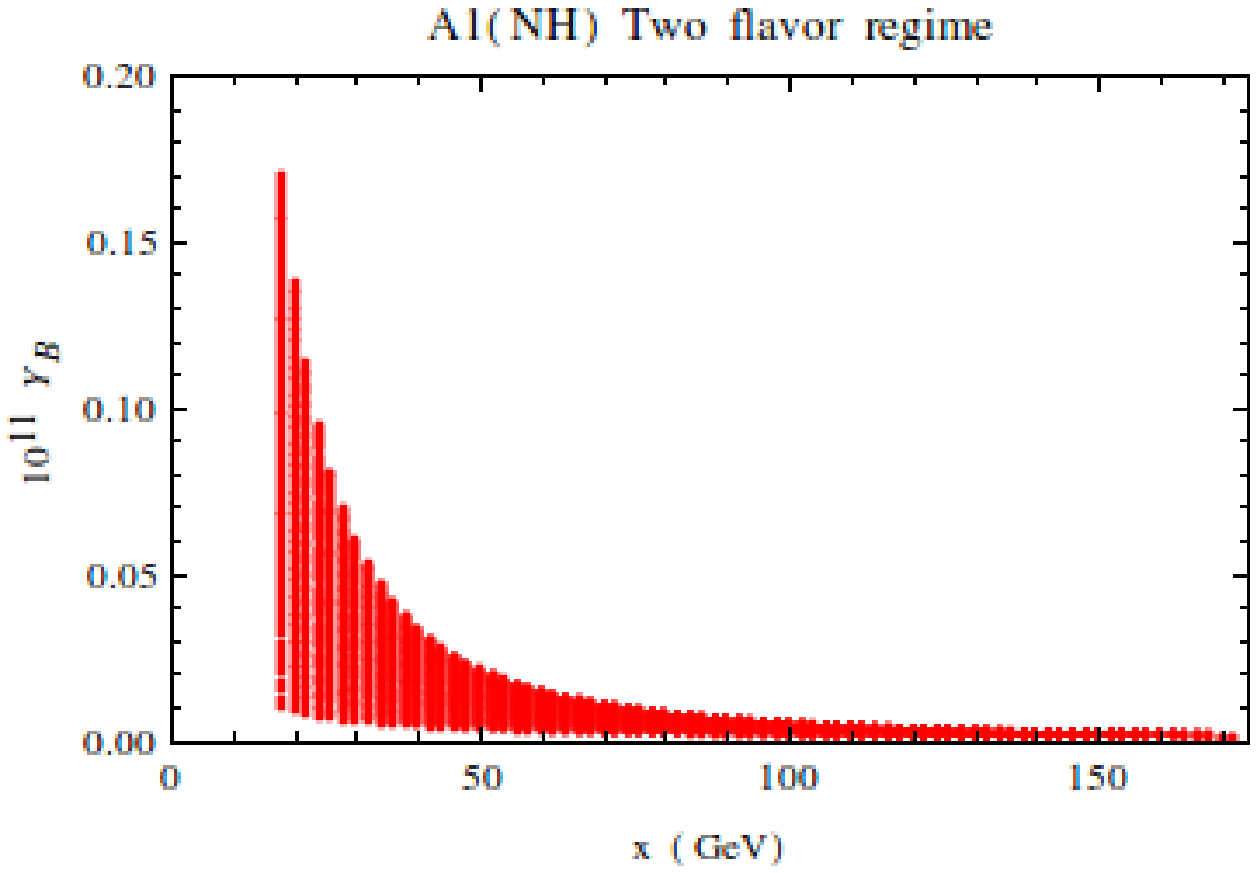}
\end{minipage}%
\begin{minipage}{.5\textwidth}
  \centering
  \includegraphics[width=1\linewidth]{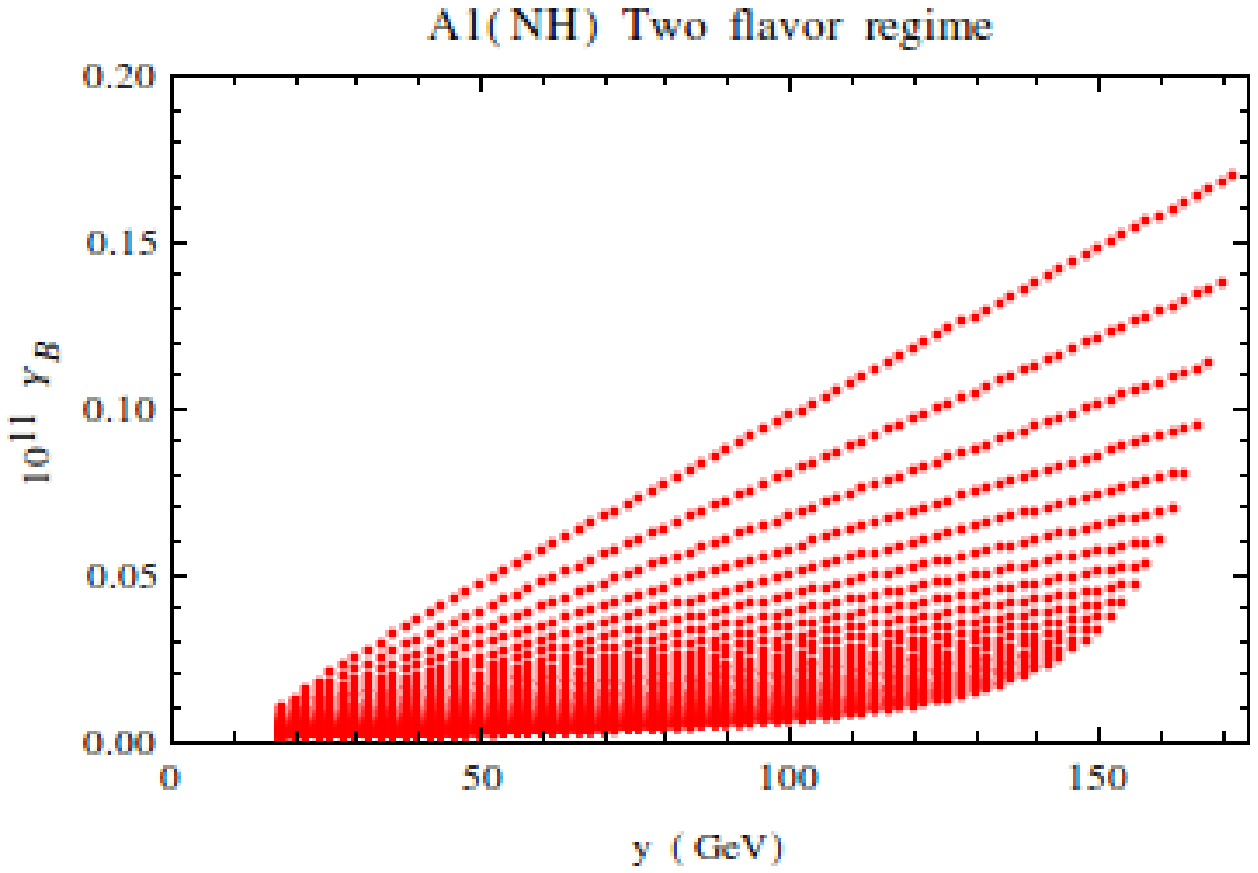}
\end{minipage}
\caption{Variation of baryon asymmetry in two flavor regime with Dirac neutrino masses for two-zero texture $A_1$.}
\label{fig14}
\end{figure}

\begin{figure}
\centering
\begin{minipage}{.5\textwidth}
  \centering
  \includegraphics[width=1\linewidth]{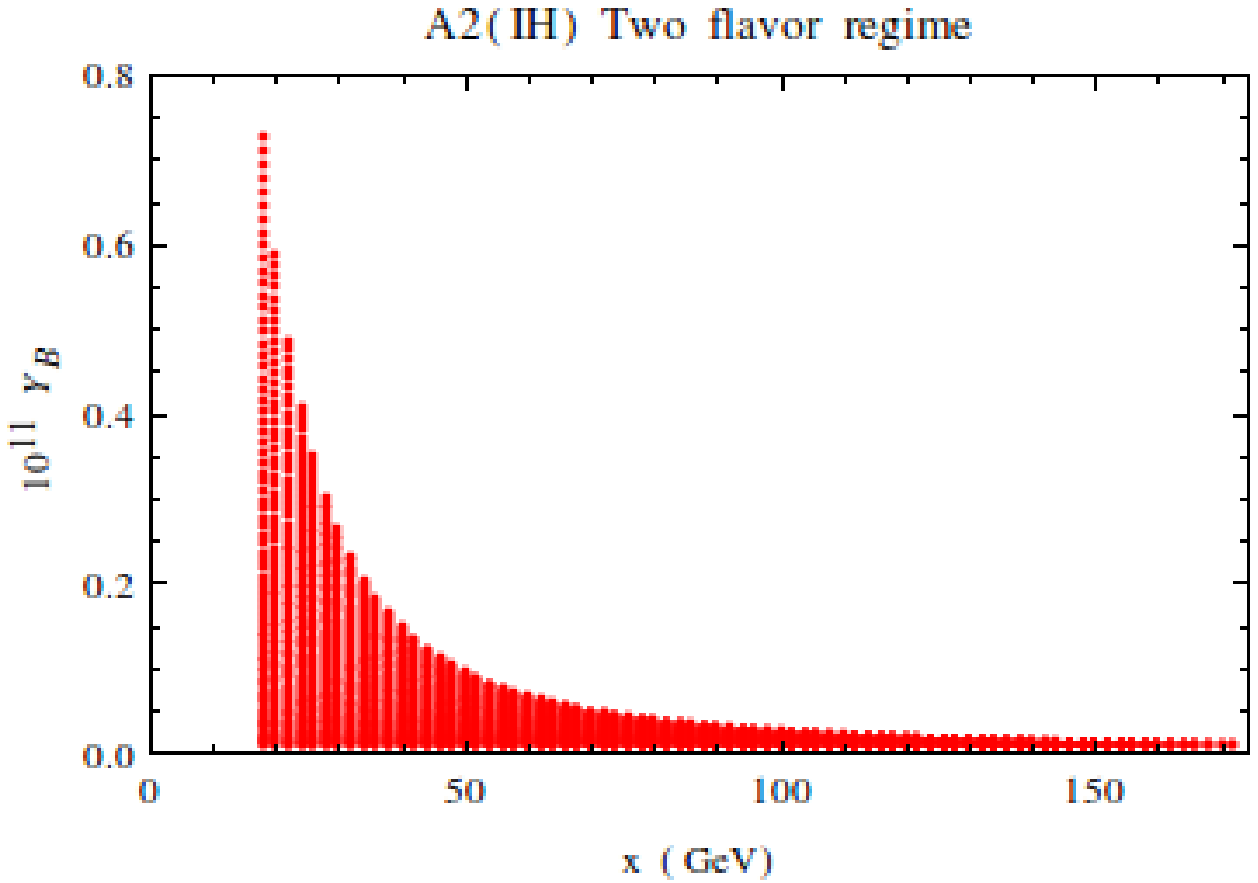}
\end{minipage}%
\begin{minipage}{.5\textwidth}
  \centering
  \includegraphics[width=1\linewidth]{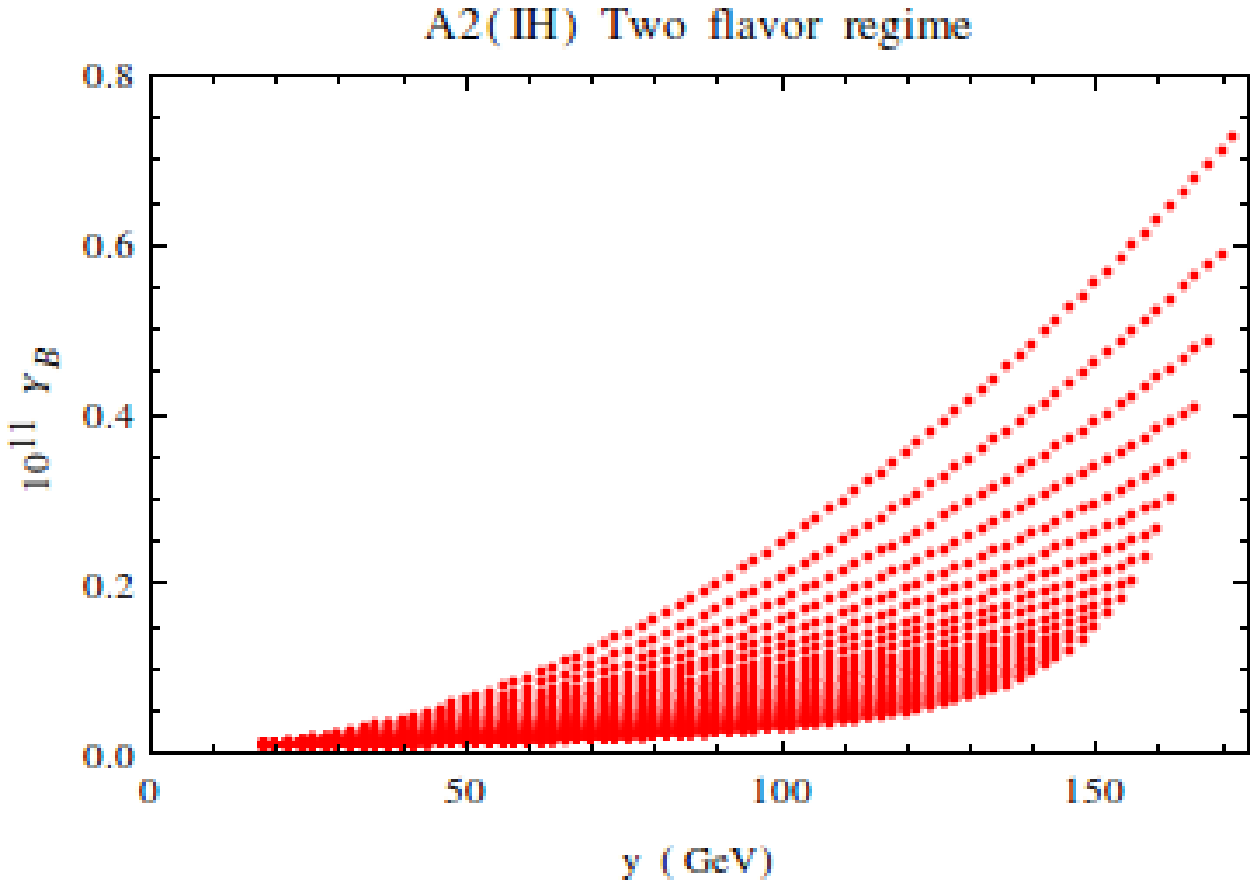}
\end{minipage}
\begin{minipage}{.5\textwidth}
  \centering
  \includegraphics[width=1\linewidth]{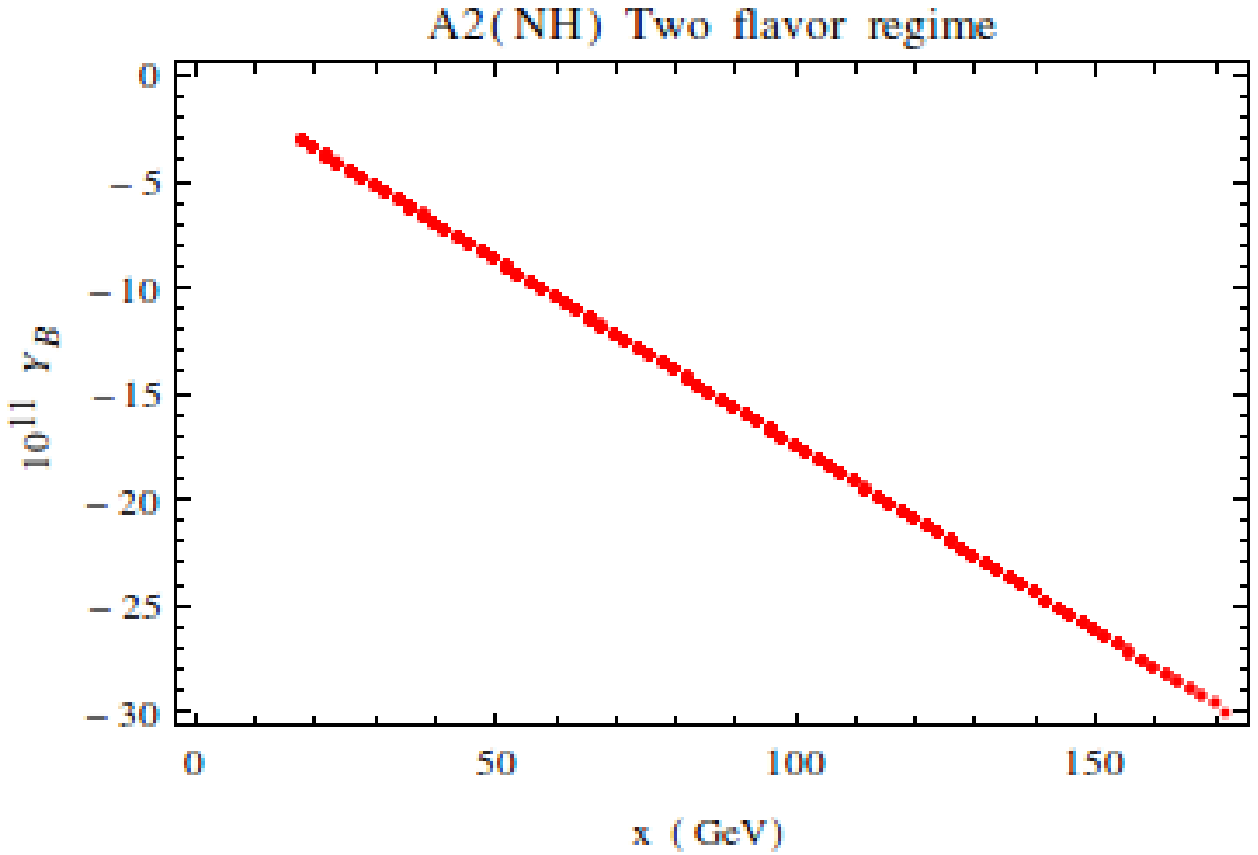}
\end{minipage}%
\begin{minipage}{.5\textwidth}
  \centering
  \includegraphics[width=1\linewidth]{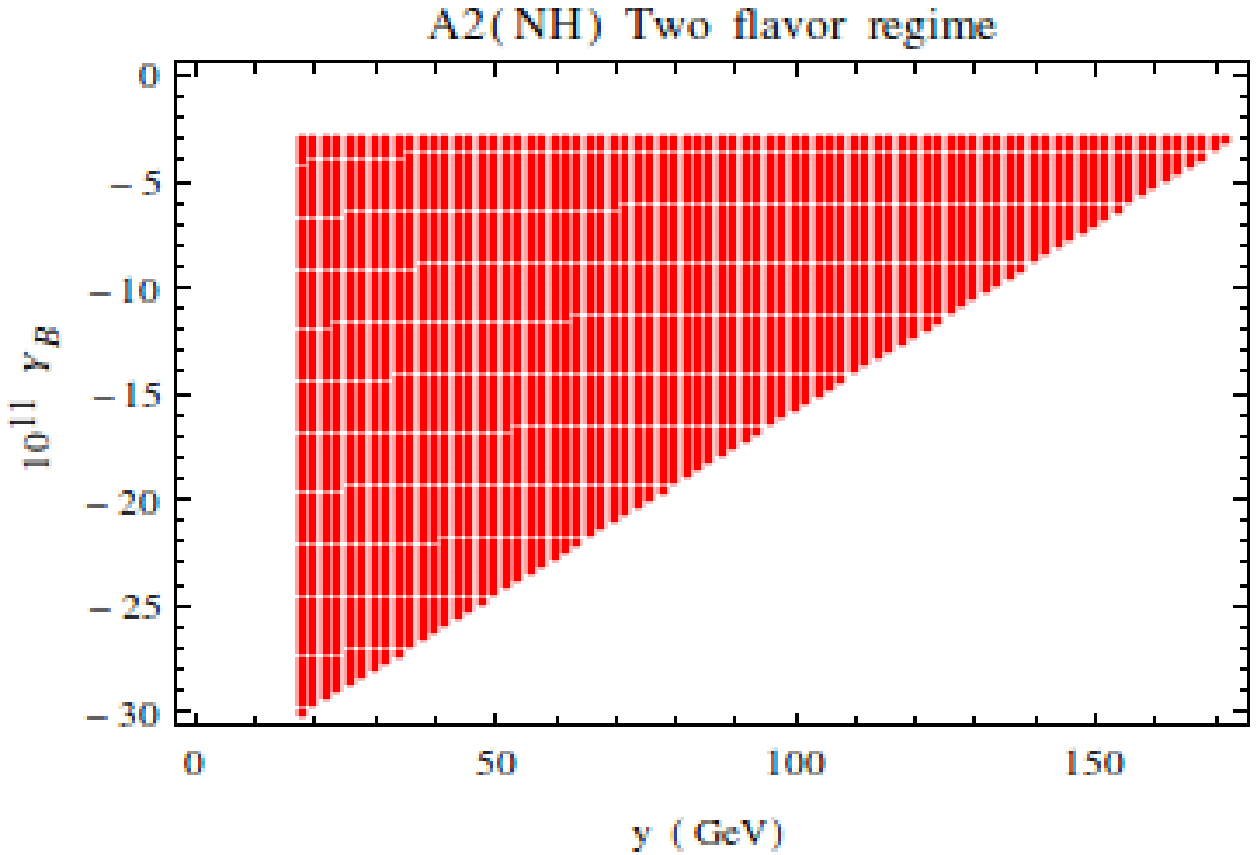}
\end{minipage}
\caption{Variation of baryon asymmetry in two flavor regime with Dirac neutrino masses for two-zero texture $A_2$.}
\label{fig15}
\end{figure}

\begin{figure}
\centering
\begin{minipage}{.5\textwidth}
  \centering
  \includegraphics[width=1\linewidth]{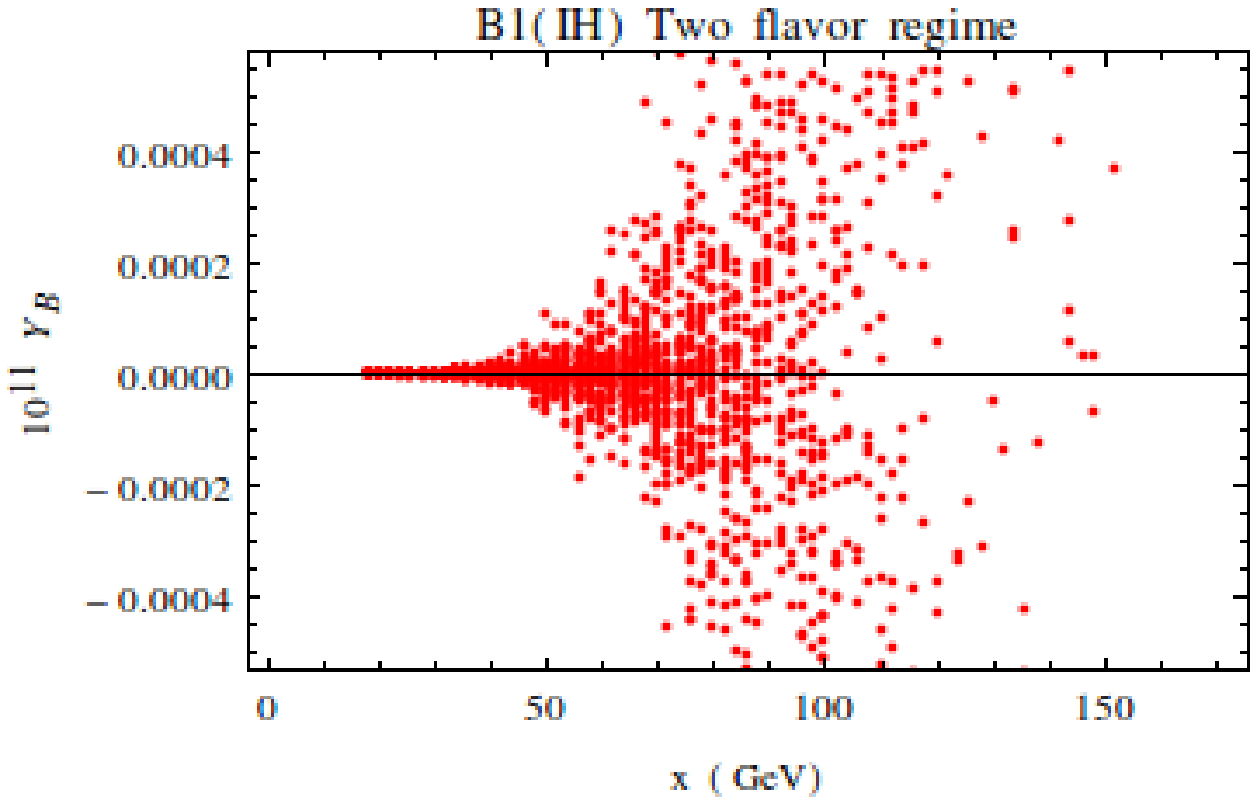}
\end{minipage}%
\begin{minipage}{.5\textwidth}
  \centering
  \includegraphics[width=1\linewidth]{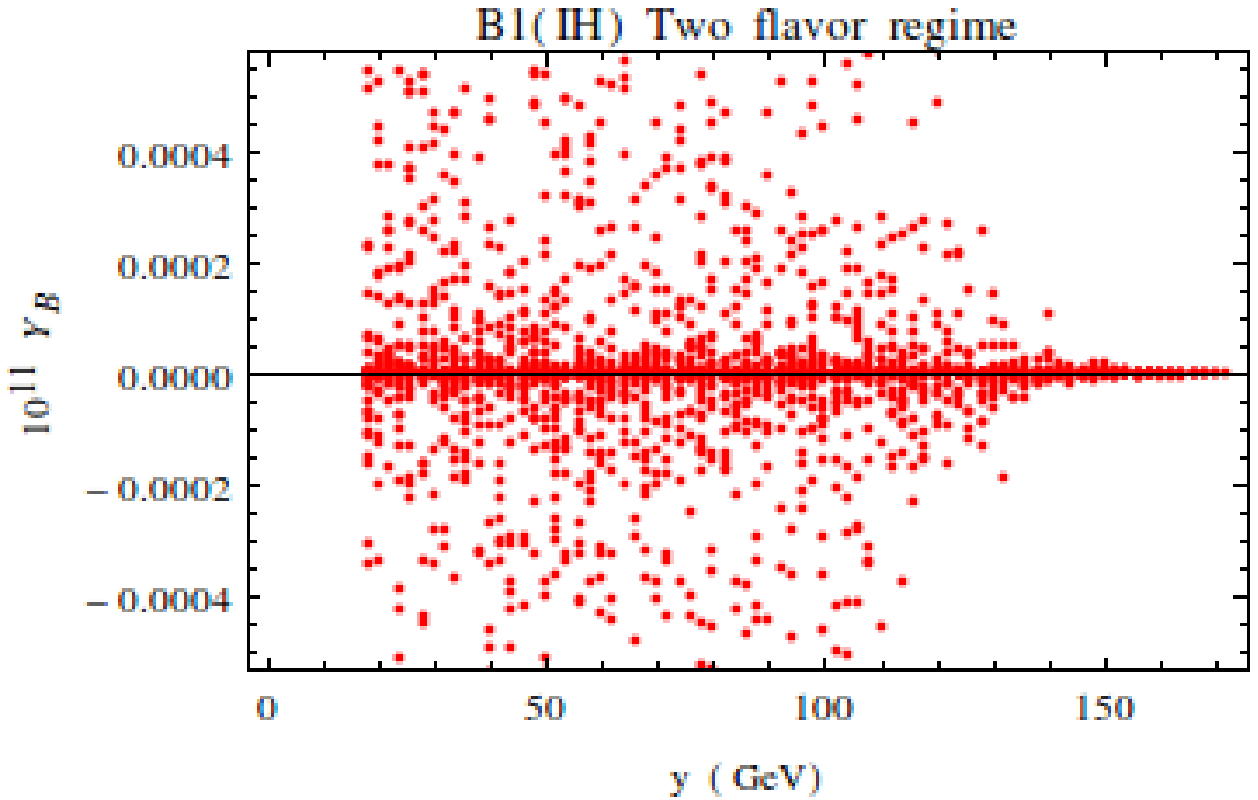}
\end{minipage}
\begin{minipage}{.5\textwidth}
  \centering
  \includegraphics[width=1\linewidth]{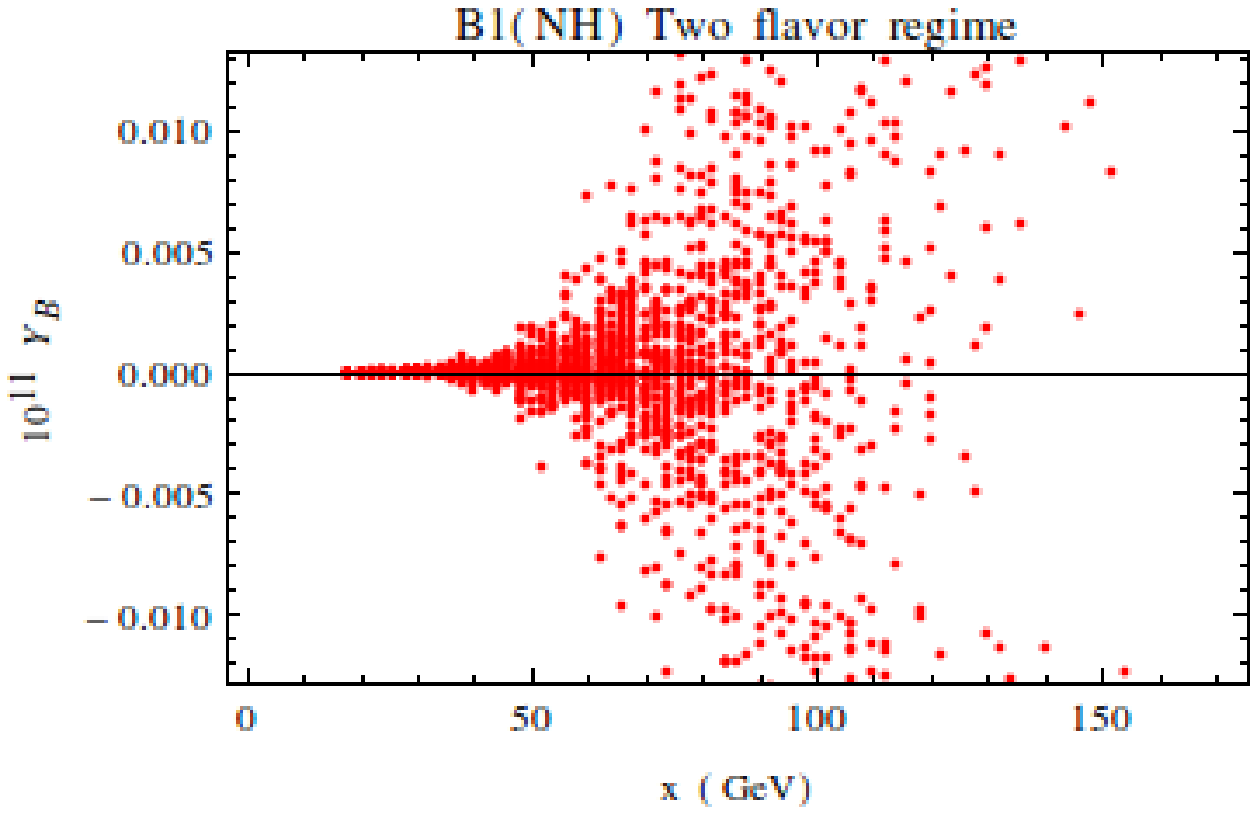}
\end{minipage}%
\begin{minipage}{.5\textwidth}
  \centering
  \includegraphics[width=1\linewidth]{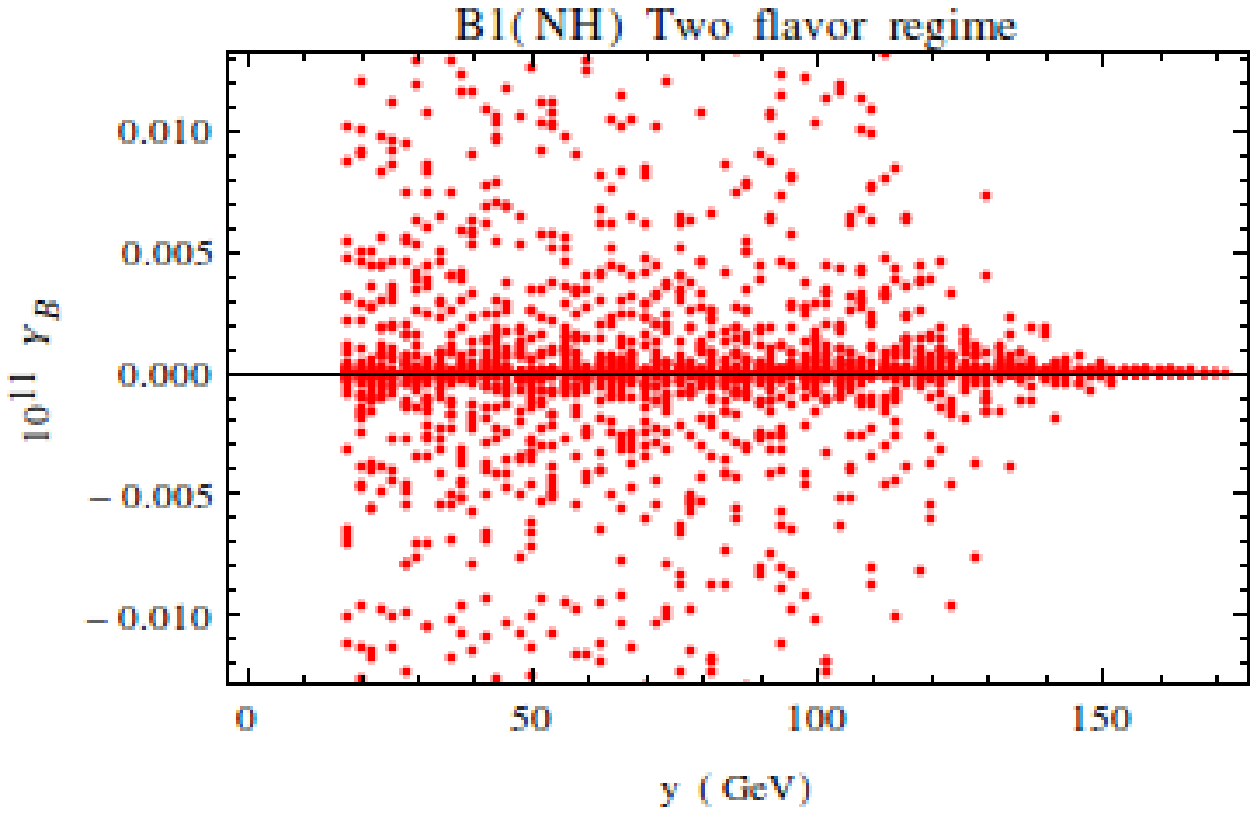}
\end{minipage}
\caption{Variation of baryon asymmetry in two flavor regime with Dirac neutrino masses for two-zero texture $B_1$.}
\label{fig16}
\end{figure}

\begin{figure}
\centering
\begin{minipage}{.5\textwidth}
  \centering
  \includegraphics[width=1\linewidth]{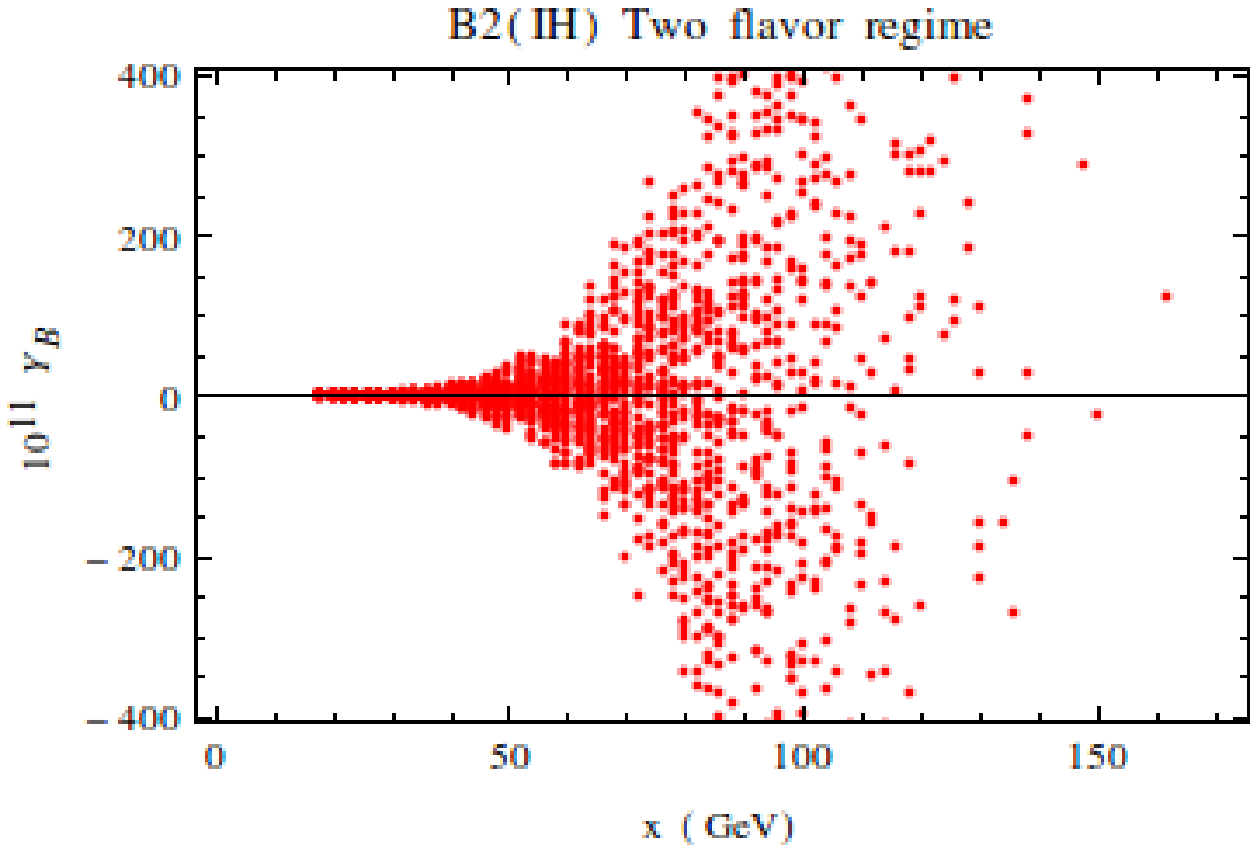}
\end{minipage}%
\begin{minipage}{.5\textwidth}
  \centering
  \includegraphics[width=1\linewidth]{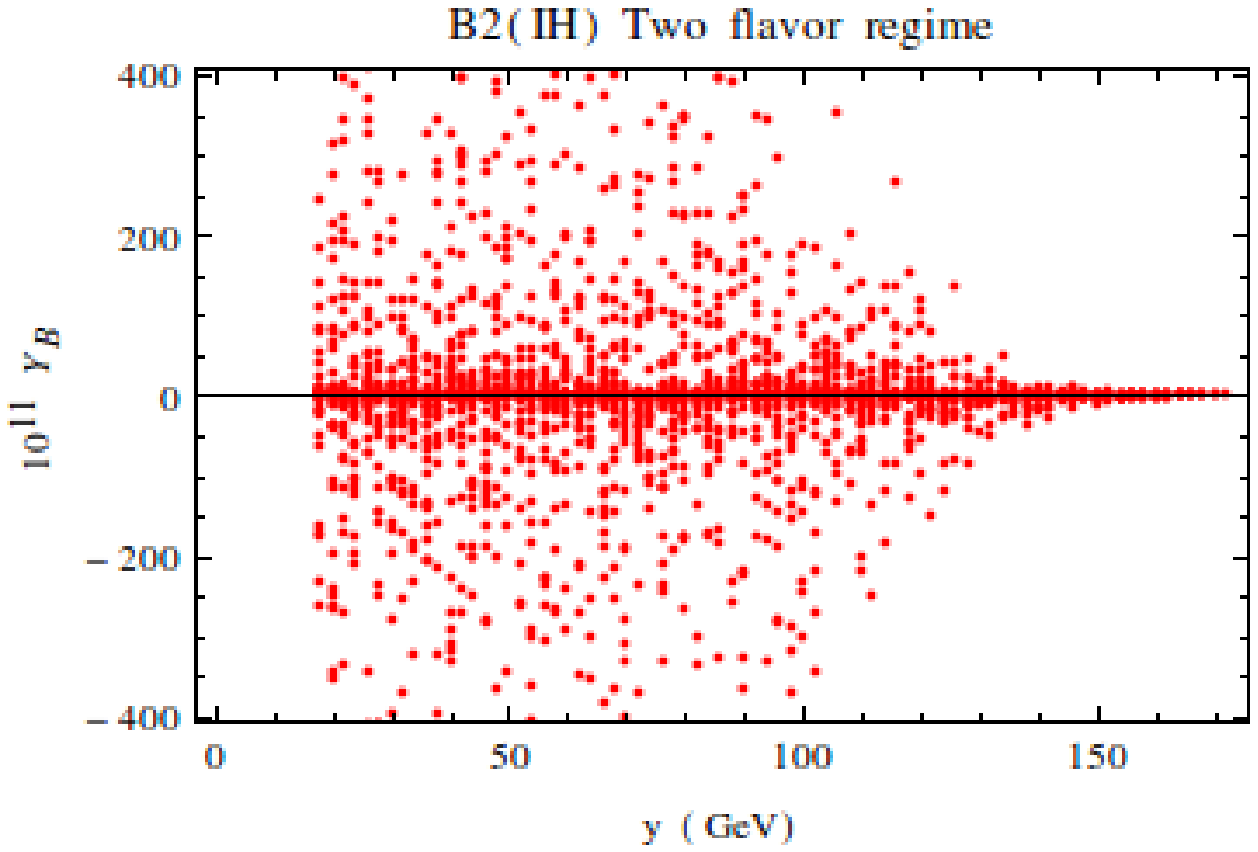}
\end{minipage}
\begin{minipage}{.5\textwidth}
  \centering
  \includegraphics[width=1\linewidth]{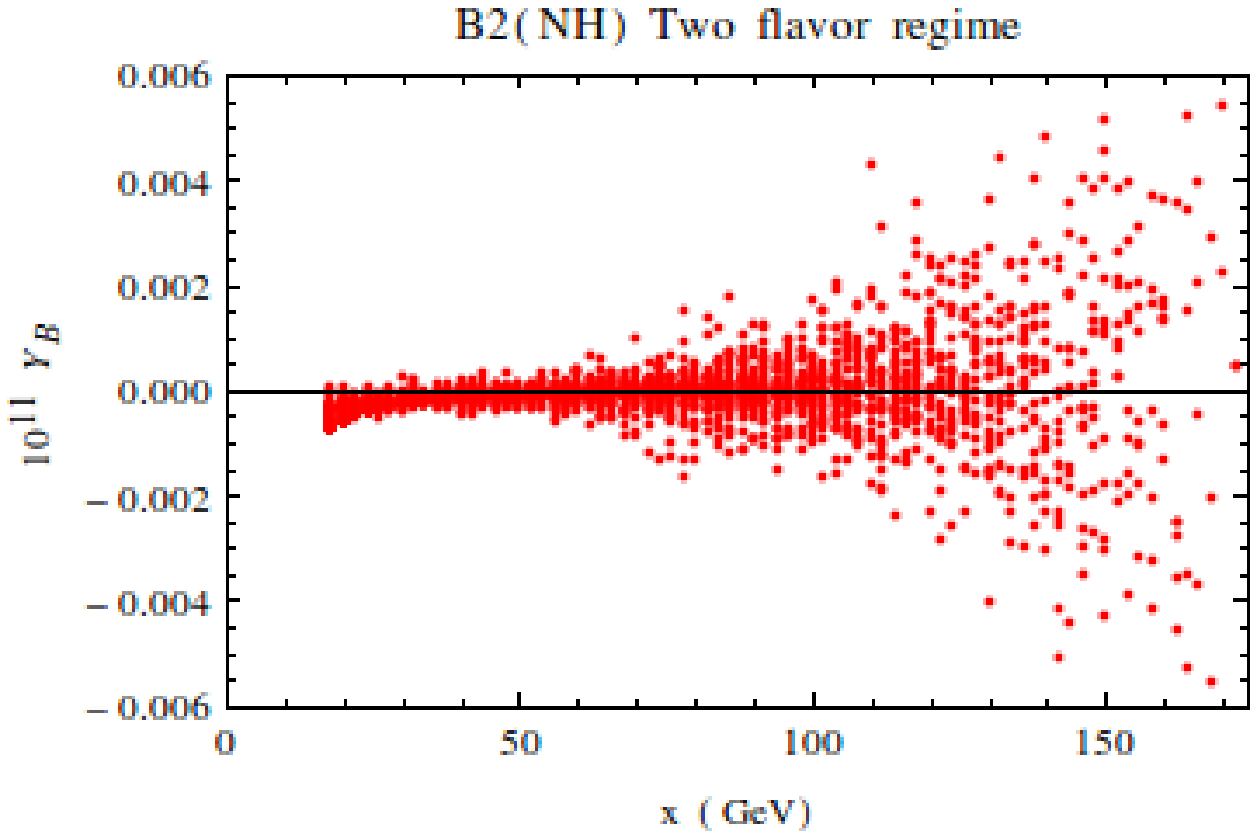}
\end{minipage}%
\begin{minipage}{.5\textwidth}
  \centering
  \includegraphics[width=1\linewidth]{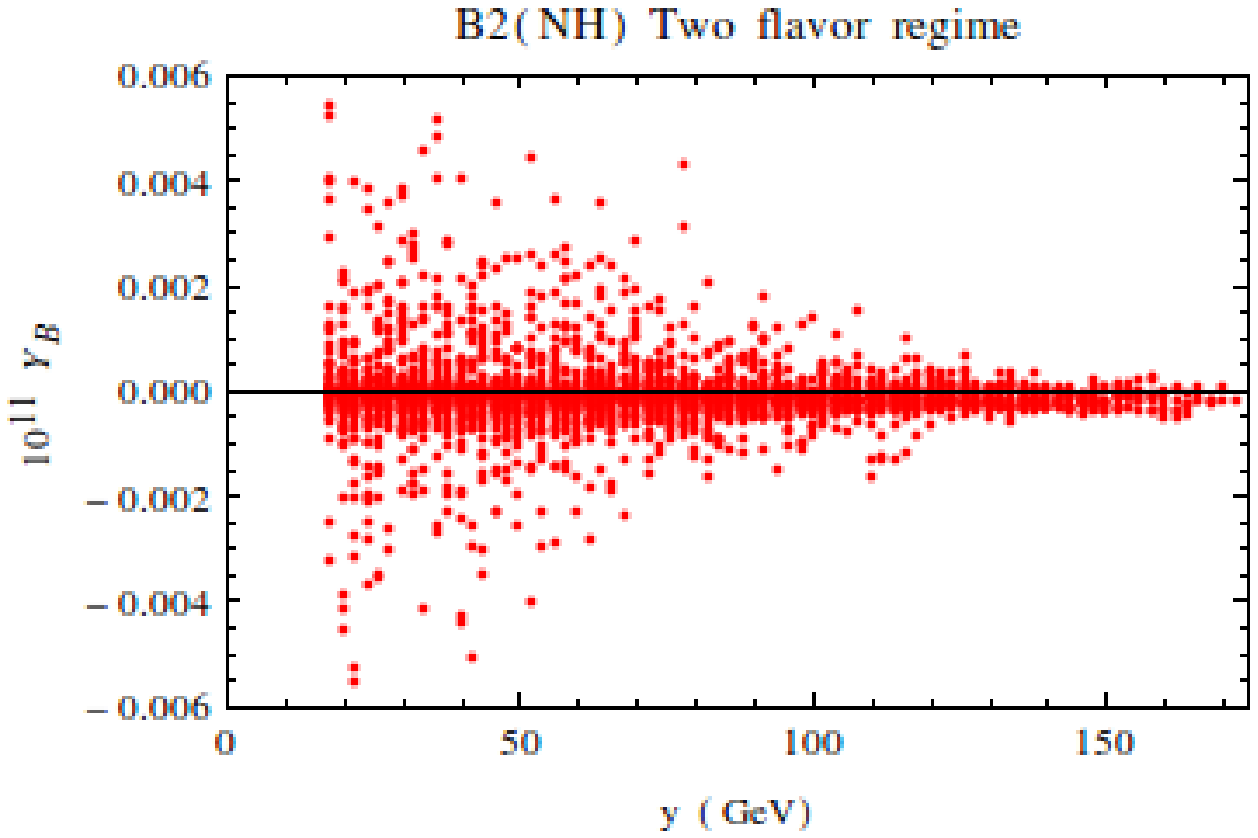}
\end{minipage}
\caption{Variation of baryon asymmetry in two flavor regime with Dirac neutrino masses for two-zero texture $B_2$.}
\label{fig17}
\end{figure}

\begin{figure}
\centering
\begin{minipage}{.5\textwidth}
  \centering
  \includegraphics[width=1\linewidth]{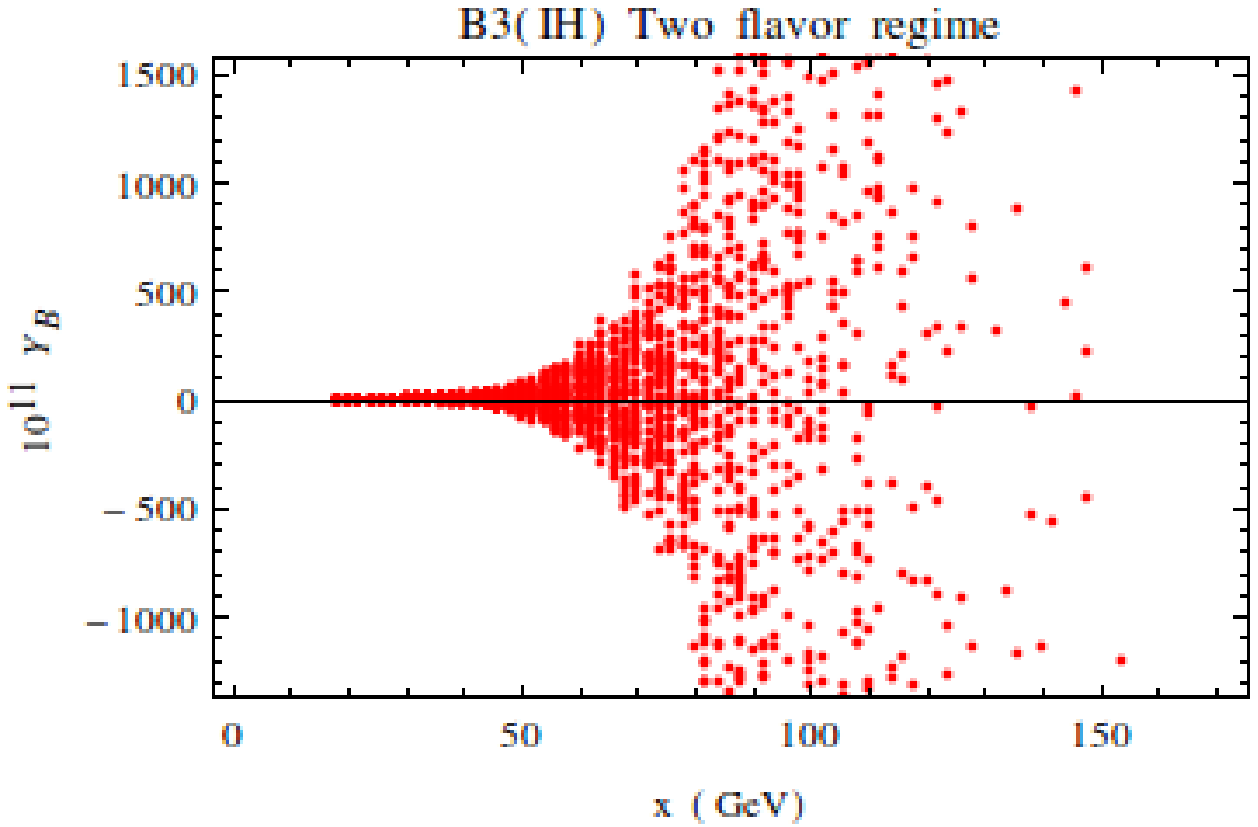}
\end{minipage}%
\begin{minipage}{.5\textwidth}
  \centering
  \includegraphics[width=1\linewidth]{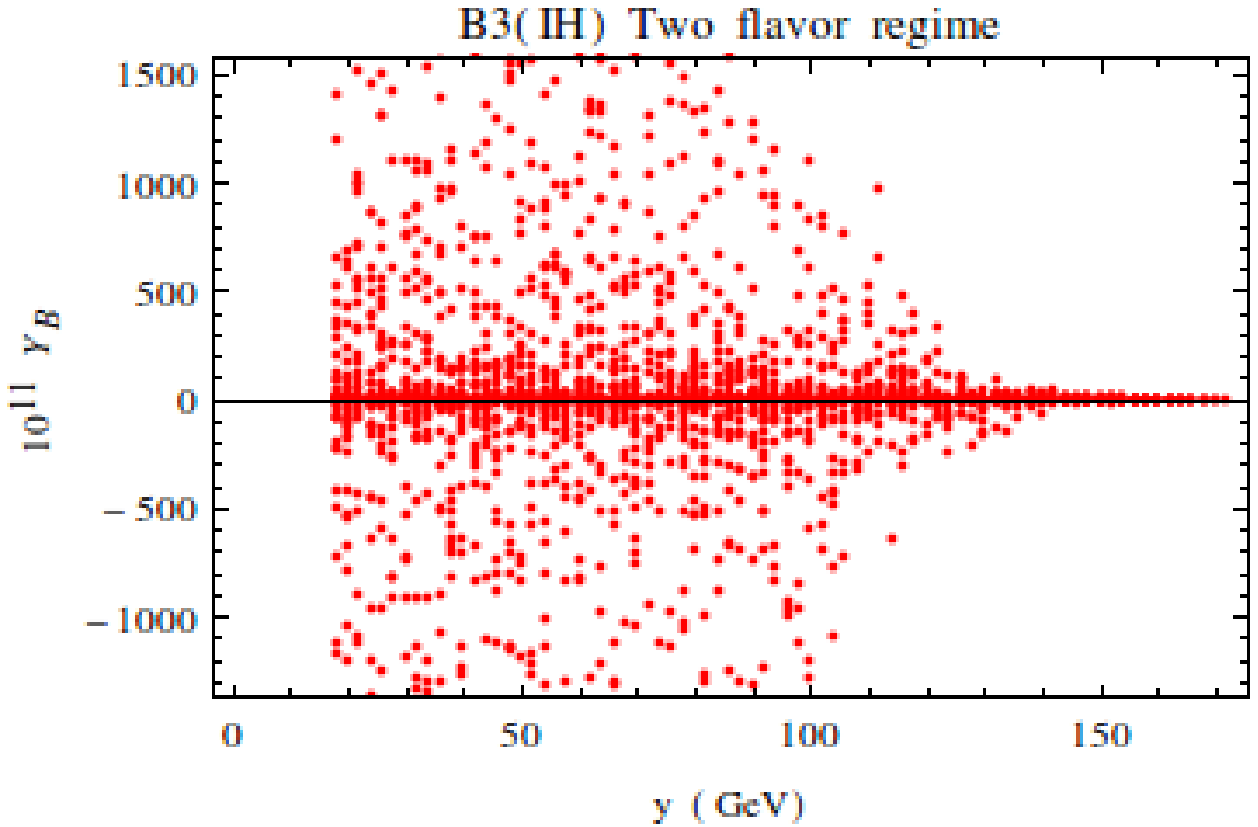}
\end{minipage}
\begin{minipage}{.5\textwidth}
  \centering
  \includegraphics[width=1\linewidth]{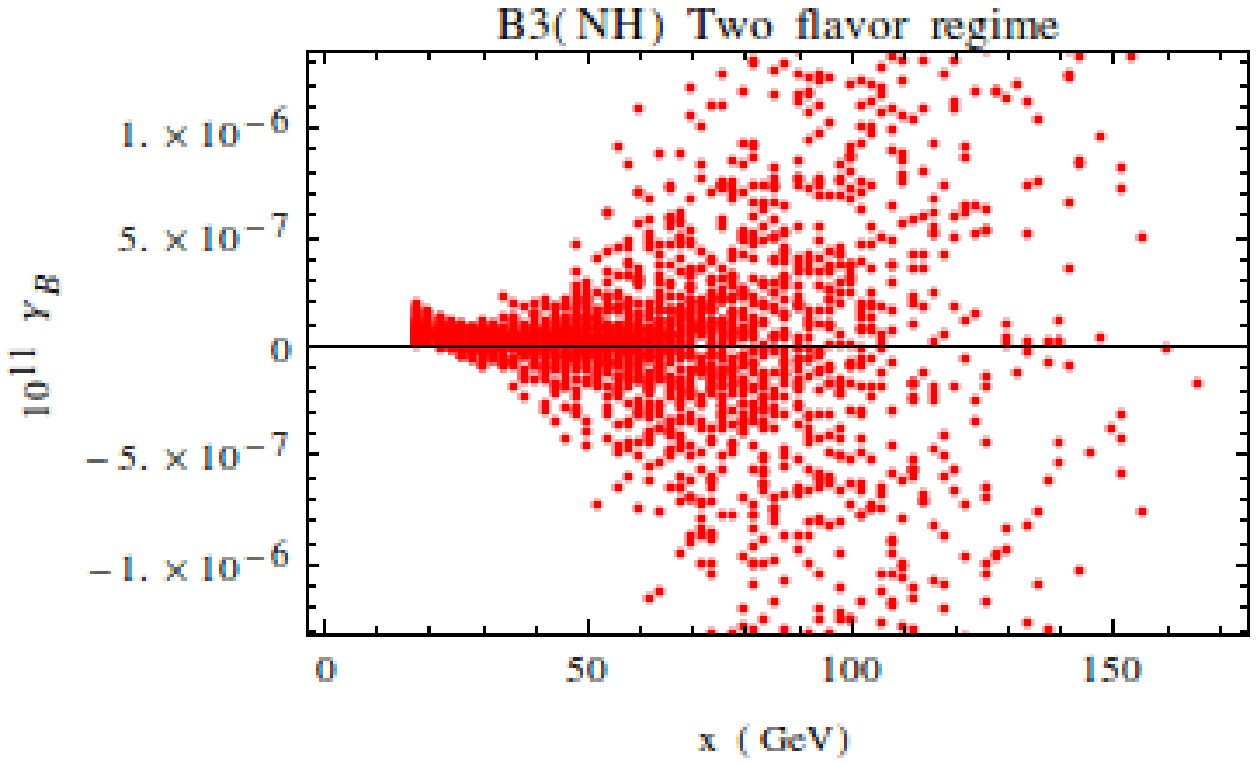}
\end{minipage}%
\begin{minipage}{.5\textwidth}
  \centering
  \includegraphics[width=1\linewidth]{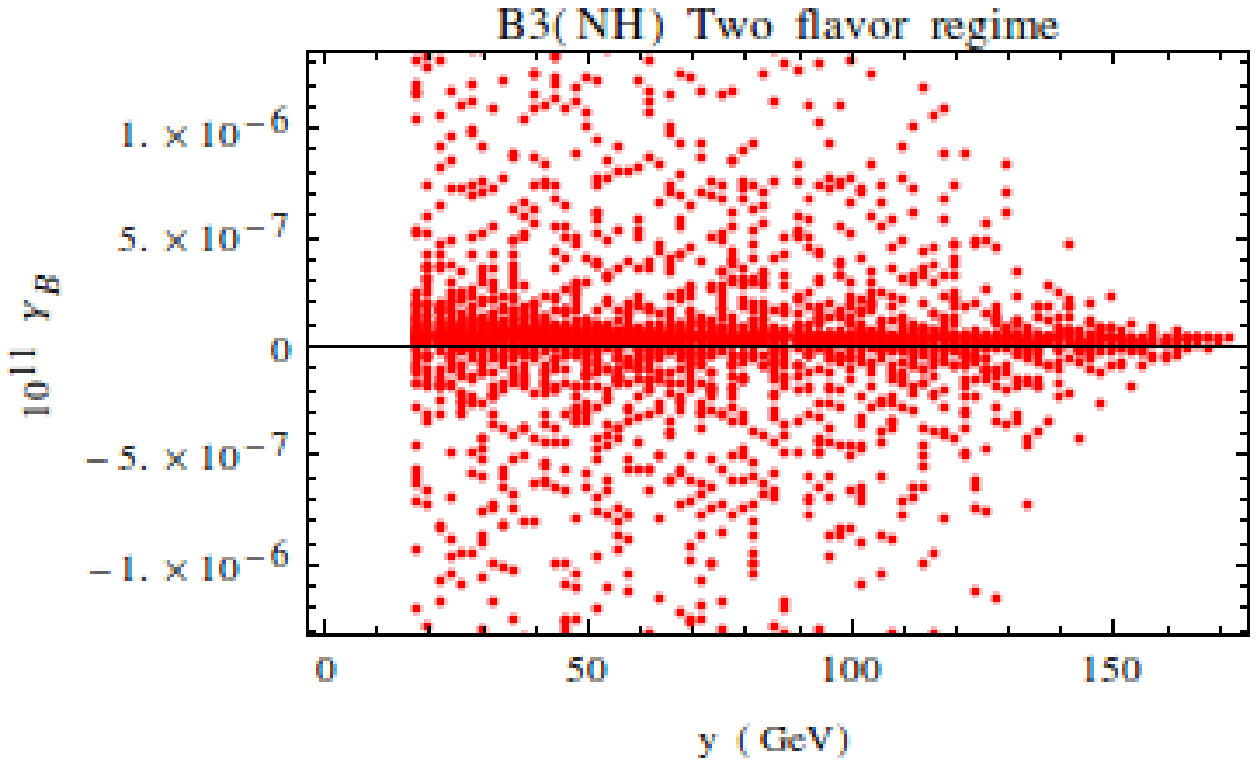}
\end{minipage}
\caption{Variation of baryon asymmetry in two flavor regime with Dirac neutrino masses for two-zero texture $B_3$.}
\label{fig18}
\end{figure}

\begin{figure}
\centering
\begin{minipage}{.5\textwidth}
  \centering
  \includegraphics[width=1\linewidth]{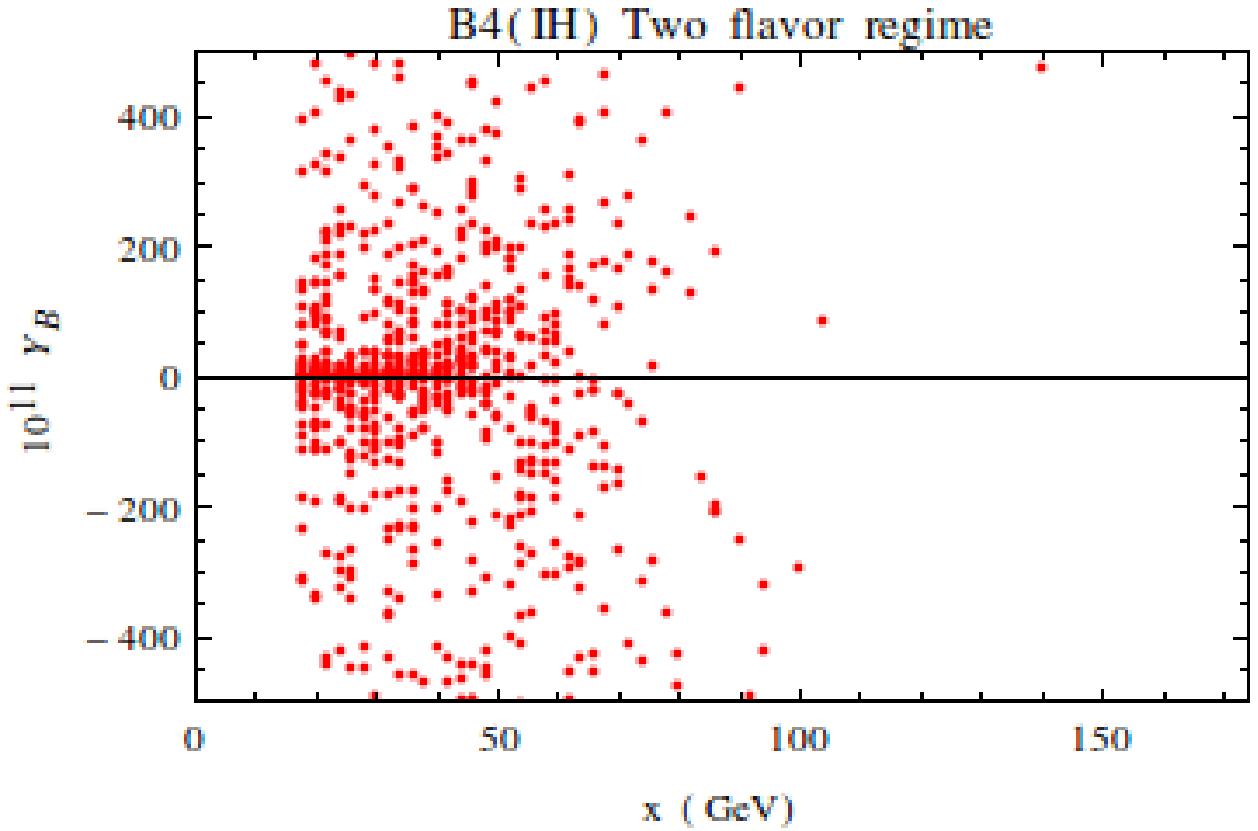}
\end{minipage}%
\begin{minipage}{.5\textwidth}
  \centering
  \includegraphics[width=1\linewidth]{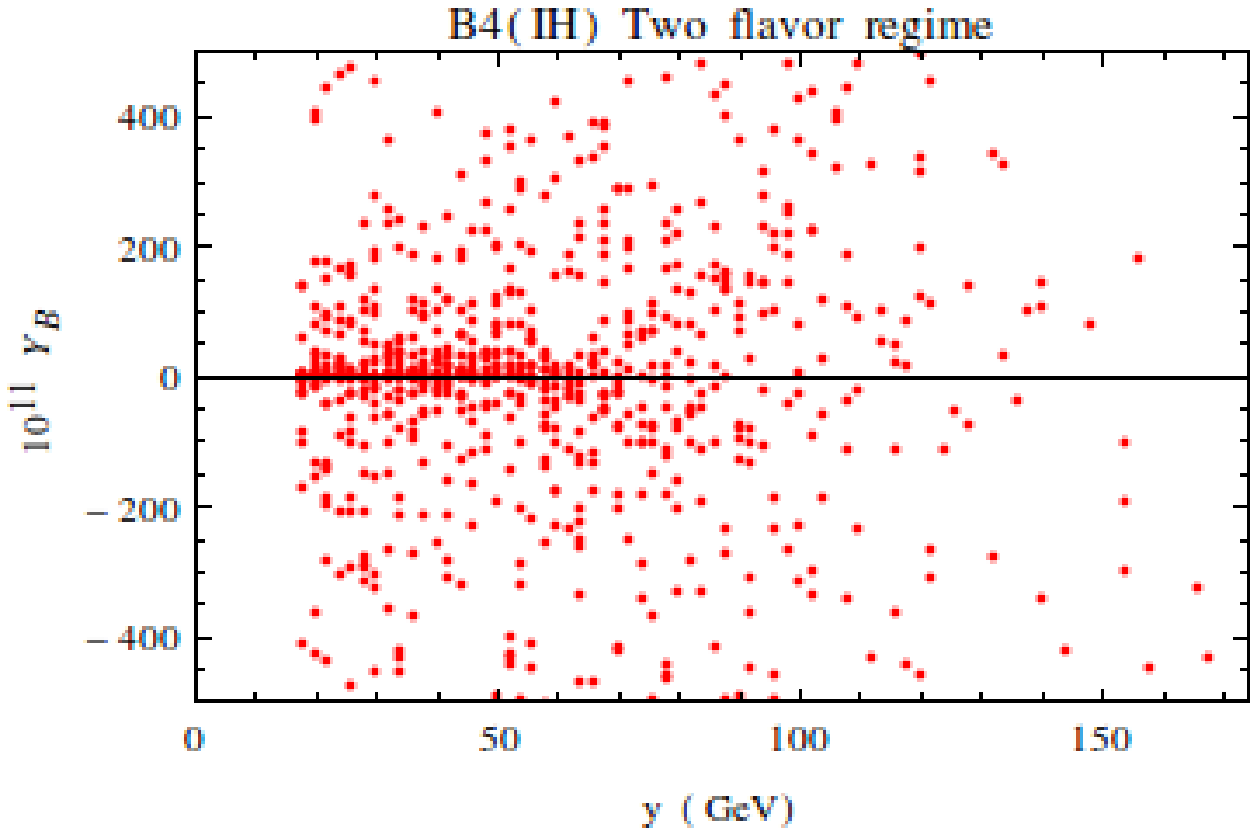}
\end{minipage}
\begin{minipage}{.5\textwidth}
  \centering
  \includegraphics[width=1\linewidth]{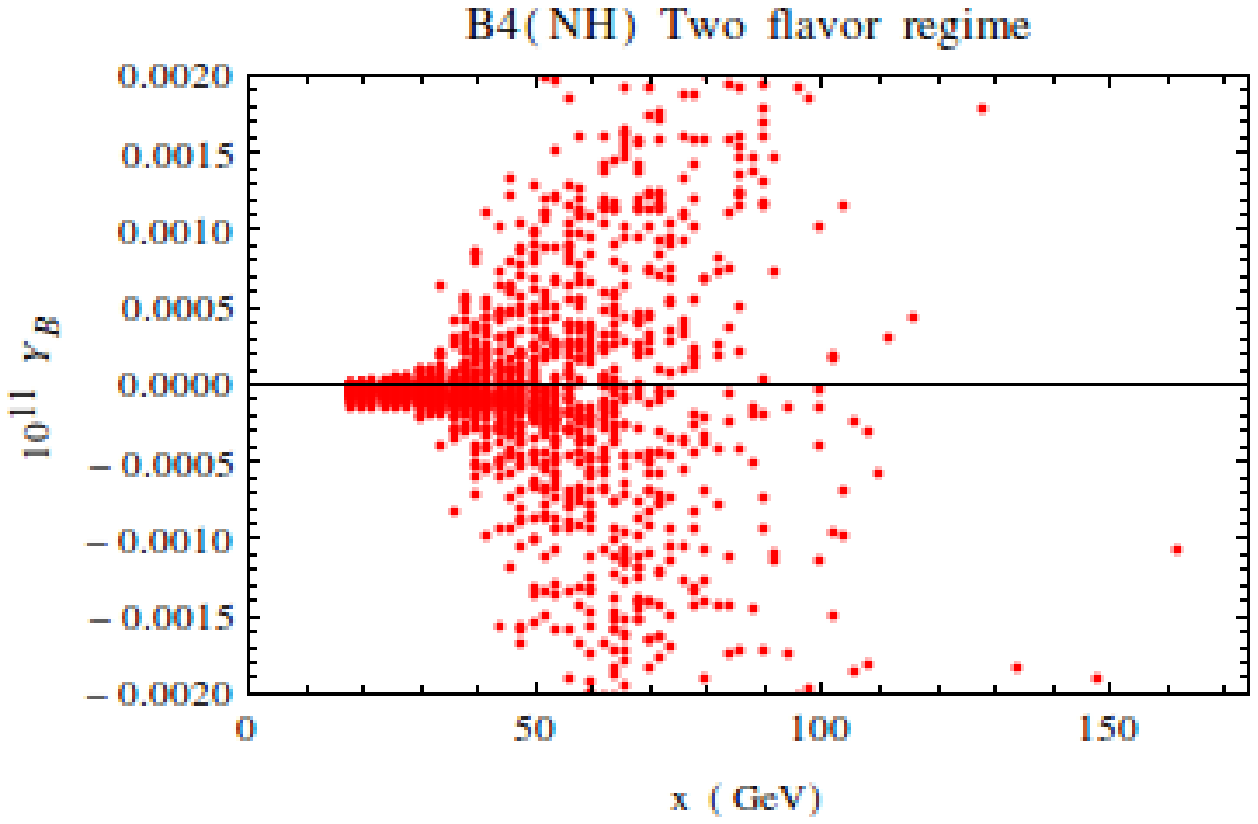}
\end{minipage}%
\begin{minipage}{.5\textwidth}
  \centering
  \includegraphics[width=1\linewidth]{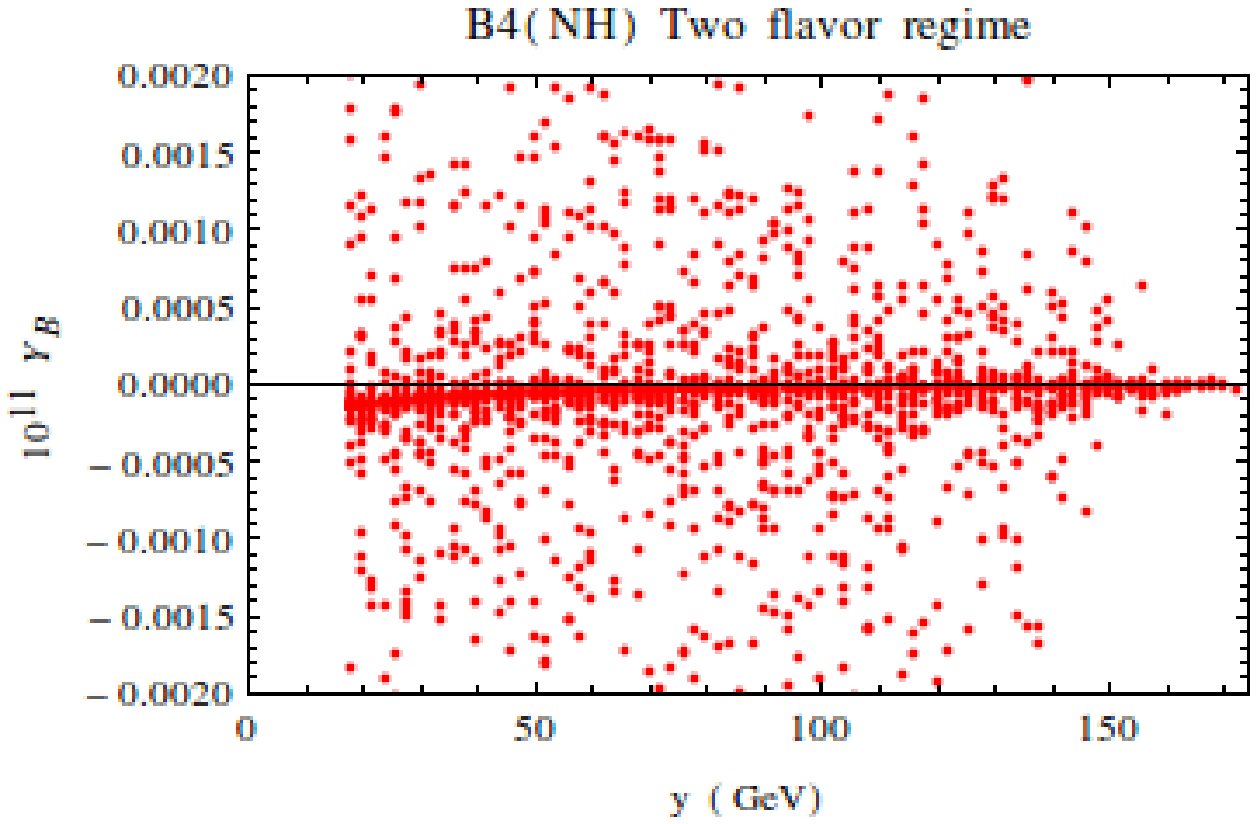}
\end{minipage}
\caption{Variation of baryon asymmetry in two flavor regime with Dirac neutrino masses for two-zero texture $B_4$.}
\label{fig19}
\end{figure}

\section{Leptogenesis}
\label{sec:lepto}
Leptogenesis is one of the most well motivated framework of producing baryon asymmetry of the Universe which creates an asymmetry in the leptonic sector first and then converts it into baryon asymmetry through $B+L$ violating electroweak sphaleron transitions. For a review of leptogenesis, please refer to \cite{Davidson:2008bu}. Although the origin of leptonic mixing and baryon asymmetry could be entirely different, leptogenesis provides a minimal setup to understand the dynamical origin of both these problems in particle physics which remain unsolved till now. There are three basic requirements to produce baryon asymmetry in our Universe which most likely, was in a baryon symmetric state initially. As pointed out first by Sakharov \cite{Sakharov:1967dj}, these three requirements are (i) Baryon number violation, (ii) C and CP violation and (iii) Departure from equilibrium. Although the standard model satisfies the first two requirements and out of equilibrium conditions in principle, can be achieved in an expanding Universe like ours, it turns out that the amount of CP violation measured in the SM quark sector is too small to account for the entire baryon asymmetry of the Universe. Since there can be more sources of CP violating phases in the leptonic sector which are not yet from experimentally determined, leptogenesis provides an indirect way of constraining these unknown phases from the requirement of producing the observed baryon asymmetry.

In a model with type I seesaw mechanism at work, the CP violating out of equilibrium decay of the lightest right handed neutrino can give rise to the required lepton asymmetry. The neutrino mass matrix in type I seesaw mechanism can be written as 
\begin{equation}
M_{\nu}=-m_{LR}M_{RR}^{-1}m_{LR}^{T}.
\label{eq:type1}
\end{equation}
where $m_{LR}$ is the Dirac neutrino mass matrix and $M_{RR}$ is the right handed singlet neutrino mass matrix. We note that the Pontecorvo-Maki-Nakagawa-Sakata (PMNS) leptonic mixing matrix is related to the diagonalizing 
matrices of neutrino and charged lepton mass matrices $U_{\nu}, U_l$ respectively, as
\begin{equation}
U_{\text{PMNS}} = U^{\dagger}_l U_{\nu}
\label{pmns0}
\end{equation}
In the diagonal charged lepton basis, $U_{\text{PMNS}}$ is same as the diagonalizing matrix $U_{\nu}$ of the neutrino mass matrix given by \eqref{eq:type1}. The PMNS mixing matrix can be parametrized as
\begin{equation}
U_{\text{PMNS}}=\left(\begin{array}{ccc}
c_{12}c_{13}& s_{12}c_{13}& s_{13}e^{-i\delta}\\
-s_{12}c_{23}-c_{12}s_{23}s_{13}e^{i\delta}& c_{12}c_{23}-s_{12}s_{23}s_{13}e^{i\delta} & s_{23}c_{13} \\
s_{12}s_{23}-c_{12}c_{23}s_{13}e^{i\delta} & -c_{12}s_{23}-s_{12}c_{23}s_{13}e^{i\delta}& c_{23}c_{13}
\end{array}\right) \text{diag}(1, e^{i\alpha}, e^{i(\beta+\delta )})
\label{matrixPMNS}
\end{equation}
where $c_{ij} = \cos{\theta_{ij}}, \; s_{ij} = \sin{\theta_{ij}}$. $\delta$ is the Dirac CP phase and  $\alpha, \beta$ are the Majorana phases.

In our work we are considering CP-violating out of equilibrium decay of heavy right handed neutrinos into Higgs and lepton 
within the framework type I seesaw mechanism. The lepton asymmetry from the decay of right handed neutrino into leptons 
and Higgs scalar is given by
\begin{equation}
\epsilon_{N_k} = \sum_i \frac{\Gamma(N_k \rightarrow L_i +H^*)-\Gamma (N_k \rightarrow \bar{L_i}+H)}{\Gamma(N_k \rightarrow L_i +H^*)
+\Gamma (N_k \rightarrow \bar{L_i}+H)}
\end{equation}
In a hierarchical pattern of three right handed neutrinos $M_{2,3} \gg M_1$, it is sufficient to consider the lepton asymmetry produced by 
the decay of the lightest right handed neutrino $N_1$. Following the notations of \cite{Joshipura:2001ya}, the lepton asymmetry arising from the decay of $N_1$ in the 
presence of type I seesaw only can be written as
\begin{eqnarray}
\epsilon^{\alpha}_1 &=& \frac{1}{8\pi v^2}\frac{1}{(m^{\dagger}_{LR}m_{LR})_{11}} \sum_{j=2,3} \text{Im}[(m^*_{LR})_{\alpha 1}
(m^{\dagger}_{LR}m_{LR})_{1j}(m_{LR})_{\alpha j}]g(x_j) \nonumber \\&& + \frac{1}{8\pi v^2}\frac{1}{(m^{\dagger}_{LR}m_{LR})_{11}} 
\sum_{j=2,3} \text{Im}[(m^*_{LR})_{\alpha 1}(m^{\dagger}_{LR}m_{LR})_{j1}(m_{LR})_{\alpha j}]\frac{1}{1-x_j}
\label{eps1}
\end{eqnarray}
where $v = 174 \; \text{GeV}$ is the vev of the Higgs bidoublets responsible for breaking the electroweak symmetry, $$ g(x) = \sqrt{x} 
\left ( 1+\frac{1}{1-x}-(1+x)\text{ln}\frac{1+x}{x} \right) $$and $x_j = M^2_j/M^2_1$. The second term in the expression for $\epsilon^{\alpha}_1$ 
above vanishes when summed over all the flavors $\alpha = e, \mu, \tau$. The sum over flavors is given by
\begin{equation}
\epsilon_1 = \frac{1}{8\pi v^2}\frac{1}{(m^{\dagger}_{LR}m_{LR})_{11}}\sum_{j=2,3} \text{Im}[(m^{\dagger}_{LR}m_{LR})^2_{1j}]g(x_j)
\label{noflavor}
\end{equation}

The corresponding baryon asymmetry is related to the lepton asymmetry as
\begin{equation}
Y_B = c \kappa \frac{\epsilon_1}{g_*}
\end{equation}
through electroweak sphaleron processes \cite{Kuzmin:1985mm}. Here, $c$ is a measure of the fraction of lepton asymmetry being 
converted into baryon asymmetry and is approximately equal to $-0.55$. $\kappa$ is the dilution factor due to wash-out processes which 
erase the produced asymmetry and can be parametrized as \cite{Kolb:1990vq,Flanz:1998kr,Pilaftsis:1998pd}
\begin{eqnarray}
-\kappa &\simeq &  \sqrt{0.1K} \text{exp}[-4/(3(0.1K)^{0.25})], \;\; \text{for} \; K  \ge 10^6 \nonumber \\
&\simeq & \frac{0.3}{K (\ln K)^{0.6}}, \;\; \text{for} \; 10 \le K \le 10^6 \nonumber \\
&\simeq & \frac{1}{2\sqrt{K^2+9}},  \;\; \text{for} \; 0 \le K \le 10.
\end{eqnarray}
where K is given as
$$ K = \frac{\Gamma_1}{H(T=M_1)} = \frac{(m^{\dagger}_{LR}m_{LR})_{11}M_1}{8\pi v^2} \frac{M_{Pl}}{1.66 \sqrt{g_*}M^2_1} $$
Here $\Gamma_1$ is the decay width of $N_1$ and $H(T=M_1)$ is the Hubble constant at temperature $T = M_1$. The factor $g_*$ is the 
effective number of relativistic degrees of freedom at $T=M_1$ and is approximately $110$.

It should be noted that the lepton asymmetry given by equation \eqref{noflavor} has been obtained by summing over all the lepton flavors $\alpha = e, \mu, \tau$. 
A non-zero lepton asymmetry can however, be obtained only when the right handed neutrino decay is out of equilibrium. Otherwise both the forward 
and the backward processes will happen at the same rate resulting in a vanishing asymmetry. Departure from equilibrium can be estimated by 
comparing the interaction rate with the expansion rate of the Universe, parametrized by the Hubble parameter. At very high temperatures $(T \geq 10^{12} \text{GeV})$ all charged 
lepton flavors are out of equilibrium and hence all of them behave similarly resulting in the one flavor regime mentioned above. However at temperatures 
$ T < 10^{12}$ GeV $(T < 10^9 \text{GeV})$, interactions involving tau (muon) Yukawa couplings enter equilibrium and flavor effects become 
important in the calculation of lepton asymmetry \cite{Barbieri:1999ma,Abada:2006fw,Abada:2006ea,Nardi:2006fx,Dev:2014laa}. The temperature regimes $10^9 < T/\text{GeV} < 10^{12}$ and $T/\text{GeV} < 10^9$ correspond to two and three flavor regimes of leptogenesis respectively. The final baryon asymmetry in the two and three flavor regimes can be written as 
\begin{equation}
Y^{2 flavor}_B = \frac{-12}{37g^*}[\epsilon_2 \eta\left (\frac{417}{589}\tilde{m_2} \right)+\epsilon^{\tau}_1\eta\left (\frac{390}{589}
\tilde{m_{\tau}}\right )] \nonumber
\end{equation}
\begin{equation}
Y^{3 flavor}_B = \frac{-12}{37g^*}[\epsilon^e_1 \eta\left (\frac{151}{179}\tilde{m_e}\right)+ \epsilon^{\mu}_1 \eta\left (\frac{344}{537}
\tilde{m_{\mu}}\right)+\epsilon^{\tau}_1\eta\left (\frac{344}{537}\tilde{m_{\tau}} \right )] \nonumber
\end{equation}
where $\epsilon_2 = \epsilon^e_1 + \epsilon^{\mu}_1, \tilde{m_2} = \tilde{m_e}+\tilde{m_{\mu}}, \tilde{m_{\alpha}} = \frac{(m^*_{LR})_{\alpha 1} 
(m_{LR})_{\alpha 1}}{M_1}$. The function $\eta$ is given by 
$$ \eta (\tilde{m_{\alpha}}) = \left [\left ( \frac{\tilde{m_{\alpha}}}{8.25 \times 10^{-3} \text{eV}} \right )^{-1}+ \left ( \frac{0.2\times 
10^{-3} \text{eV}}{\tilde{m_{\alpha}}} \right )^{-1.16} \right ]^{-1} $$

For the calculation of baryon asymmetry, we choose the basis where right handed singlet neutrino mass matrix $M_{RR}$ takes the diagonal form
\begin{equation}
U^*_R M_{RR} U^{\dagger}_R = \text{diag}(M_1, M_2, M_3)
\label{mrrdiag}
\end{equation}
In this diagonal $M_{RR}$ basis, the Dirac neutrino mass matrix also changes to 
\begin{equation}
m_{LR} = m^0_{LR} U_R
\label{mlrdiag}
\end{equation}
where $m^0_{LR}$ is the Dirac neutrino mass matrix given. If the Dirac neutrino mass matrix is assumed to be diagonal, it can be parametrized by 
\begin{equation}
m^d_{LR}=\left(\begin{array}{ccc}
\lambda^m & 0 & 0\\
0 & \lambda^n & 0 \\
0 & 0 & 1
\end{array}\right)m_f
\label{mLR1}
\end{equation}
where $\lambda = 0.22$ is the standard Wolfenstein parameter and $(m,n)$ are positive integers. We choose the integers $(m,n)$ in such a way which keeps the lightest right handed neutrino mass in the appropriate flavor regime.

\begin{figure}
\centering
\begin{minipage}{.5\textwidth}
  \centering
  \includegraphics[width=1\linewidth]{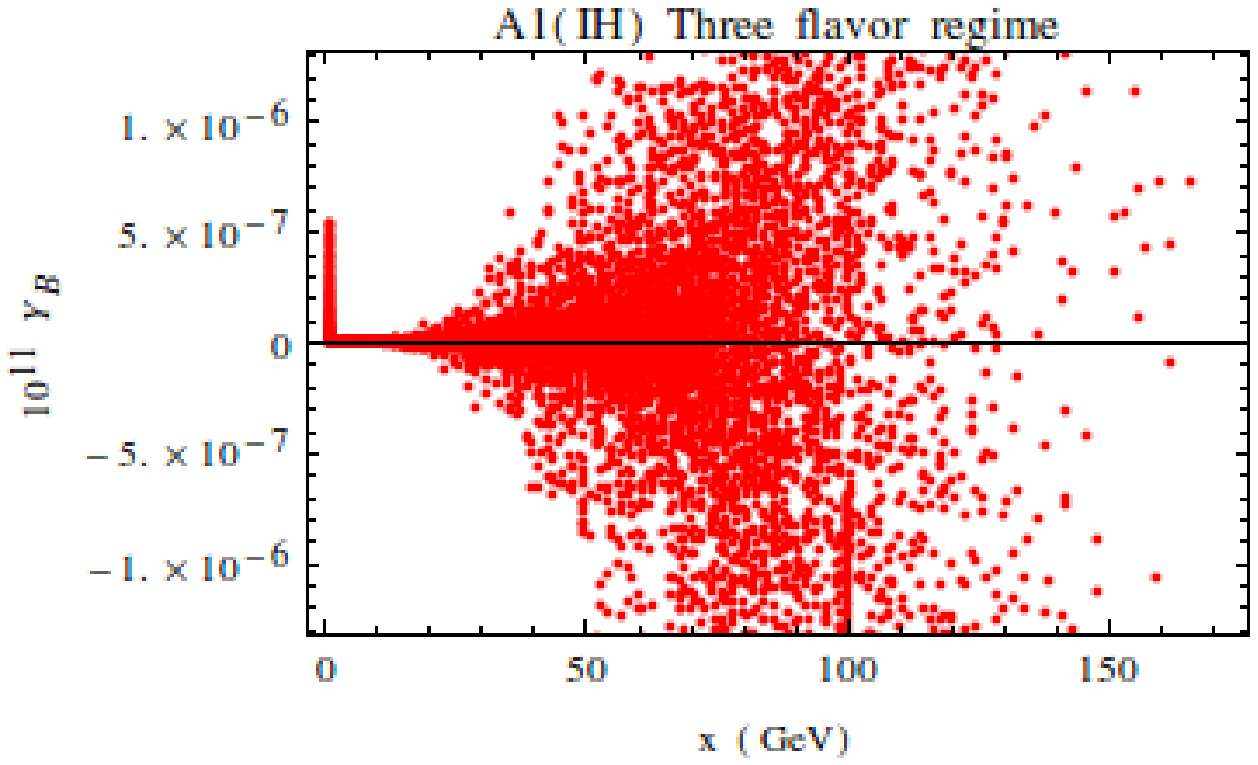}
\end{minipage}%
\begin{minipage}{.5\textwidth}
  \centering
  \includegraphics[width=1\linewidth]{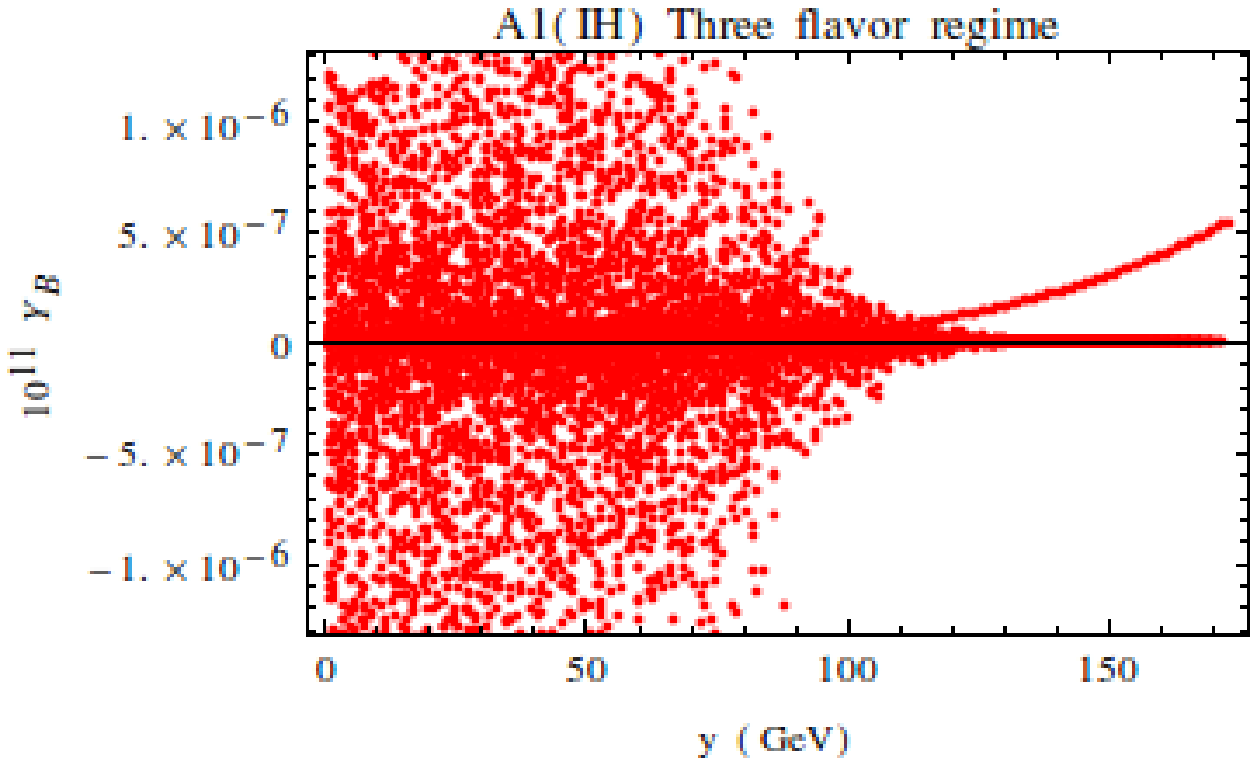}
\end{minipage}
\begin{minipage}{.5\textwidth}
  \centering
  \includegraphics[width=1\linewidth]{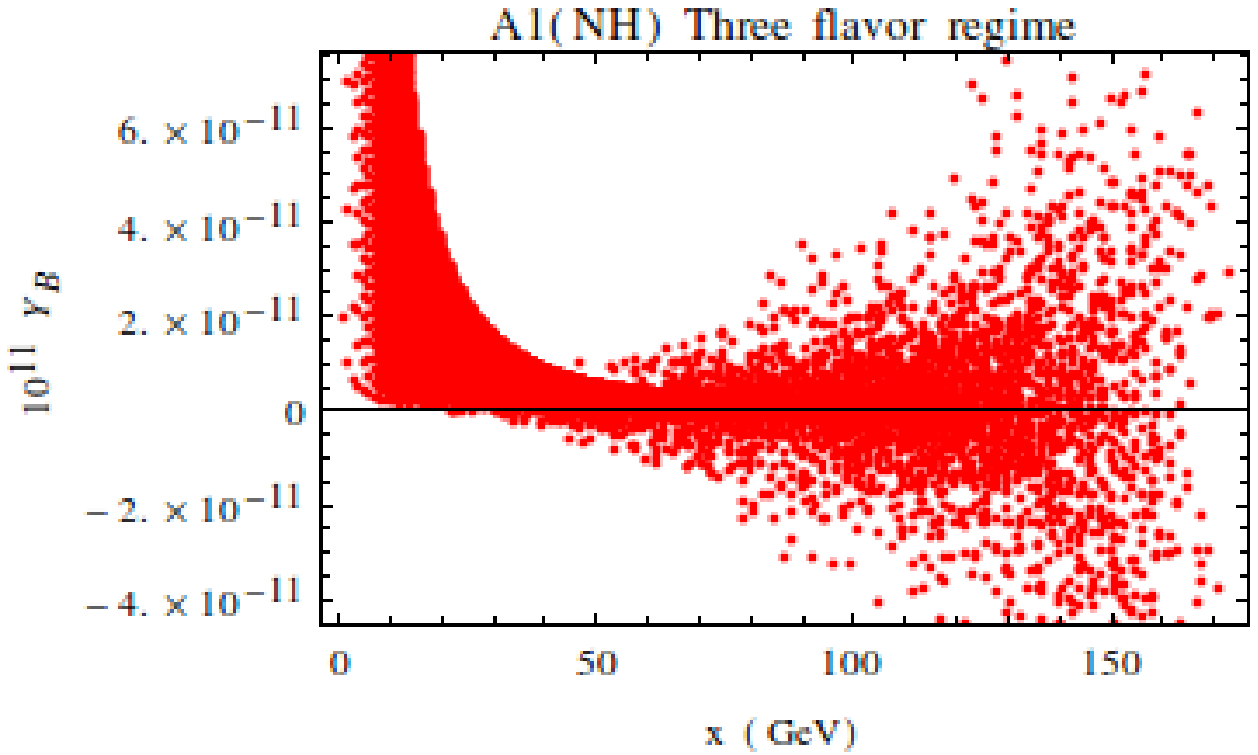}
\end{minipage}%
\begin{minipage}{.5\textwidth}
  \centering
  \includegraphics[width=1\linewidth]{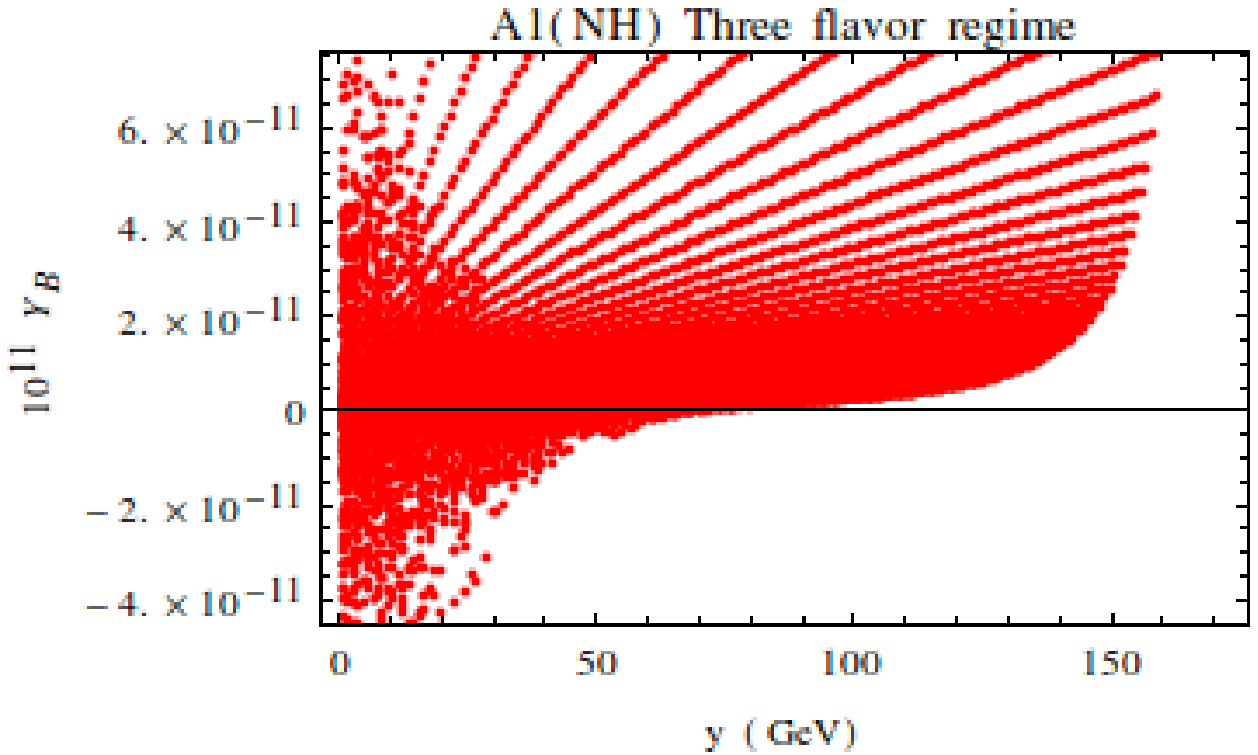}
\end{minipage}
\caption{Variation of baryon asymmetry in three flavor regime with Dirac neutrino masses for two-zero texture $A_1$.}
\label{fig20}
\end{figure}

\begin{figure}
\centering
\begin{minipage}{.5\textwidth}
  \centering
  \includegraphics[width=1\linewidth]{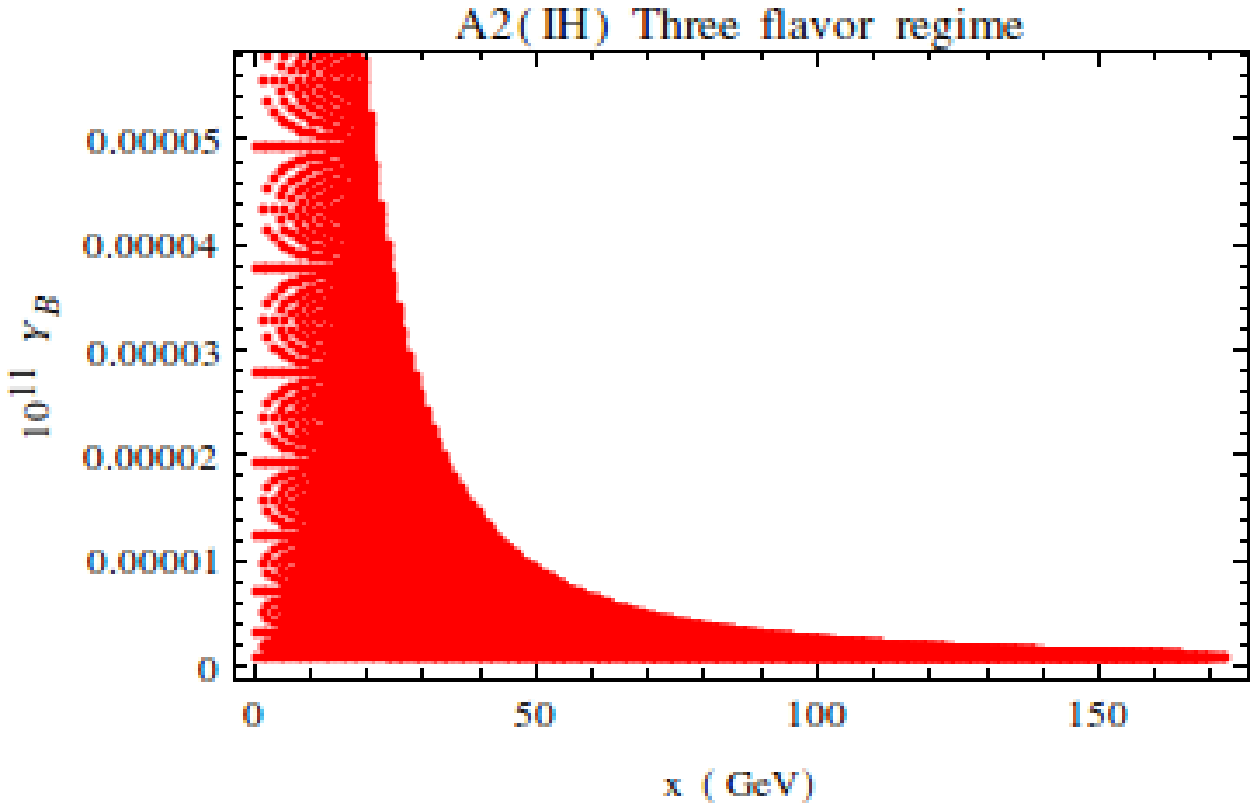}
\end{minipage}%
\begin{minipage}{.5\textwidth}
  \centering
  \includegraphics[width=1\linewidth]{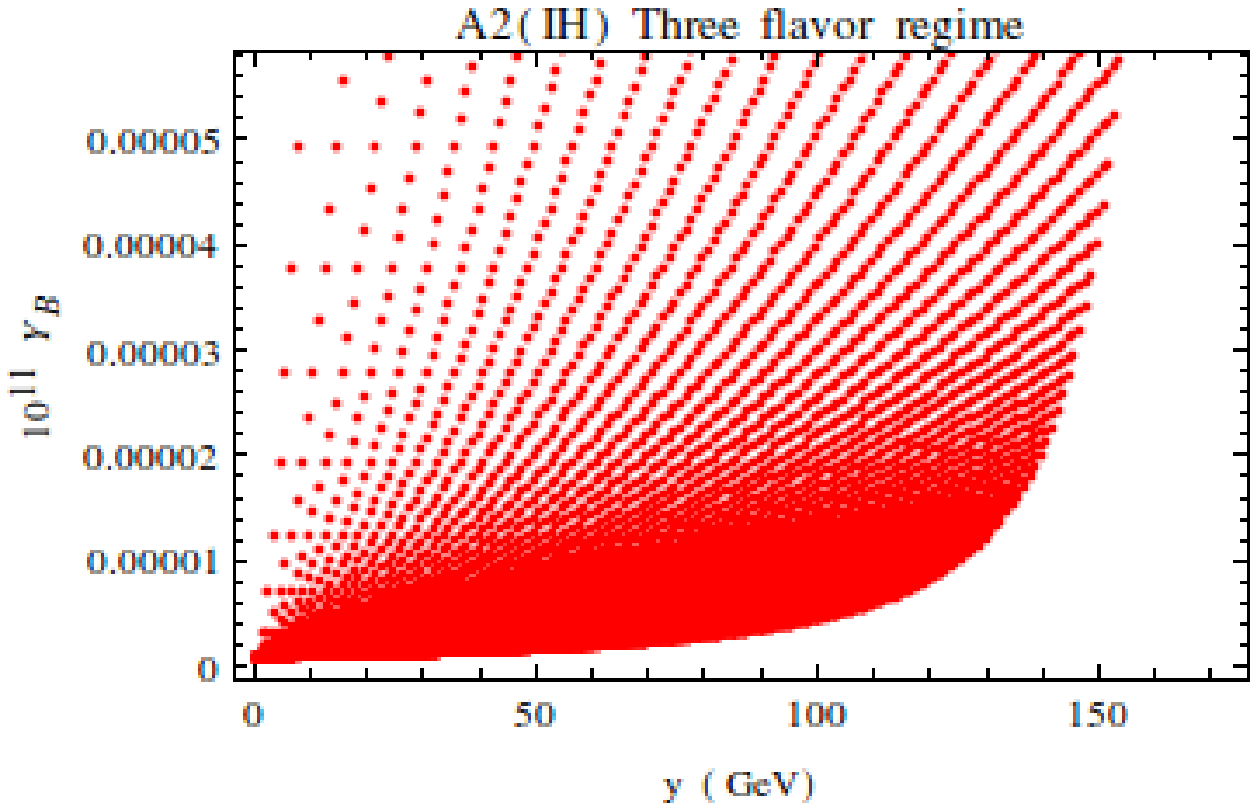}
\end{minipage}
\begin{minipage}{.5\textwidth}
  \centering
  \includegraphics[width=1\linewidth]{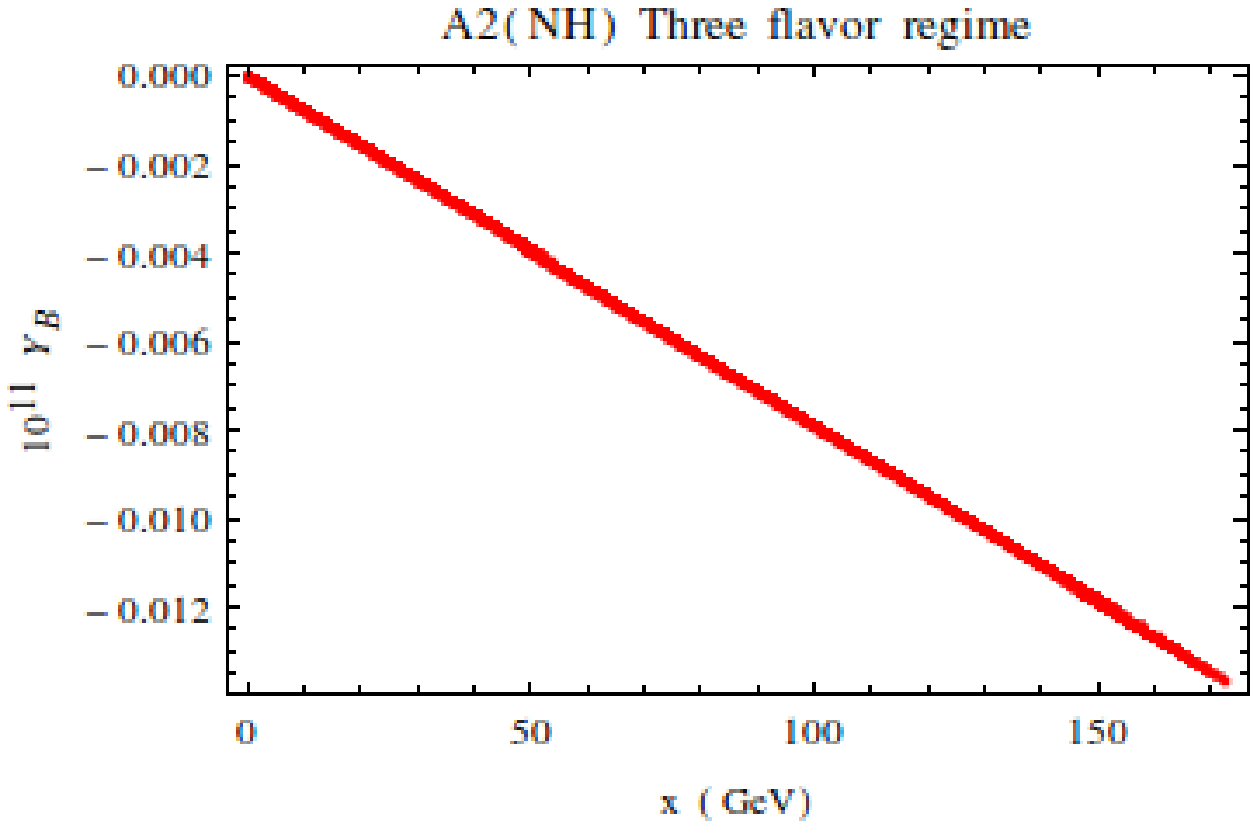}
\end{minipage}%
\begin{minipage}{.5\textwidth}
  \centering
  \includegraphics[width=1\linewidth]{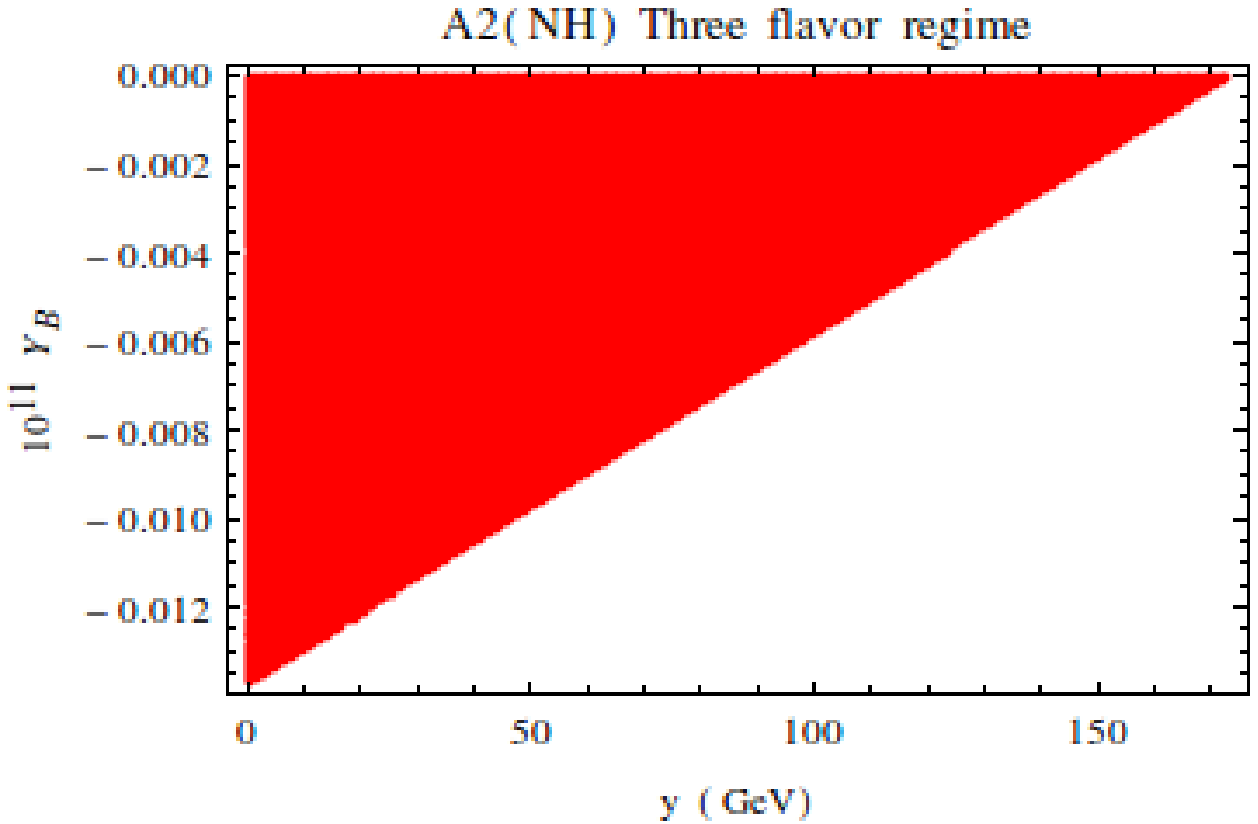}
\end{minipage}
\caption{Variation of baryon asymmetry in three flavor regime with Dirac neutrino masses for two-zero texture $A_2$.}
\label{fig21}
\end{figure}

\begin{figure}
\centering
\begin{minipage}{.5\textwidth}
  \centering
  \includegraphics[width=1\linewidth]{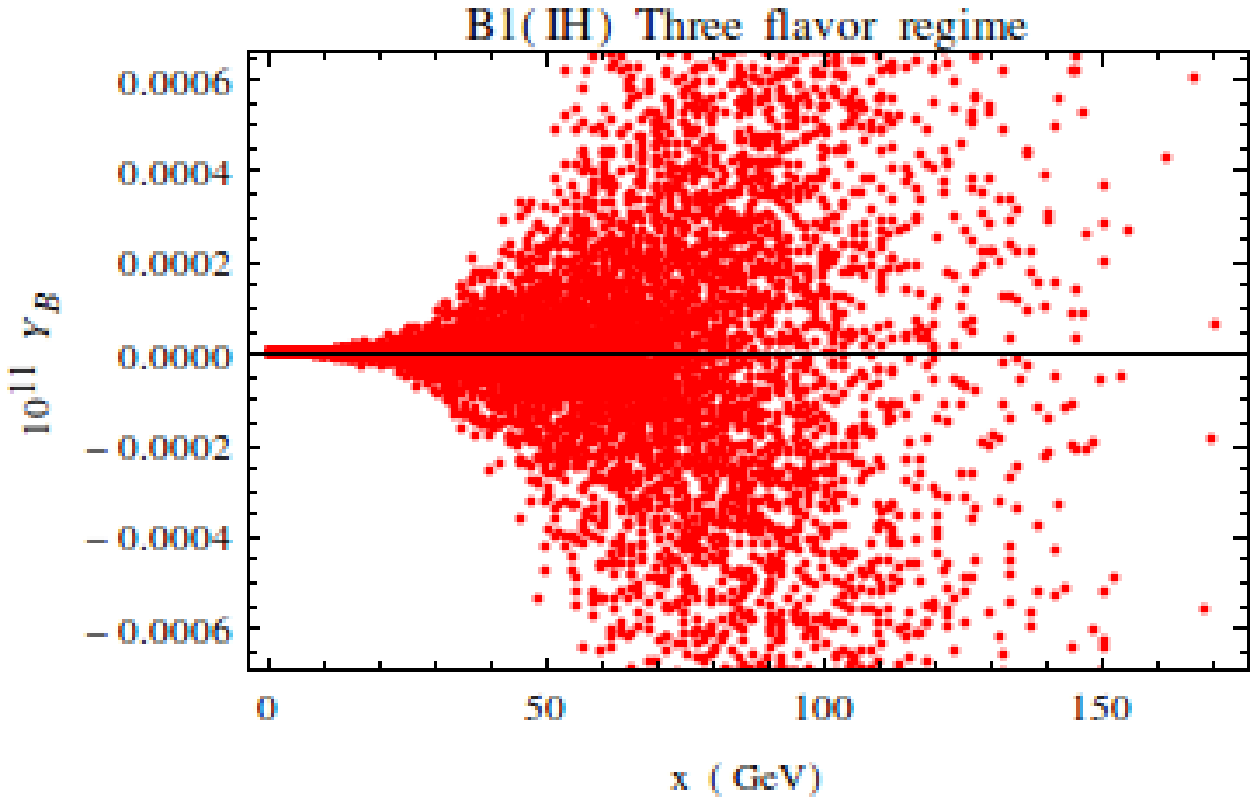}
\end{minipage}%
\begin{minipage}{.5\textwidth}
  \centering
  \includegraphics[width=1\linewidth]{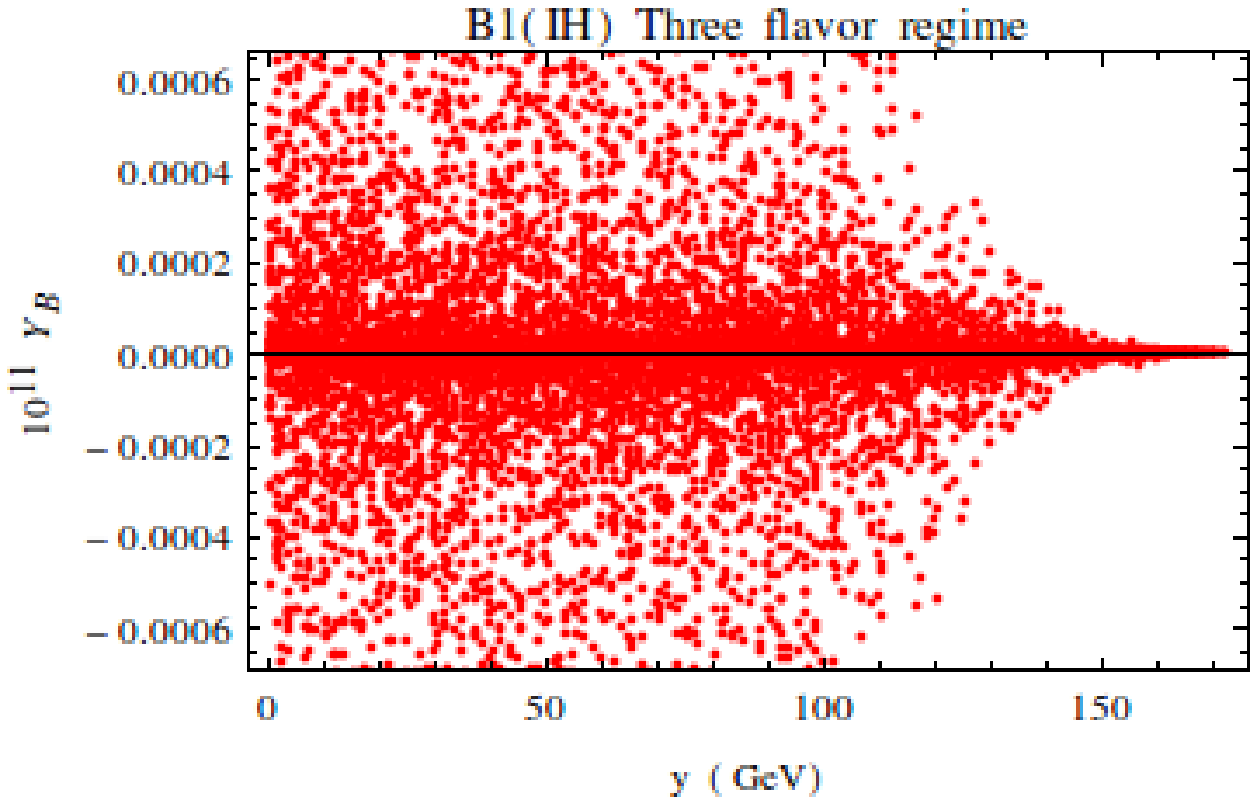}
\end{minipage}
\begin{minipage}{.5\textwidth}
  \centering
  \includegraphics[width=1\linewidth]{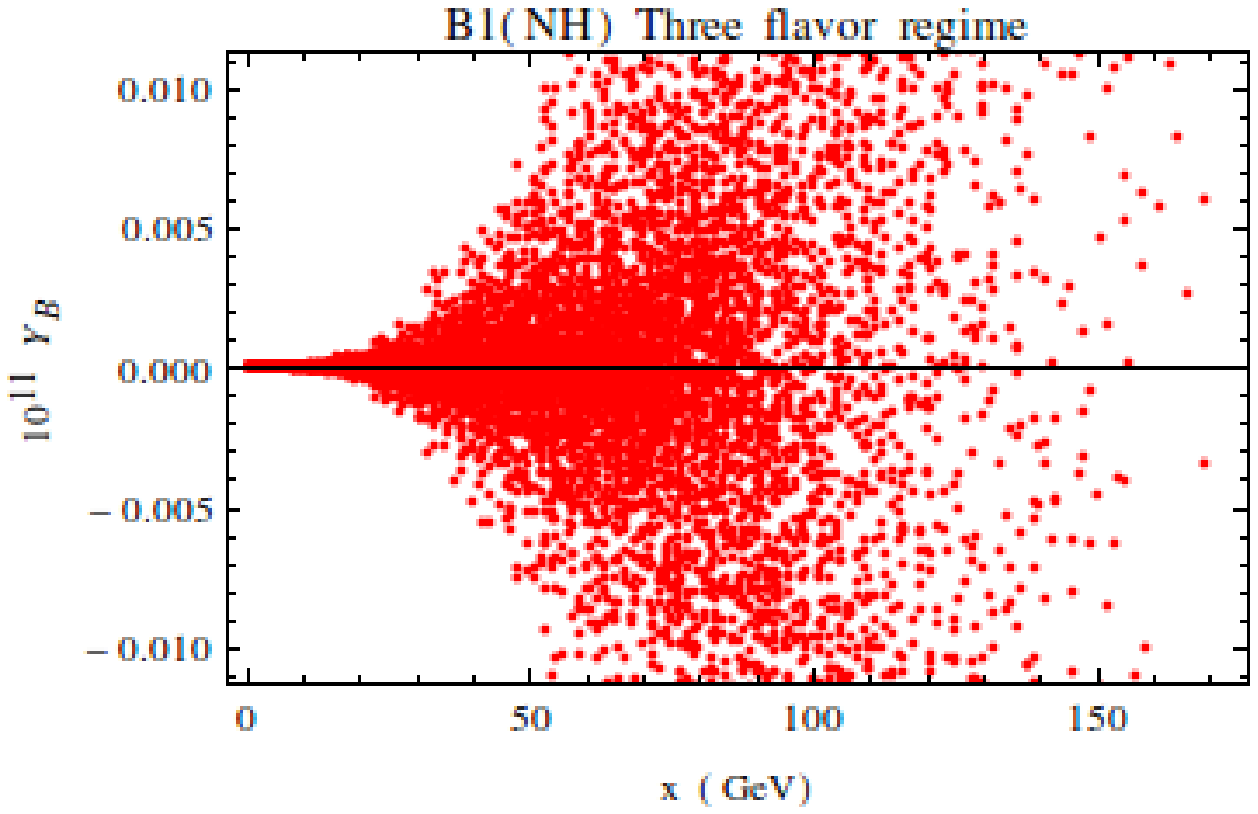}
\end{minipage}%
\begin{minipage}{.5\textwidth}
  \centering
  \includegraphics[width=1\linewidth]{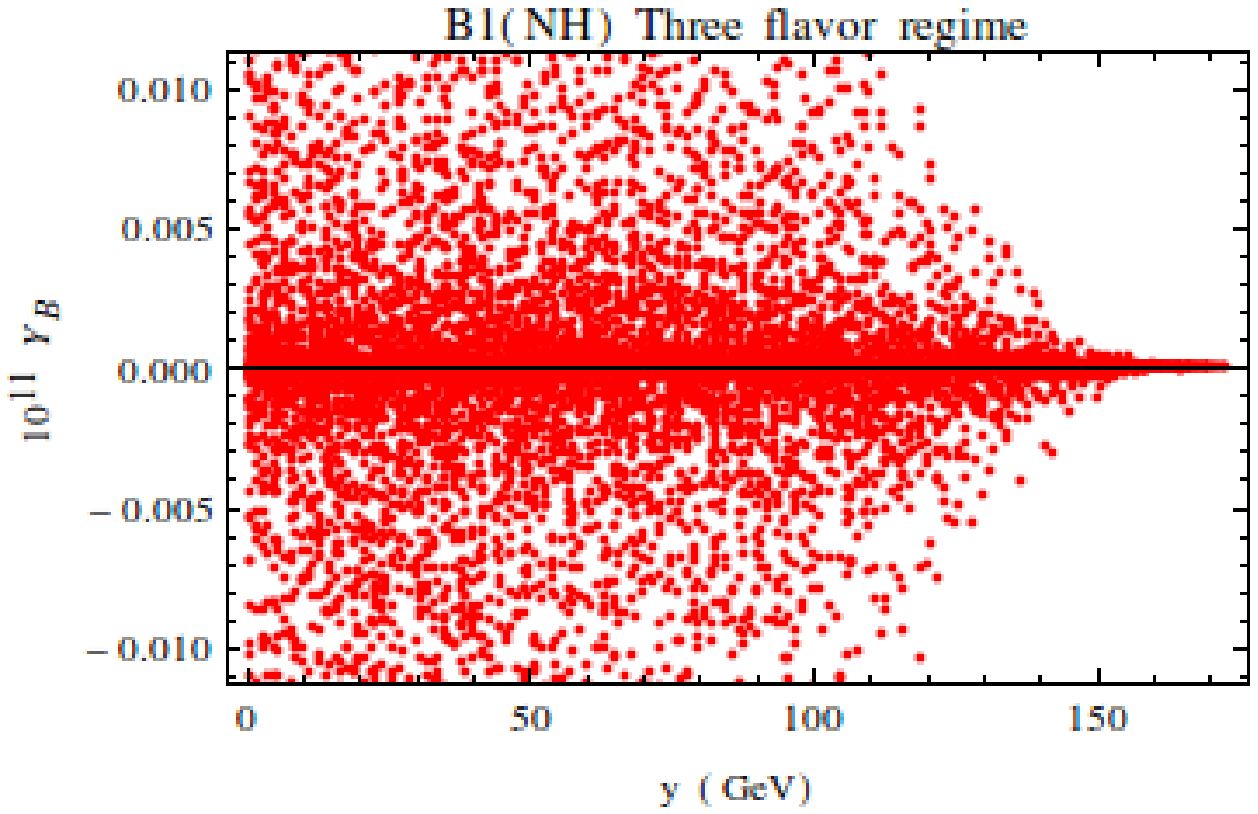}
\end{minipage}
\caption{Variation of baryon asymmetry in three flavor regime with Dirac neutrino masses for two-zero texture $B_1$.}
\label{fig22}
\end{figure}

\begin{figure}
\centering
\begin{minipage}{.5\textwidth}
  \centering
  \includegraphics[width=1\linewidth]{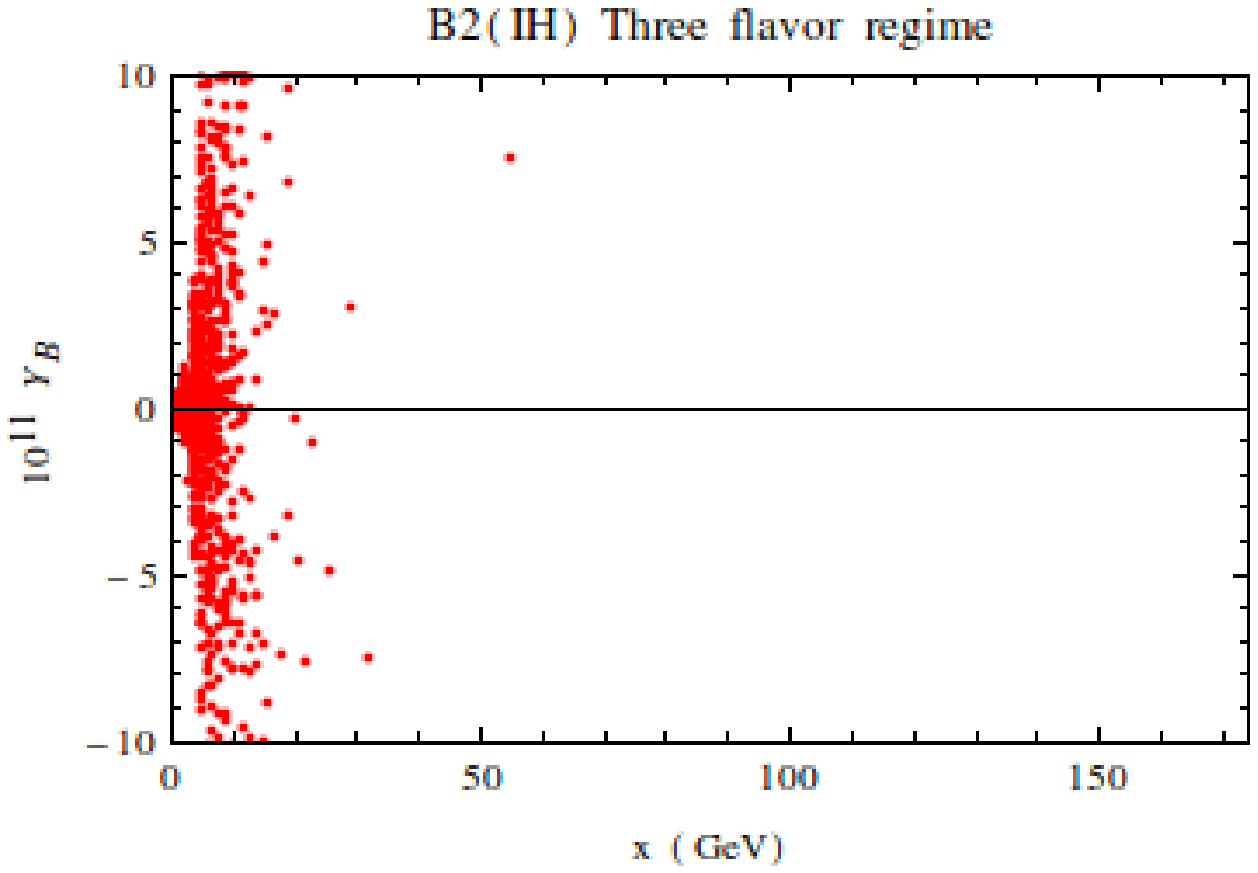}
\end{minipage}%
\begin{minipage}{.5\textwidth}
  \centering
  \includegraphics[width=1\linewidth]{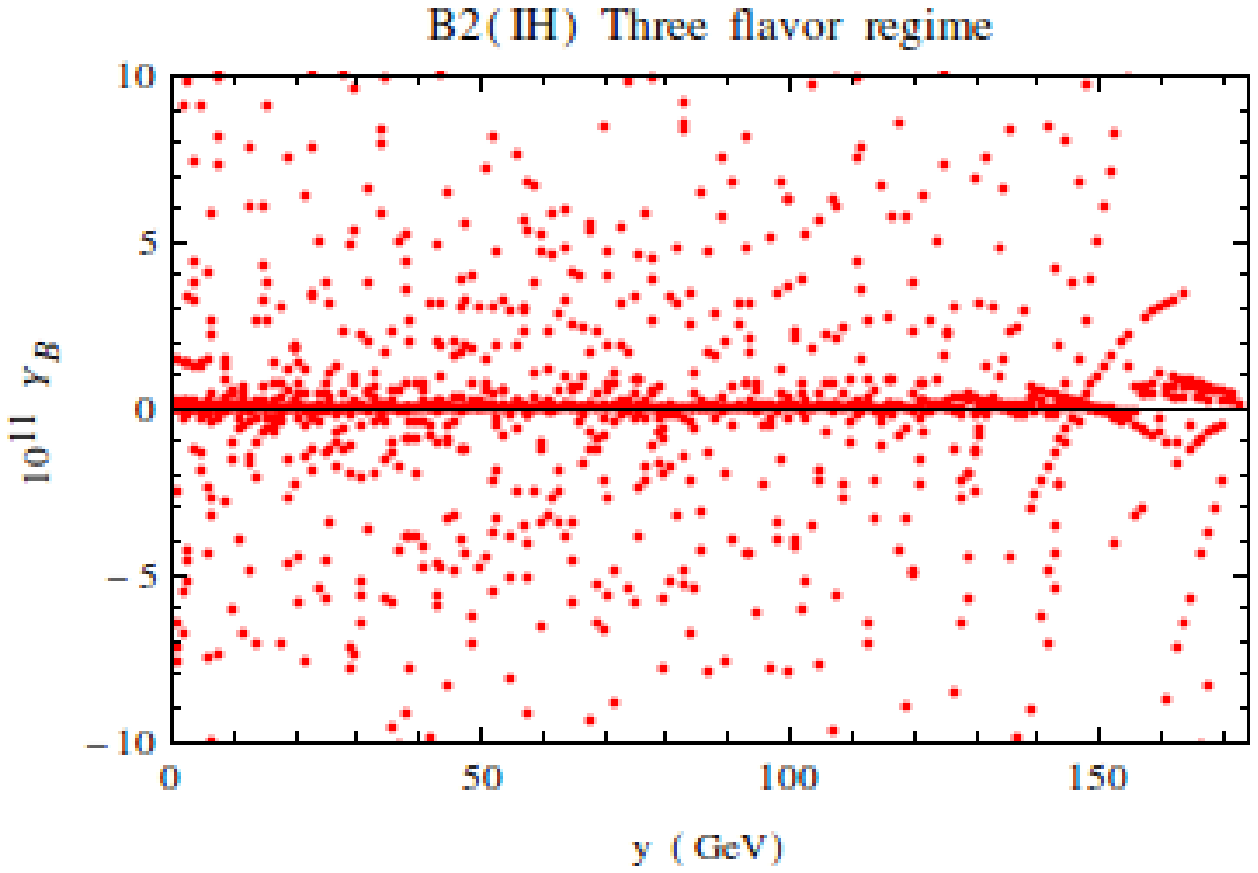}
\end{minipage}
\begin{minipage}{.5\textwidth}
  \centering
  \includegraphics[width=1\linewidth]{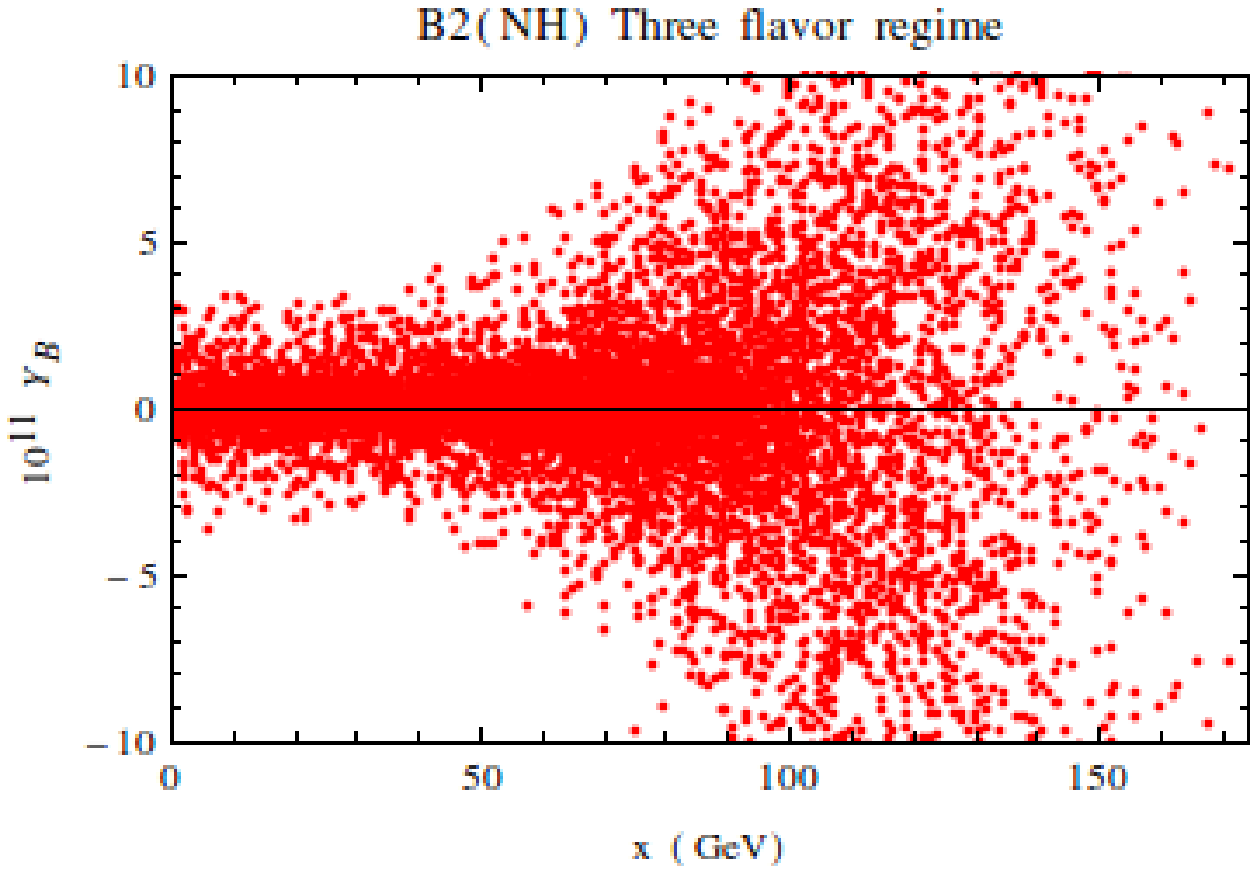}
\end{minipage}%
\begin{minipage}{.5\textwidth}
  \centering
  \includegraphics[width=1\linewidth]{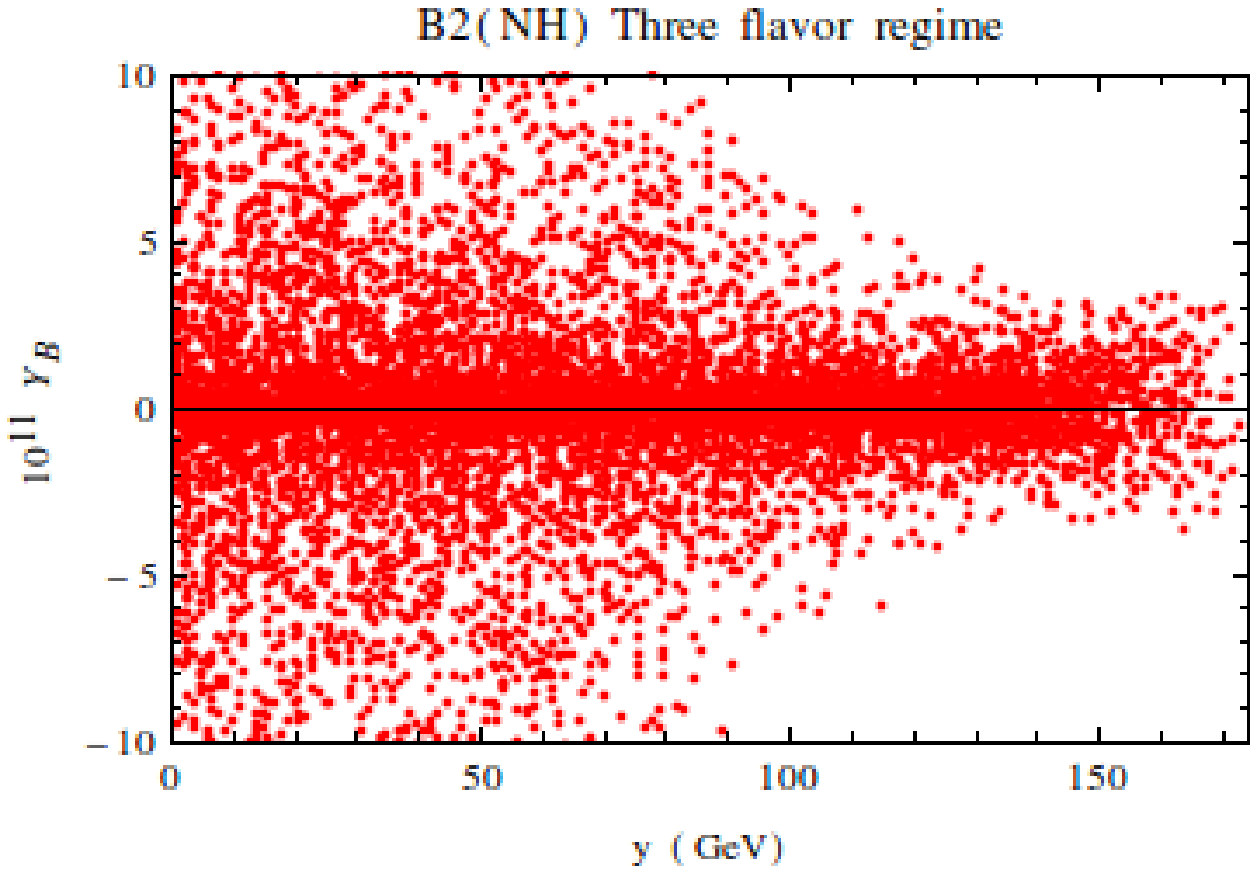}
\end{minipage}
\caption{Variation of baryon asymmetry in three flavor regime with Dirac neutrino masses for two-zero texture $B_2$.}
\label{fig23}
\end{figure}

\begin{figure}
\centering
\begin{minipage}{.5\textwidth}
  \centering
  \includegraphics[width=1\linewidth]{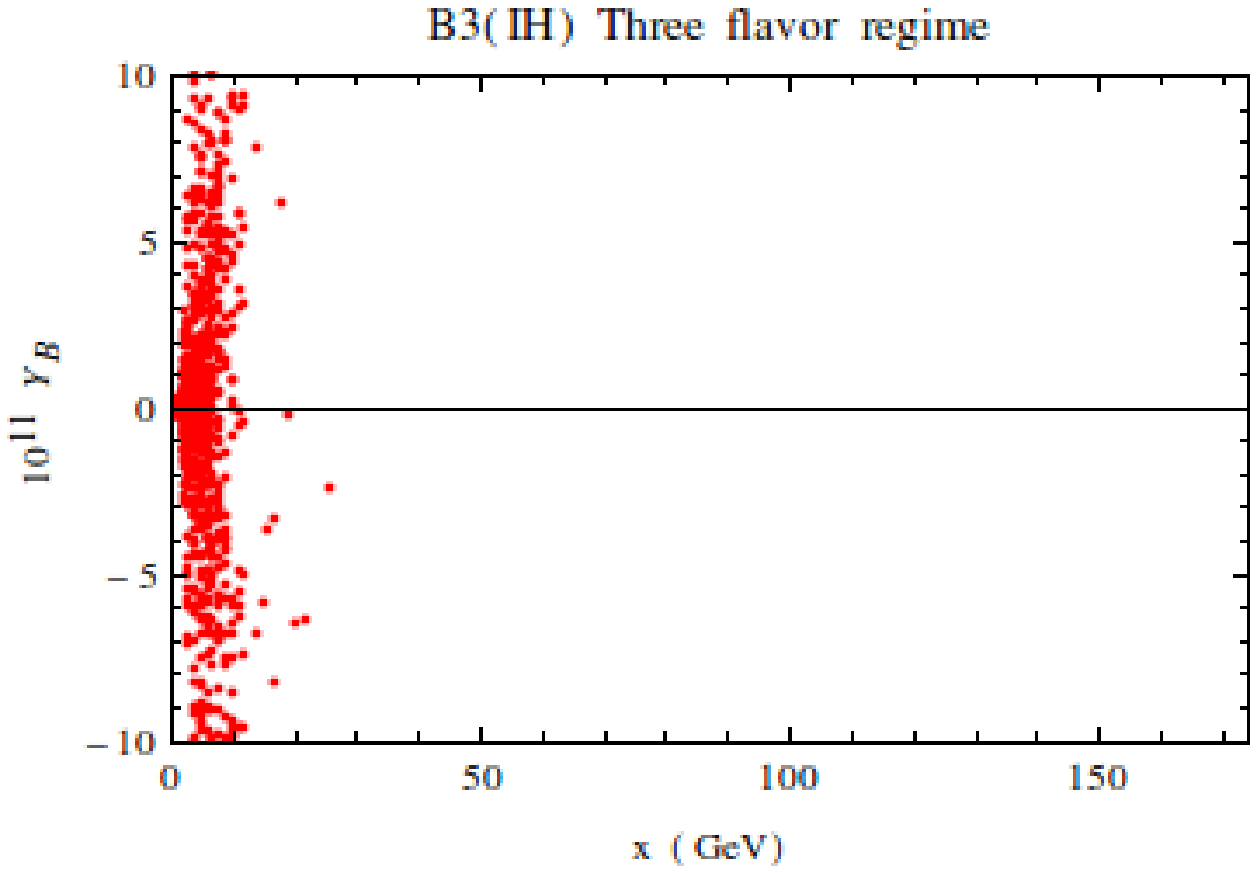}
\end{minipage}%
\begin{minipage}{.5\textwidth}
  \centering
  \includegraphics[width=1\linewidth]{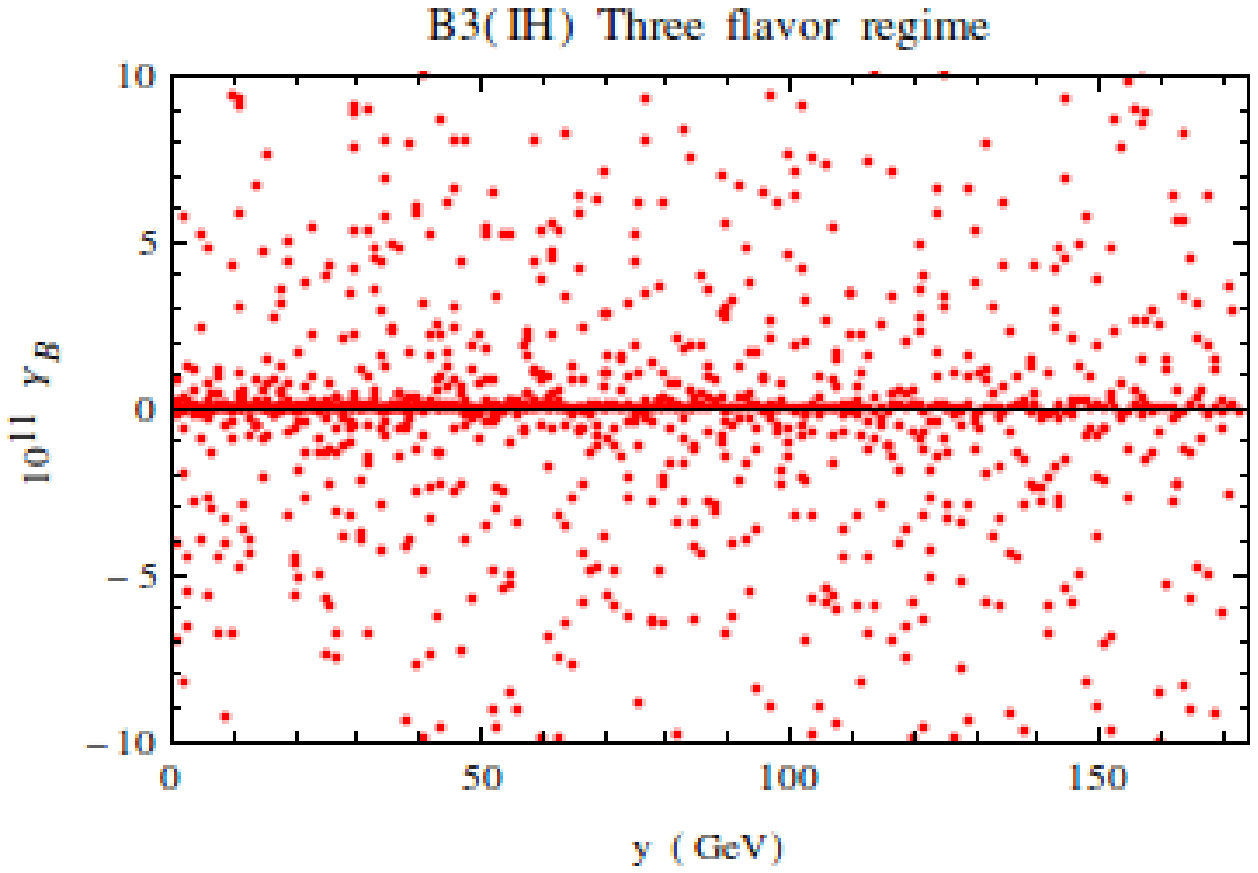}
\end{minipage}
\begin{minipage}{.5\textwidth}
  \centering
  \includegraphics[width=1\linewidth]{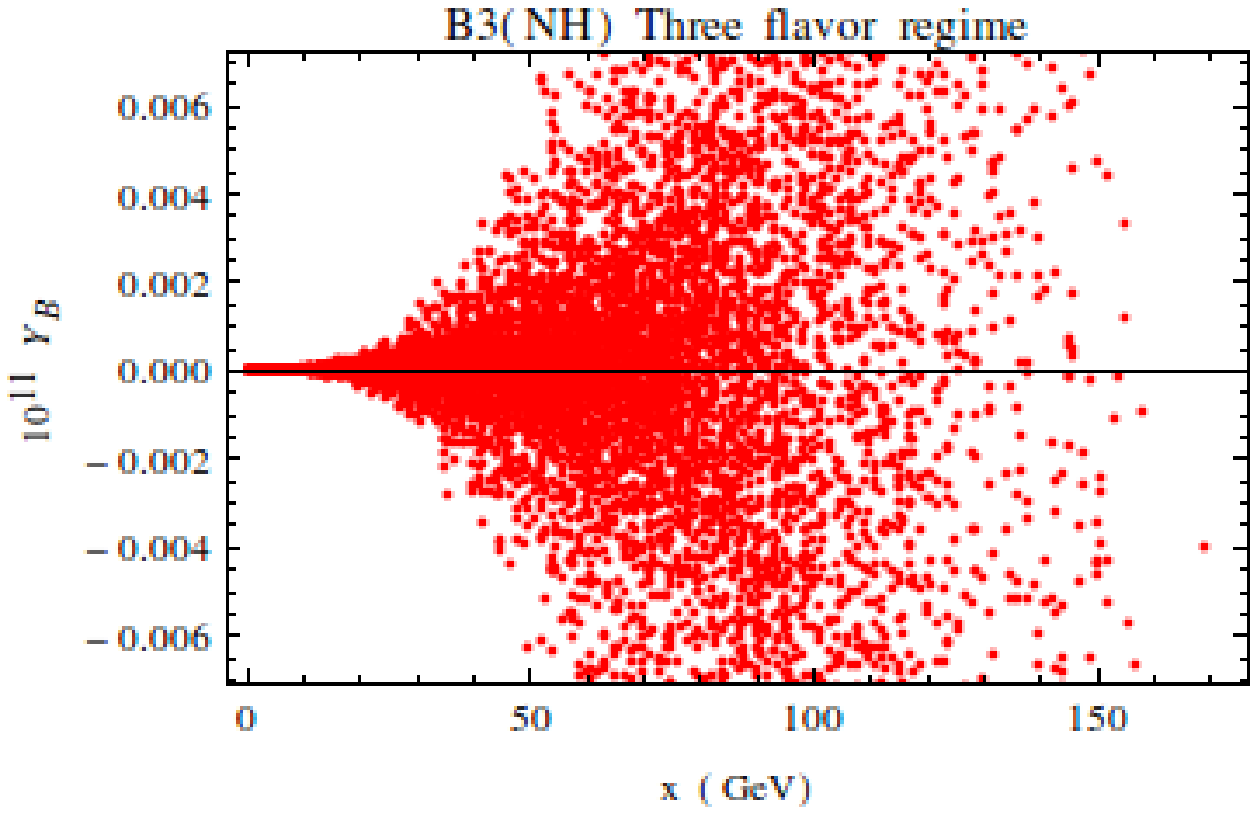}
\end{minipage}%
\begin{minipage}{.5\textwidth}
  \centering
  \includegraphics[width=1\linewidth]{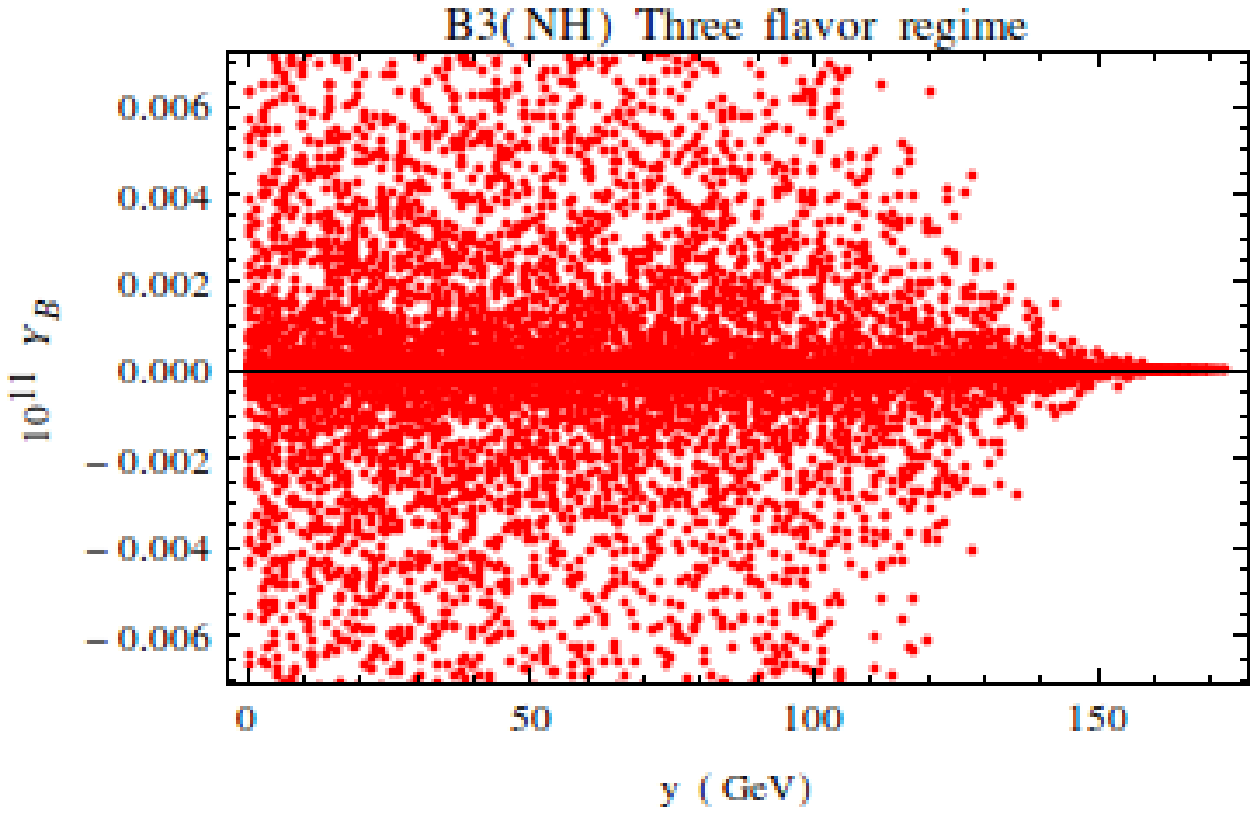}
\end{minipage}
\caption{Variation of baryon asymmetry in three flavor regime with Dirac neutrino masses for two-zero texture $B_3$.}
\label{fig24}
\end{figure}

\begin{figure}
\centering
\begin{minipage}{.5\textwidth}
  \centering
  \includegraphics[width=1\linewidth]{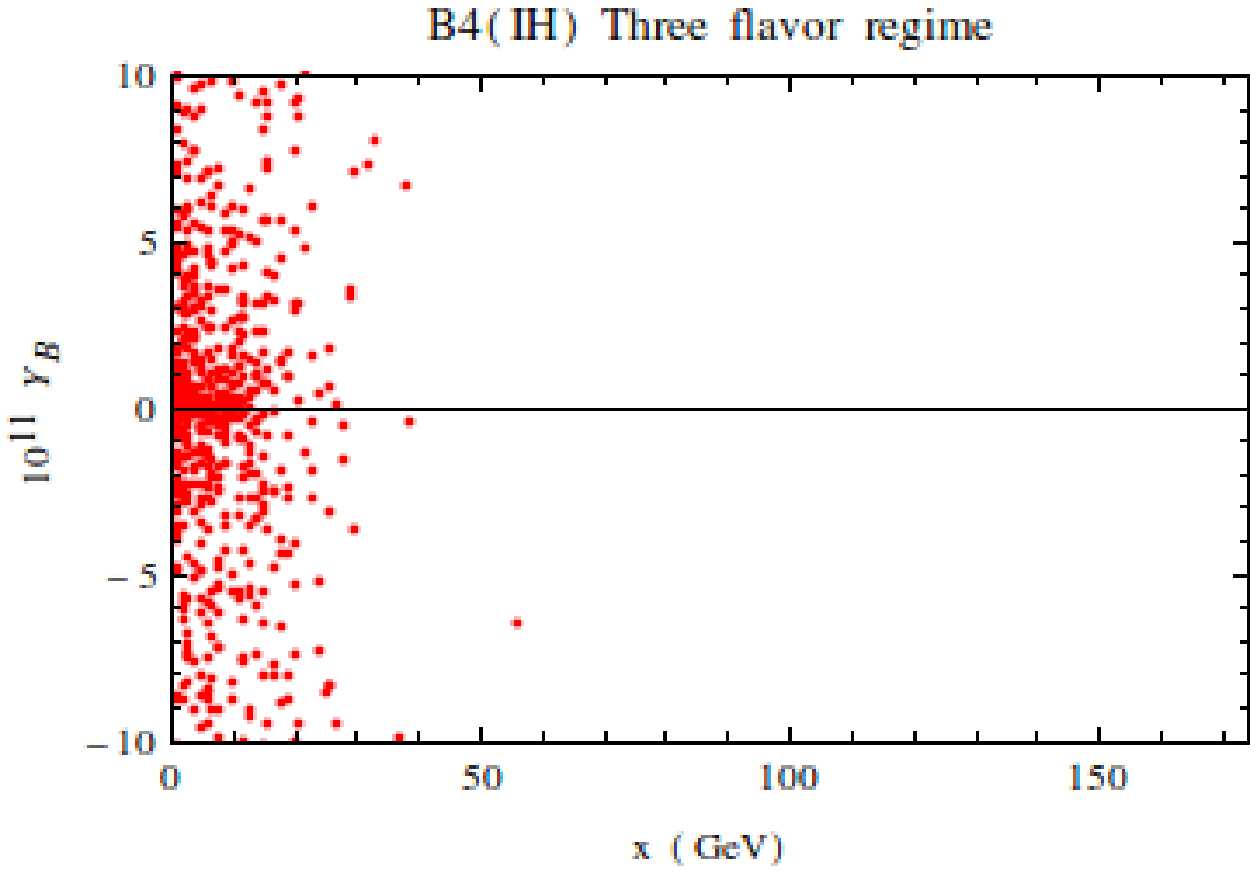}
\end{minipage}%
\begin{minipage}{.5\textwidth}
  \centering
  \includegraphics[width=1\linewidth]{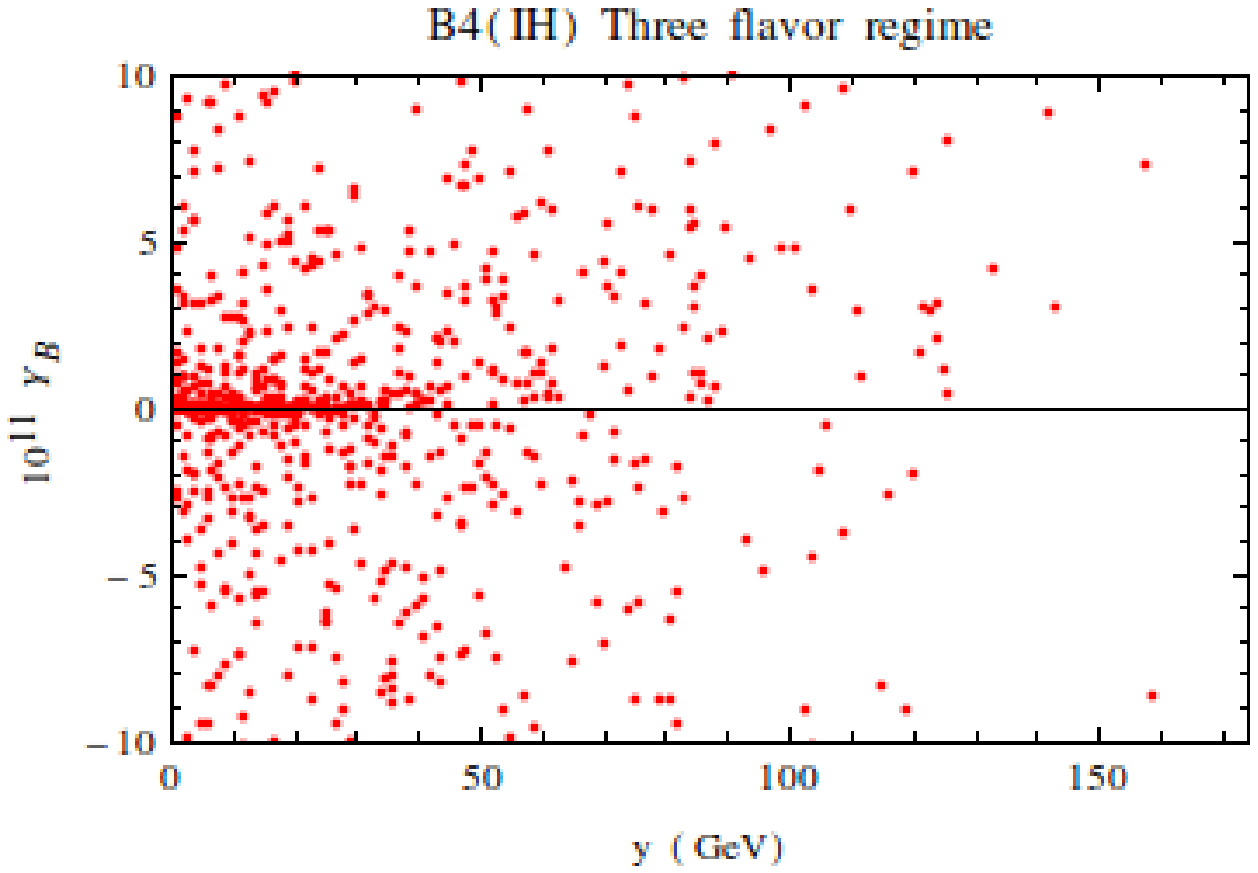}
\end{minipage}
\begin{minipage}{.5\textwidth}
  \centering
  \includegraphics[width=1\linewidth]{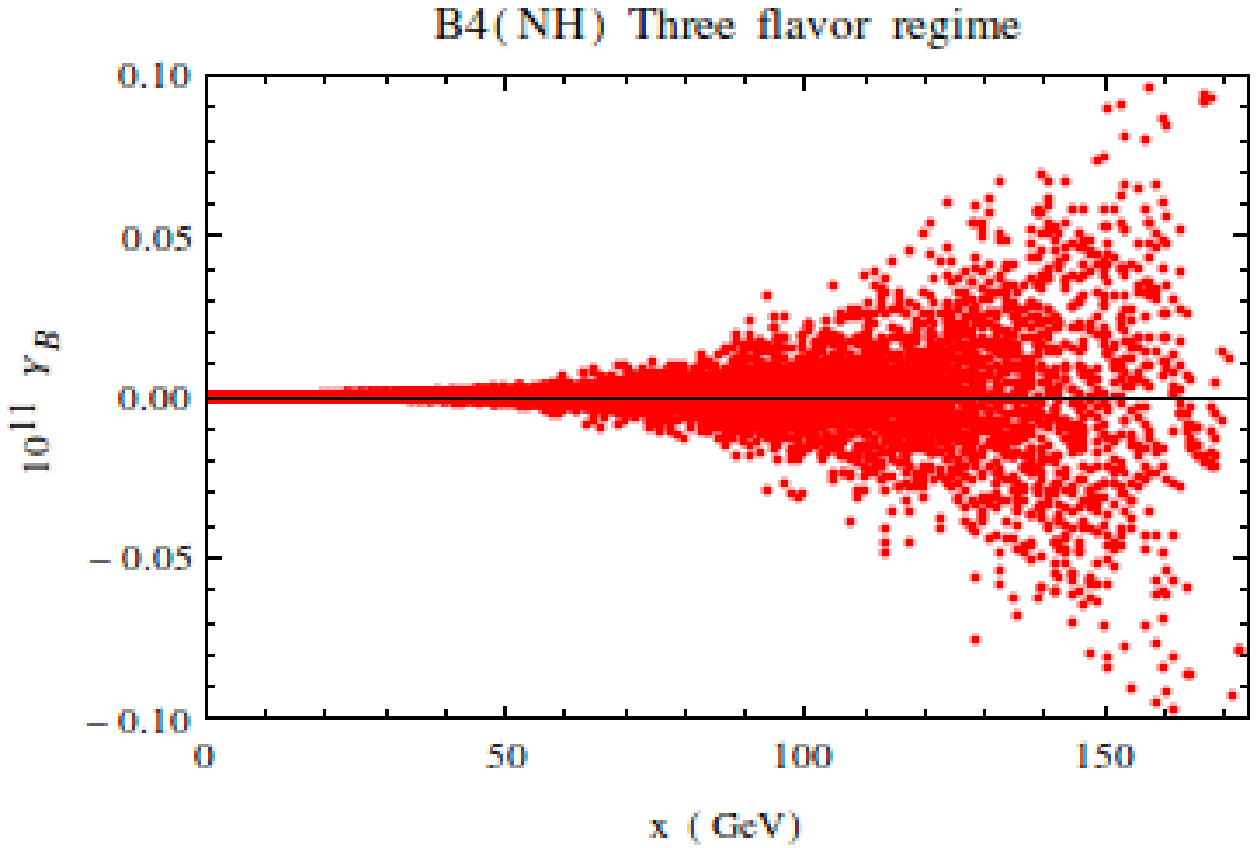}
\end{minipage}%
\begin{minipage}{.5\textwidth}
  \centering
  \includegraphics[width=1\linewidth]{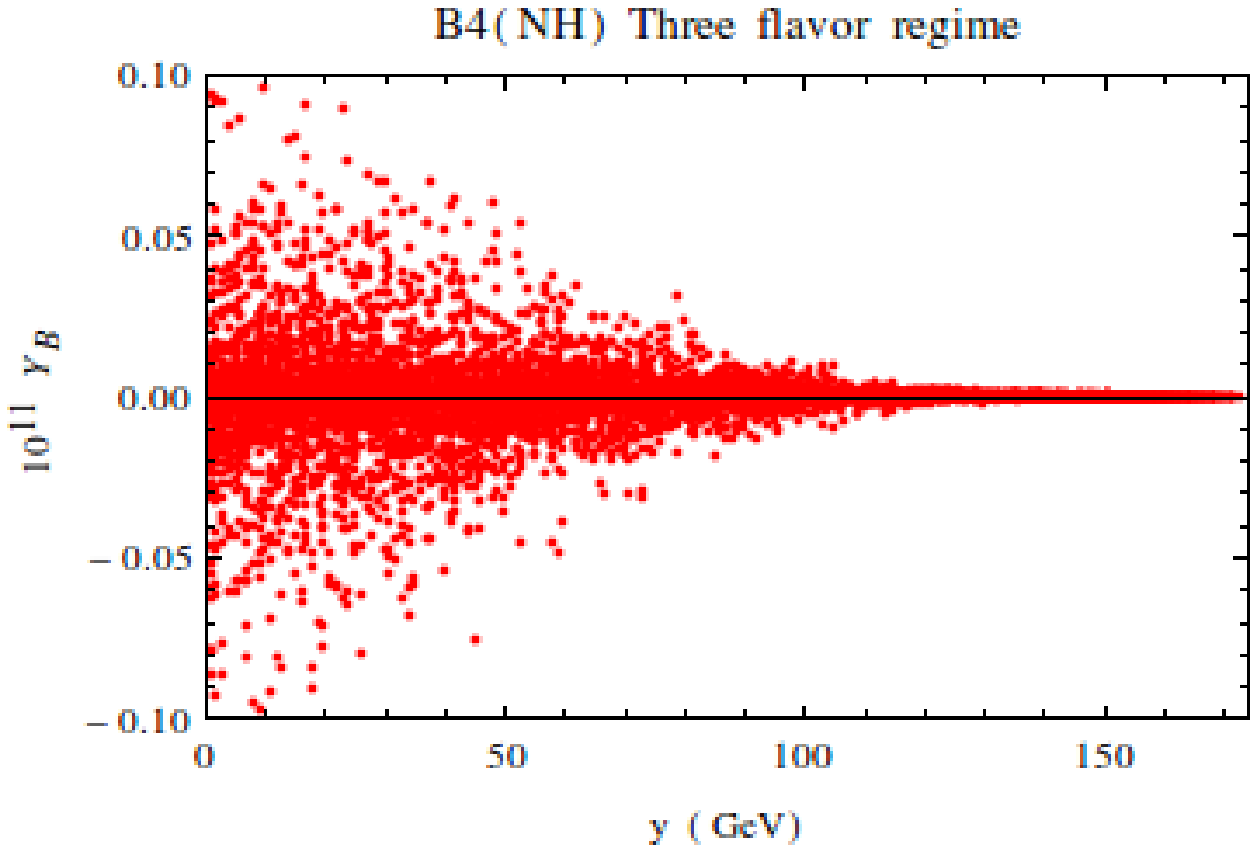}
\end{minipage}
\caption{Variation of baryon asymmetry in three flavor regime with Dirac neutrino masses for two-zero texture $B_4$.}
\label{fig25}
\end{figure}


\section{Numerical Analysis}
\label{sec:numeric}
Using the parametric form of PMNS matrix shown in \eqref{matrixPMNS}, the Majorana neutrino mass matrix $M_{\nu}$ can be found as
\begin{equation}
 M_{\nu}=U_{\text{PMNS}} M^{\text{diag}}_{\nu}U^T_{\text{PMNS}}
 \label{numatrix}
\end{equation}
where
\begin{equation}
M^{\text{diag}}_{\nu}=\left(\begin{array}{ccc}
m_1& 0&0\\
0& m_2& 0  \\
0& 0 &m_3
\end{array}\right),
\end{equation}
where $m_1, m_2$ and $m_3$ are the three neutrino mass eigenvalues. As mentioned earlier, here we assume that the diagonalizing matrix of the neutrino mass matrix $M_{\nu}$ is same as the PMNS mixing matrix due to the chosen charged lepton mass matrix in the diagonal form.

For the case of normal hierarchy (NH), the three neutrino mass eigenvalues can be written as $m_{\text{diag}} 
= \text{diag}(m_1, \sqrt{m^2_1+\Delta m_{21}^2}, \sqrt{m_1^2+\Delta m_{31}^2})$, while for the case of inverted hierarchy (IH), it can be written as 
$m_{\text{diag}} = \text{diag}(\sqrt{m_3^2+\Delta m_{23}^2-\Delta m_{21}^2}, \sqrt{m_3^2+\Delta m_{23}^2}, m_3)$.
For illustrative purposes, we consider two different order of magnitude values for the lightest neutrino mass $m_1$ for NH and $m_3$ for IH.
In the first case, we assume $m_{\text{lightest}}$ as large as possible so that the sum of the absolute neutrino masses lie just below the 
cosmological upper bound and it turns out to be $0.07$ eV and $0.065$ eV for NH and IH respectively. This gives rise to a quasi-degenerate type of neutrino mass spectrum. Secondly.
we choose the lightest mass eigenvalue to be $10^{-6}$ eV for both NH and IH cases so that
we have a hierarchical pattern of neutrino masses. The PMNS mixing matrix is evaluated by taking the best fit values of the neutrino mixing angles given in 
Table \ref{tab:data1}. After using the best fit values of two mass squared differences and three mixing angles, the most general neutrino mass matrix given by \eqref{numatrix} contain four parameters: the lightest neutrino mass, Dirac CP phase and two Majorana phases. Comparing the most general neutrino mass matrix to the texture zero mass matrices, we can either relate two or more terms in the mass matrix or equate them to zero. Depending upon the number of constraints for a specific texture zero mass matrix, we can either write down some free parameters in the most general neutrino mass matrix in terms of the others or we can find the exact numerical values of the free parameters. We briefly discuss the procedure we adopt for numerical analysis involving different types of Majorana texture zero mass matrices in the following subsections \ref{ponezero} and \ref{ptwo}.

\subsection{Parametrization of One-zero Texture}
\label{ponezero}
In the case of one-zero texture mass matrices discussed in subsection \ref{one}, there is only one independent zero and hence we have only one complex equation as constraint resulting in two real equations relating $m_1 (m_3), \delta, \alpha, \beta$. To simplify the analysis, we assume equality of two Majorana phases $\alpha = \beta$. Using the constraints, we write down the Majorana phases $\alpha = \beta$ as well as lightest neutrino mass in terms of Dirac CP phase $\delta$. However, for a specific type of one-zero texture mass matrix denoted by $G_1$, the zero appears in the $(1,1)$ term and in the most general neutrino mass matrix \eqref{numatrix}, the $(1,1)$ term does not depend upon the Dirac CP phase. Therefore, in this case the Majorana phase is independent of the Dirac phase, but depends upon the value of lightest neutrino mass. The lightest neutrino mass for normal hierarchy is found to be $0.0062$ eV, whereas for inverted hierarchy, we do not get any real solution for lightest neutrino mass, satisfying the constraint. For one-zero texture $G_2$ also, we do not get any real solution for lightest neutrino mass in the case of inverted hierarchy. For $G_{3,4,5,6}$ with inverted hierarchy, the variation of Majorana CP phase with Dirac CP phase is shown in figure \ref{fig001}. Similarly, the dependence of lightest neutrino mass on Dirac CP phase is shown in the first panel of figures \ref{fig1}, \ref{fig2}, \ref{fig3} and \ref{fig4} respectively. For normal hierarchy, we do not get any real solution for lightest neutrino mass for the one-zero textures $G_{4,5,6}$. For $G_{2,3}$, the dependence of Majorana CP phase with Dirac CP phase is shown in figure \ref{fig01}. For $G_1$, the lightest neutrino mass is exactly determined whereas for $G_{2,3}$ its dependence on $\delta$ can be seen in the first panel of figure \ref{fig6} and \ref{fig7}.

\subsection{Parametrization of Two-zero Texture}
\label{ptwo}
In two-zero texture mass matrices discussed in subsection \ref{two}, the Majorana neutrino mass matrix contains two independent zeros. Therefore, we have two complex and hence four real constraint equations to relate the four independent parameters. We numerically solve these four equations to find lightest neutrino mass, Dirac CP phase $\delta$ and Majorana CP phases $\alpha, \beta$. A set of such solutions are shown in table \ref{table-2zero1} and \ref{table-2zero2}.

\subsection{Calculation of Baryon Asymmetry}
To calculate the baryon asymmetry in the appropriate flavor regime, we choose the diagonal Dirac neutrino mass matrix in such a way that the lightest right handed singlet neutrino mass lies in the same flavor regime. Similar to the discussion in earlier works \cite{Borah:2014fga,Borah:2014bda,Kalita:2014mga}, we choose $m_f=82.43$ GeV in the Dirac neutrino mass matrix given by \eqref{mLR1}. We also choose $(m,n) = (1,1), (3,1)$ and $(5,3)$ to keep the lightest right handed neutrino mass in one, two and three flavor regimes respectively. The resulting baryon asymmetry as a function of Dirac CP phase for different patterns of one-zero texture in the Majorana neutrino mass matrix are shown in figures \ref{fig1}, \ref{fig2}, \ref{fig3}, \ref{fig4}, \ref{fig5}, \ref{fig6} and \ref{fig7}.

In case of two-zero texture mass matrices, since all the neutrino parameters are fixed, we compute the baryon asymmetry by varying the Dirac neutrino mass matrix. We choose the Dirac neutrino mass matrix to be of the form
\begin{equation}
m^d_{LR}=\left(\begin{array}{ccc}
m_{11} & 0 & 0\\
0 & m_{22} & 0 \\
0 & 0 & m_{33}
\end{array}\right)
\label{mLR1}
\end{equation}
We fix $m_{11}$ such that the lightest right handed neutrino mass falls in the appropriate flavor regime, and vary $m_{22}=x, m_{22}=y \geq x$ and calculate the amount of baryon asymmetry. The resulting baryon asymmetry as a function of $m_{22}=x, m_{22}=y \geq x$ are shown in figures from \ref{fig8} to figure \ref{fig25}.

 \begin{table}[!h]
\begin{tabular}{cccc}
 \hline

  \hline
Patterns     & One flavor & Two flavor & Three flavor\\
        \hline
\mbox{$G_1$}     & $\times$     & $\times$   & $\times$\\
\mbox{$G_2$}     & $\times$     & $\times$   & $\times$\\
\mbox{$G_3$}     & $\checkmark$     & $\checkmark$   & $\times$\\
\mbox{$G_4$}     & $\checkmark$      & $\checkmark$   & $\checkmark$ \\ 
\mbox{$G_5$}     & $\checkmark$      & $\times$ & $\checkmark$ \\
\mbox{$G_6$}     & $\checkmark$     & $\checkmark$ & $\checkmark$ \\
        \hline
\end{tabular} 
\caption{Summary of results for one-zero texture with inverted hierarchy. 
The symbol $\checkmark$ ($\times$) is used when the baryon asymmetry $Y_B$ is in (not in) range.}
\label{table3}
\end{table}

\begin{table}[!h]
\begin{tabular}{cccc}
 \hline

  \hline
Patterns     & One flavor & Two flavor & Three flavor \\
        \hline
\mbox{$G_1$}     & $\checkmark$     & $\checkmark$   & $\checkmark$\\
\mbox{$G_2$}     & $\checkmark$      & $\times$   & $\times$ \\ 
\mbox{$G_3$}     & $\checkmark$      & $\times$ & $\times$ \\
\mbox{$G_4$}     & $\times$     & $\times$   & $\times$\\
\mbox{$G_5$}     & $\times$     & $\times$   & $\times$\\
\mbox{$G_6$}     & $\times$     & $\times$   & $\times$\\

        \hline
\end{tabular} 
\caption{Summary of results for one-zero texture with normal hierarchy. 
The symbol $\checkmark$ ($\times$) is used when the baryon asymmetry $Y_B$ is in (not in) range.}
\label{table4}
\end{table}

\begin{table}[!h]
\begin{tabular}{cccc}
 \hline

Patterns     &One flavor IH (NH) & Two flavor IH (NH)& Three flavor IH (NH)\\
        \hline \hline
\mbox{$A_1$}     & $\checkmark$($\checkmark$)    & $\times$($\times$)        &$\times$($\times$) \\
\mbox{$A_2$}     & $\checkmark$($\checkmark$)    & $\times$($\times$)        & $\times$($\times$)\\ 
\mbox{$B_1$}     & $\checkmark$($\times$)        & $\times$($\times$)        &$\times$($\times$)\\
\mbox{$B_2$}     & $\checkmark$($\checkmark$)    & $\checkmark$($\times$)    &$\checkmark$($\checkmark$)\\
\mbox{$B_3$}     & $\times$($\times$)            & $\checkmark$($\times$)    &$\checkmark$($\times$)\\
\mbox{$B_4$}     & $\checkmark$($\times$)        & $\checkmark$($\times$)    &$\checkmark$($\times$)\\
        \hline
\end{tabular}
 
\caption{Summary of results for two-zero texture with inverted and normal hierarchy. 
The symbol $\checkmark$ ($\times$) is used when the baryon asymmetry $Y_B$ is in (not in) range.}
\label{table5}
\end{table}

\section{Results and Conclusion}
\label{sec:conclude}
Assuming the charged lepton mass matrix to be diagonal, we have studied all possible texture zeros in the Majorana neutrino mass matrix that are allowed by latest neutrino oscillation data as well as the Planck bound on the sum of absolute neutrino masses and constrain them further from the requirement of producing correct baryon asymmetry of the Universe through the mechanism of leptogenesis. The allowed Majorana texture zeros broadly fall into two categories: one-zero texture and two-zero texture. There are six different one-zero textures all of which are allowed by latest oscillation and cosmology data and hence we consider all of them in our analysis. Out of fifteen possible two-zero textures, only six are compatible with oscillation data and the Planck bound, as pointed out by \cite{Fritzsch:2011qv,Meloni:2014yea}. 

We have first derived the most general Majorana neutrino mass matrix in terms of the neutrino best fit values as well as the free parameters: lightest neutrino mass, Dirac CP phase and two Majorana phases. Comparing this mass matrix to a specific type of texture zero mass matrix we arrive at one or two complex constraints relating some or all of the free parameters. Since in case of one-zero texture we have only one complex and hence two real constraints but four free parameters, we assume the equality between two Majorana phases so that we can write them as a function of Dirac CP phase. In case of two-zero textures, we have two complex and hence four real constraints that allow us to find all the four free parameters numerically. We get several solutions for $(m_{\text{lightest}}, \delta, \alpha, \beta)$ all of which give $m_{\text{lightest}}$ of the same order of magnitude but different possible values of phases. We list one such set of solutions for each two-zero texture in table \ref{table-2zero1} and \ref{table-2zero2}. Since all the free neutrino parameters are numerically determined in this case, any future measurement of Dirac CP phase in neutrino experiments will verify or falsify some of these sets of solutions.

The summary of our baryon asymmetry results in one-zero texture models is given in table \ref{table3} and \ref{table4}. It can be seen from these tables that in the one-flavor regime, all one-zero textures can give rise to correct baryon asymmetry depending upon the hierarchy of light neutrino masses. However, in two flavor regime, only $G_1$ with NH and $G_{3,4,6}$ with IH can give rise to correct baryon asymmetry. In the three flavor regime, $G_1$ with NH and $G_{4,5,6}$ with IH can give rise to correct baryon asymmetry.

The summary of results in two-zero texture models are shown in table \ref{table5}. In one flavor regime, all the two-zero texture mass matrices except $B_3$ can give rise to correct $Y_B$, depending on the neutrino mass hierarchy. For two flavor regime, only $B_{2,3,4}$ with IH can give rise to the observed baryon asymmetry. Similarly, in three flavor regime $B_2$ with both IH, NH and $B_{3,4}$ with only IH can produce correct $Y_B$. Thus if $M_1 < 10^{12}$ GeV, then all the allowed two-zero textures except $B_{3,4,5}$ are disfavored in the light of baryon asymmetry. For the two-zero texture mass matrices that give correct baryon asymmetry, we also constrain the entries in the diagonal Dirac neutrino mass matrices as can be seen from figures \ref{fig8} to \ref{fig25}.

In all the tables mentioned above, the symbol $\checkmark$ ($\times$) is used when the baryon asymmetry $Y_B$ for a particular case in (not in) the range given by the Planck experiment. As we mention above, here we have tried to discriminate between all possible Majorana neutrino textures by demanding the observed baryon asymmetry to arise from leptogenesis through the CP violating decay of the lightest right handed neutrino. We have not only constrained the number of texture zero mass matrices, but also constrained the parameters of the neutrino mass matrix which are yet undetermined in experiments. We should however, note that although a certain number of texture zero mass matrices with a particular light neutrino mass hierarchy do not give rise to correct baryon asymmetry, this does not rule out that particular texture as there could be some other source of baryon asymmetry in the Universe. Our analysis in this work only provide a guideline for future works related to model building in neutrino physics attempting to understand the dynamical origin of neutrino mass and mixing.

\section*{Acknowledgments}
The work of M K Das is partially supported by the grant no. 42-790/2013(SR) from University Grants Commission, Government of India. 

\end{document}